\documentclass[twocolumn]{aastex631}
\usepackage{amsmath}
\usepackage{todonotes}

\makeatletter
\let\frontmatter@title@above=\relax

\newcommand*{\defeq}{\mathrel{\rlap{%
			\raisebox{0.3ex}{$\m@th\cdot$}}%
			\raisebox{-0.3ex}{$\m@th\cdot$}}%
			=}
\makeatother

\received{April 7, 2023}
\revised{May 30, 2023}
\accepted{June 18, 2023}
\published{August 8, 2023}
\shorttitle{Major Mergers during BCG assembly}
\shortauthors{Kluge \& Bender}

\begin{document}

\title{Minor Mergers are not enough:\\The importance of Major Mergers during Brightest Cluster Galaxy assembly}

\author[0000-0002-9618-2552]{Matthias Kluge}
\thanks{E-mail: mkluge@mpe.mpg.de}
\affil{University-Observatory, Ludwig-Maximilians-University, Scheinerstrasse 1, D-81679 Munich, Germany}
\affil{Max Planck Institute for Extraterrestrial Physics, Giessenbachstrasse, D-85748 Garching, Germany}

\author[0000-0001-7179-0626]{Ralf Bender}
\affil{University-Observatory, Ludwig-Maximilians-University, Scheinerstrasse 1, D-81679 Munich, Germany}
\affil{Max Planck Institute for Extraterrestrial Physics, Giessenbachstrasse, D-85748 Garching, Germany}

\begin{abstract}

We investigate the roles of major and minor mergers during brightest cluster galaxy (BCG) assembly using surface brightness profiles, line indices, and fundamental plane relations. Based on our own sample and consistently reanalyzed Sloan Digital Sky Survey data, we find that BCGs and luminous normal ellipticals (LNEs) have similar central velocity dispersions, central absorption line strengths, and central surface brightnesses. However, BCGs are more luminous due to their much larger radial extent. These properties result in a flattening of the Faber--Jackson and Mg$_{\rm b}$--luminosity relations above 10$^{10.6}$ L$_{\odot,g'}$. We use this effect to estimate an amount of 60--80\% of accreted and merged light in BCGs relative to LNEs, which agrees with results from cosmological simulations. We determine the contribution of this excess light (EL) at each radius from the difference between the surface flux profiles of BCGs and LNEs. It is small in the center but increases steeply to 50\% at $\sim$3\,kpc radius. The shape of these profiles suggests that BCGs could be formed from LNEs in three major merger processes. This is also consistent with the mild increase of the S\'ersic indices from $n\approx4$ to $n\approx6$, as confirmed in merger simulations. We note that minor mergers cannot be the dominant origin of the BCG's EL because they deposit too few stars at intermediate radii $r\lesssim20$\,kpc. The shape of the EL profile also explains a detected offset of 0.14\,dex of the fundamental planes for BCGs and LNEs relative to each other.

\end{abstract}


\section{Introduction} \label{sec:introduction}

Galaxies populate the 3D parameter space spanned by effective (half-light) radius $r_{\rm e}$, average effective surface intensity $\langle I_{\rm e}\rangle$ within $r_{\rm e}$, and central stellar velocity dispersion $\sigma_0$ on a thin 2D plane, called the fundamental plane \citep[FP;][]{Djorgovski1987,Dressler1987,Bender1992}:

\begin{equation}
	r_{\rm e}\propto \sigma_0^{\alpha} \langle I_{\rm e}\rangle^{\beta}.
\end{equation}

This relationship can be understood using (a) the definition of the average effective surface intensity

\begin{equation}
	\langle I_{\rm e}\rangle = \frac{L}{2\pi r_{\rm e}^2},
\end{equation}

where $L$ is the total luminosity of the galaxy, and (b) the relationship between central stellar velocity dispersion and total mass $M$ of the galaxy

\begin{equation}
	\sigma_0^2 = c \frac{M}{r_{\rm e}},
\end{equation}

which can be derived from the virial theorem. We define $c$ to include the gravitational constant $G$ and corrections for rotation, orbit anisotropy, and velocity dispersion gradients. Solving for $r_{\rm e}$ gives

\begin{equation}
	r_{\rm e} = \frac{1}{2\pi c'} \sigma_0^2 \langle I_{\rm e}\rangle^{-1}, \label{eq:fptheory}
\end{equation}

where $c'=c\frac{M}{L}$. In the case of constant $c'$, the predicted FP slopes are $\alpha=2$ and $\beta=-1$. In reality, we observe $1.2\lesssim\alpha\lesssim1.4$ and $-0.8\lesssim\beta\lesssim-0.7$. Hence, there must be systematic trends in the homology parameter $c$ and/or mass-to-light ratio $M/L$.

Previous FP studies have split their samples into brightest cluster galaxies (BCGs) and non-BCG ellipticals and discovered steeper FP slopes for the BCG subsamples \citep{vonderLinden2007,Samir2020}. BCGs are special, usually being the most central and most extended cluster galaxy. Their exceptionally large effective radii are due to their embedding in the intracluster light (ICL). The ICL is defined as the stellar component in a cluster, which is unbound from galaxies. Most of the ICL surrounds the central galaxy and the radial transition from stars bound to the central galaxy to high-velocity stars bound only to the cluster is smooth and photometrically not identifiable \citep{Kluge2021}. Therefore, for operational simplicity, we here consider the ICL to be part of the BCG, unless we say otherwise.

A promising explanation why BCGs are found to follow different FP relations is the presence of ICL, which makes these galaxies nonhomologous to normal Es. However, the previously discovered FP differences must be interpreted cautiously, because the structural parameters in the above-mentioned studies have been calculated using a de Vaucouleurs model \citep{DeVaucouleurs1948}, which severely underestimates the ICL. For this reason, we readdress this question with improved photometry.

Our current understanding of ICL formation is that it happens during the second phase of a two-phase scenario (e.g., \citealt{Naab2009,Oser2010,vanDokkum2010,Rodriguez2016}). First, the BCG's main body is formed at high redshift $z\gtrsim2$ by in-situ star formation. Later on below redshift $z<1$, the ICL is accreted by mergers, disruption of dwarf galaxies, tidally stripping the outskirts of intermediate-mass galaxies, and pre-processing in groups (see recent reviews by \citealt{Contini2021,Arnaboldi2022,Montes2022}).

There is consensus that the ICL is accreted from other cluster members, but the predicted progenitor stellar masses are debated. They range from $10^8$ to $10^{11}$\,M$_\odot$. Based on the low observed ICL metallicity, resolved stellar population analyses suggest lower progenitor masses of $\sim$10$^8$\,M$_\odot$ \citep{Williams2007,Lee2016} and spectroscopic analysis indicates $10^8-10^9$\,M$_\odot$ \citep{Gu2020}. On the other hand, semianalytic models imply that $10^{10}-10^{11}$\,M$_\odot$ galaxies contribute most to the ICL \citep{Contini2014}. In any case, these events are minor mergers when we acknowledge the high total ICL stellar mass of $\sim$10$^{12}$\,M$_\odot$ \citep{Kluge2021}.

We emphasize that these processes increase the size and luminosity of BCGs but not necessarily deposit all of the progenitor light into the ICL. Some stars bind to the original E, which is embedded in the ICL. Using photometry alone without detailed dynamical analysis over large radii, the relative contributions in BCGs from in-situ stars, accreted bound stars, and (accreted) ICL stars cannot be separated unambiguously \citep{Bender2015,Remus2017}.

Nevertheless, estimating the amount and distribution of the excess light (EL) relative to luminous normal Es is useful. The EL is what makes BCGs special and provides a lower limit on the accreted and merged-in stellar component.

If the two-phase formation scenario is correct, then the central BCG regions are only mildly affected by the second formation phase. Neither minor nor major mergers increase the surface brightness (SB) inside 0.02\,$r_{\rm e}$ \citep{Hilz2013}. Moreover, minor mergers do not impact the central velocity dispersion \citep{Hilz2012}, which makes it a good tracer of the original E. Hence, we propose to utilize a projection of the FP, the Faber--Jackson relation \citep{Faber1976}, to estimate the amount of EL by comparing the luminosities of BCGs to normal Es with the same central dispersion.

On the other hand, multiple major mergers can increase the central velocity dispersion by up to 50\% \citep{Hilz2012}. In that case, it would not qualify as a good tracer of the original E. Hence, we additionally use as an alternative probe the stellar populations in the galaxy center, which we expect to remain preserved during EL accretion at larger radii.

This paper is organized as follows. In Section \ref{sec:sample}, we define our galaxy samples. Our photometric measurements on Sloan Digital Sky Survey (SDSS) data of normal Es are detailed in Section \ref{sec:lowermass}. Section \ref{sec:obsstrat} describes the new spectroscopic observations of BCGs, data reduction, and measurements of the line-of-sight velocity distribution (LOSVD) as well as lick indices. Consistency checks are provided in Appendices \ref{sec:robustness} and \ref{sec:applick}. The FP fitting procedure is outlined in Section \ref{sec:fp}. A more detailed description can be found in Appendix \ref{sec:fitting3d}. We present our results in Section \ref{sec:results}, discuss them in Section \ref{sec:discussion} and conclude in Section \ref{sec:conclusions}.

Throughout the paper, we assume a flat cosmology with $H_0=69.6$ km s$^{-1}$ Mpc$^{-1}$ and $\Omega_{\rm m}=0.286$ \citep{Bennett2014}. Distances and angular scales were calculated using the web tool from \cite{Wright2006}. Virgo infall is not considered. Three types of flux corrections were applied: (1) dust extinction using the maps from \cite{Schlafly2011}, (2) K corrections following \cite{Chilingarian2010} and \cite{Chilingarian2012}, and (3) cosmic $(1+z)^4$ SB dimming. Magnitudes are always given in the AB system.

\section{Sample} \label{sec:sample}

Spectroscopic observations of 75 BCGs are obtained with the Low Resolution Spectrograph 2 (LRS2) on the Hobby Eberly Telescope (HET). The sample is selected from the photometric catalog of 170 low-redshift ($z\lesssim0.08$) BCGs of \cite{Kluge2020}. Galaxies are selected for observability with a minor bias toward brighter nuclei in order to improve the spectral signal-to-noise ratio (S/N). We extend the sample with 40 BCGs that have SDSS spectra available. For an overlapping subsample of 23 BCGs, both LRS2 and SDSS spectra exist. This enables consistency checks on the measured central stellar velocity dispersions $\sigma_0$ (see Appendix \ref{sec:robustness}). The total spectroscopic BCG sample consists of 115 galaxies. Photometric BCG parameters are not adopted from \cite{Kluge2020} but measured anew along the effective axis instead of the major axis. The effective axis is defined as $r=\sqrt{ab}$, where $a$ is the semimajor axis radius and $b$ is the semiminor axis radius. The reason for the modification is that we need to be consistent in the analysis of all galaxies in this work. Radially varying ellipticity profiles can bias the SB profile curvature and other S\'ersic profile \citep{Sersic1968} parameters when they are measured along a different axis.

The main goal of this paper is to investigate the role of major and minor mergers in the assembly of BCGs using SB profiles, central velocity dispersions, line indices, and FP analysis. Therefore, we need a reference sample of non-BCGs, which we refer to as ``normal Es". \cite{Zhu2010} provided a catalog of 1923 SDSS-DR6-selected elliptical galaxies. Their selection constraints are as follows:

\begin{itemize}
	\item spectroscopically confirmed redshifts $z<0.05$,
	\item brighter than $M<-19$ $r$ mag,
	\item central velocity dispersion $\sigma_0 > 70$ km s$^{-1}$ (limited by the same spectral resolution of the SDSS spectrograph),
	\item bulge-to-total light ratio $B/T>0.7$,
	\item ellipticity $\epsilon<0.6$,
	\item location on the red sequence, and
	\item featureless appearance.
\end{itemize}

The environments are diverse, with 347 elliptical galaxies in the field, 682 in poor groups, and 706 in rich groups.

We query the SDSS-DR17 \citep{Abdurrouf2021} photometric parameters and velocity dispersions using the python tool {\tt astroquery} \citep{Ginsburg2019}. Furthermore, we also download the SDSS DR17 images and spectra and refit the parameters with our algorithms. After discarding problematic cases (see Section \ref{sec:photprocedure}), we end up with a sample of 1420 SDSS galaxies. Visual inspection reveals a very small contamination by $\sim$1--2\% by late-type Sa galaxies and by $\sim$1\% by merging galaxies.

\section{Photometry of Normal Ellipticals}\label{sec:lowermass}

\begin{figure*}
	\includegraphics[width=\linewidth]{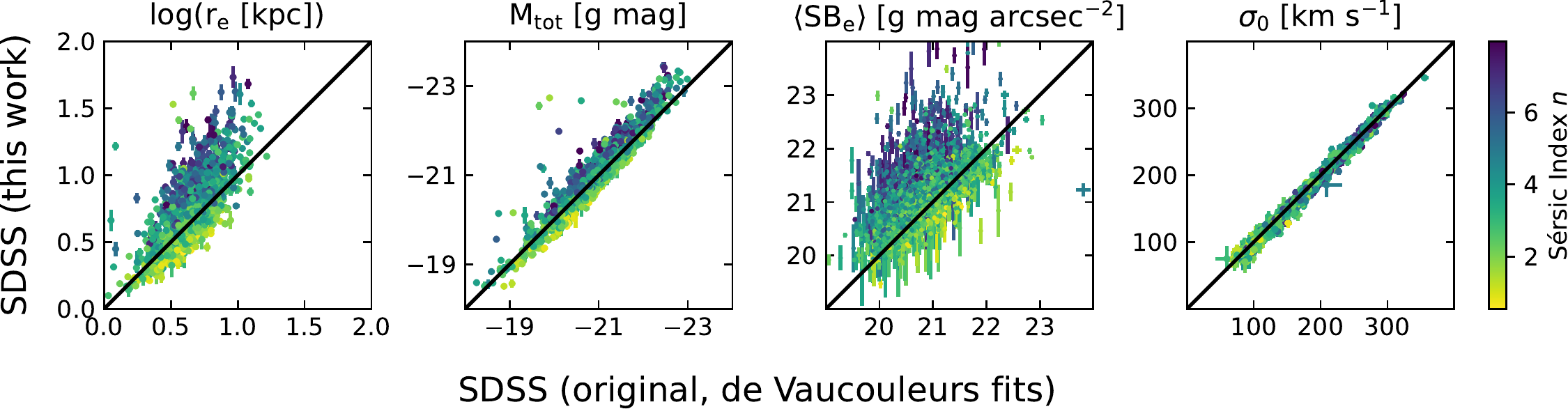}
	\caption{Comparison between photometric and kinematic parameters of 1420 normal Es. From left to right: effective radius $r_{\rm e}$, total brightness $M_{\rm tot}$, average effective surface brightness $\langle{\rm SB_e}\rangle$, and central (uncorrected; see Section \ref{sec:sigmae8}) stellar velocity dispersion $\sigma_0$. Photometric parameters based on SDSS De Vaucouleurs models are plotted along the $x$-axes. Our nonparametric measurements on the same data are plotted along the y-axes. We calculate them by directly integrating the light profiles. S\'ersic functions are only used for extrapolation. Best-fit S\'ersic indices $n$ are color-coded. Galaxies with $n>8$ are discarded (see Section \ref{sec:photprocedure}). The velocity dispersions in the fourth panel are published SDSS values (see Appendix \ref{sec:sigmacomparison}) along the $x$-axis and our fits to the same spectra using pPXF are plotted along the $y$-axis. \label{fig:devsdss_mysersic}}
\end{figure*}

\begin{figure*}
	\includegraphics[width=\linewidth]{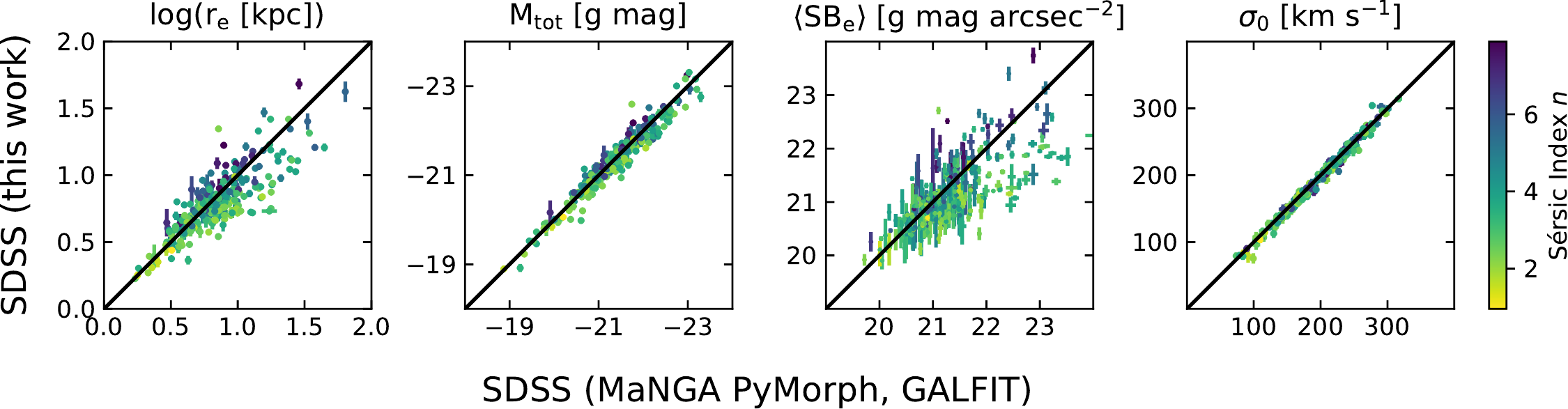}
	\caption{Same as Figure \ref{fig:devsdss_mysersic}, but the SDSS de Vaucouleurs models are replaced by parametric MaNGA S\'ersic models. The sample consists of 245 normal Es. \label{fig:devsdss_mysersic_manga}}
\end{figure*}

\begin{figure*}
	\includegraphics[width=\linewidth]{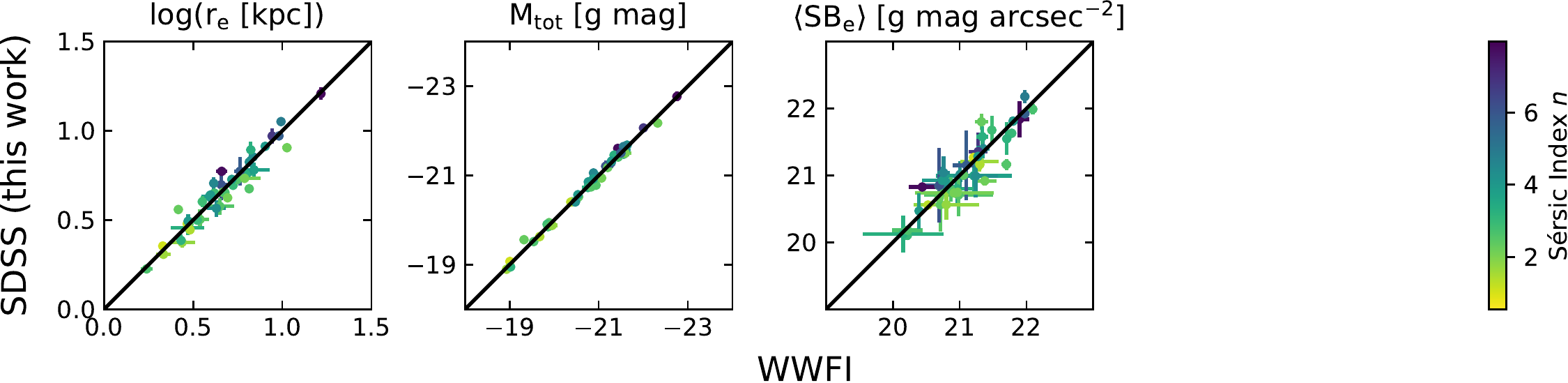}
	\caption{Same as Figure \ref{fig:devsdss_mysersic}, but we compare our nonparametric measurements of 54 normal Es based on deep WWFI ($x$-axis) and shallower SDSS ($y$-axis) imaging data.  \label{fig:devsdss_mysersic_wwfi}}
\end{figure*}

The SDSS photometric catalogs are often utilized as a resource to study the FP or other galaxy scaling relations (e.g., \citealt{Bernardi2003,Saulder2013,Joachimi2015,Samir2020,Singh2021}). After it became known that SDSS data before DR8 had severe sky oversubtraction issues (\citealt{Bernardi2007b,Lauer2007a,vonderLinden2007,AdelmanMcCarthy2008,Abazajian2009}, amongst others), various authors have tried to correct these effects \citep{Bernardi2007a,vonderLinden2007,Hyde2009a} or limited their studies to the bright, inner galaxy regions (\citealt{vonderLinden2007}: ${\rm SB}<23~r~{\rm mag~arcsec^{-2}}$; \citealt{Liu2008}: ${\rm SB}<25~r~{\rm mag~arcsec^{-2}}$). For SDSS DR8, the data reduction was improved significantly \citep{Aihara2011,Blanton2011}.\footnote{The reduced images remain unchanged after DR8. For DR9, only an astrometry error was fixed and for DR13, the photometric calibration was improved on the order of 0.01\,mag. This affects only the catalog data.} But still, the problem is not fully solved. The official photometric galaxy parameters are measured using a de Vaucouleurs model \citep{Stoughton2002} and continue to over- or underestimate intrinsic galaxy parameters \citep{Bernardi2013,DSouza2015,Fischer2017,Miller2021}. We remeasure the SB profiles of all SDSS Es from the selected sample and fit them using S\'ersic profiles (for details, see Section \ref{sec:photprocedure}). The structural parameters (apart from the S\'ersic index $n$) are calculated by integrating the (extrapolated) SB profiles. Hence, they are largely model-independent. Figure \ref{fig:devsdss_mysersic} shows a comparison between the SDSS de Vaucouleurs model parameters on the $x$-axis and our parameters on the $y$-axis. Galaxies are color-coded based on their best-fit S\'ersic index $n$. There is good agreement for galaxies, whose SB profile is consistent with a de Vaucouleurs profile ($n=4$, green), but the radii and brightnesses are overestimated for galaxies with $n<4$ (yellow) and underestimated for galaxies with $n>4$ (blue). This obvious behavior had already been demonstrated \citep{Bernardi2007a,Hyde2009a}.

For this reason, many authors performed better S\'ersic fits with independent algorithms \citep{Bernardi2007a,Bernardi2020,LaBarbera2010b} or corrected Petrosian quantities to approximate the parameters derived from S\'ersic fits \citep{Desroches2007}. These fits are usually performed using 2D image fitting with a single seeing-convolved S\'ersic model \citep{LaBarbera2008a,Magoulas2012,Zahid2015,Bernardi2020}. Other than being inadequate in the presence of ellipticity gradients or miscentering, such a parametric fitting also assumes that the SB profile follows the model all the way to the center. Here is where SB profiles of elliptical galaxies usually deviate the most from S\'ersic profiles \citep{Kormendy2009}. Two-component fits \citep{Fischer2017,Dominguez2021} can partly account for these effects. Another problem is that the formal uncertainties highly underestimate the real parameter uncertainties (\citealt{Peng2010}; Figures \ref{fig:devsdss_mysersic}--\ref{fig:devsdss_mysersic_wwfi}).

\subsection{Procedure} \label{sec:photprocedure}

The imaging dataset for the normal E sample is the SDSS DR17 $g$-band images. On them, we apply a mostly nonparametric approach of calculating the galaxies' structural parameters \citep{Kluge2020,Kluge2023}. It consists of four steps.

\begin{figure*}[t]
	\includegraphics[width=\linewidth]{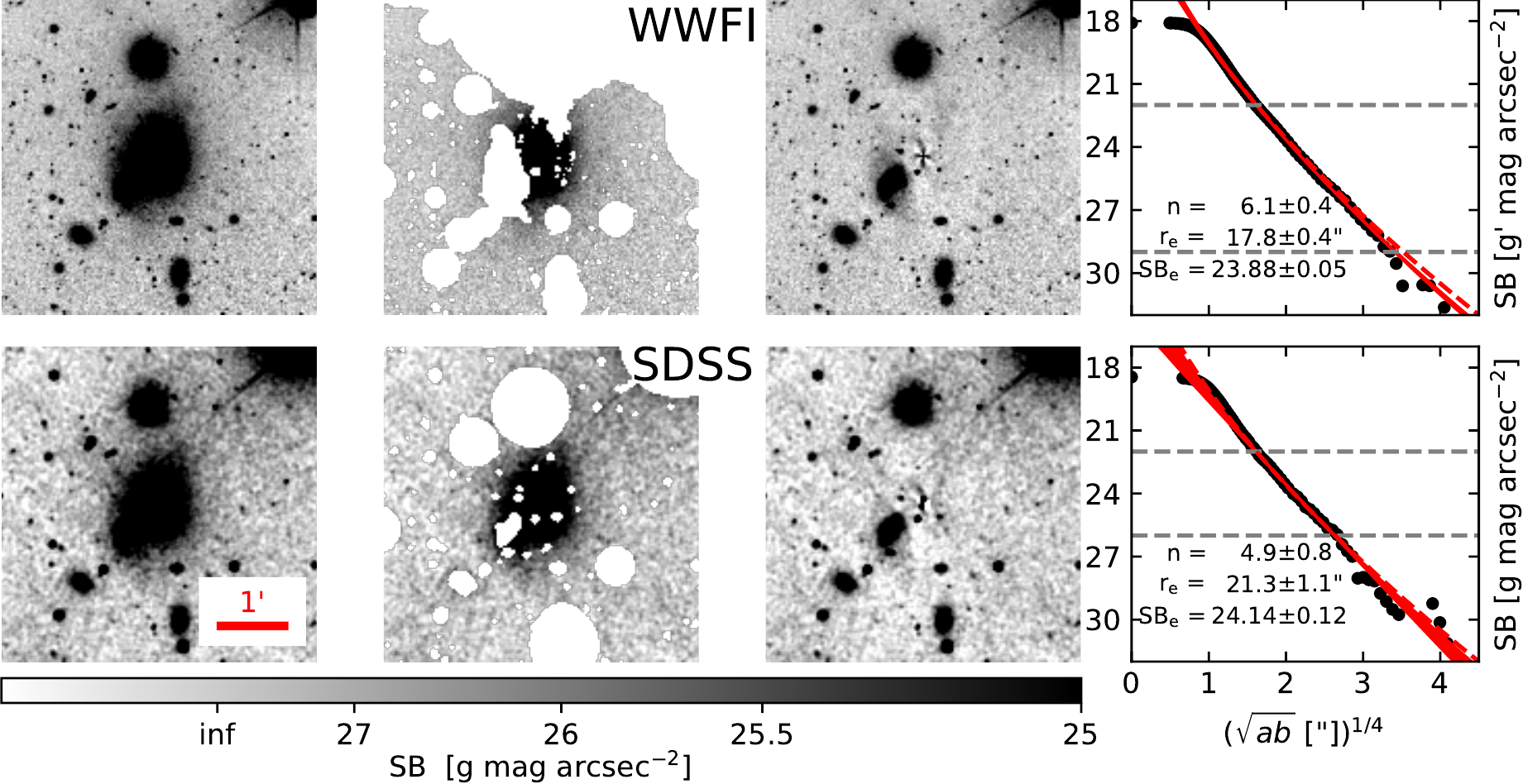}
	\caption{SB profile measurement for MCG+07-34-043 at redshift $z=0.029$. Top row: deep WWFI data; bottom row: shallower SDSS data. Left panel: observed image of the galaxy. The full image size is $3\times3$ times larger than shown. The images are smoothed using a 2D Gaussian kernel with 3 pixel standard deviation. This corresponds to 0.6\arcsec~for the WWFI images and 1.2\arcsec~for the SDSS images. North is up, east is left. Second panel: automatically masked image combined with a manually refined mask only for the WWFI data. Third panel: residual after subtracting an isophote model. Residuals at $\sim0.5\arcmin$ from the center toward the bottom left and top right arise from real isophotal distortions due to interactions with neighboring galaxies. Right panel: measured SB profile (black points) and best-fit S\'ersic function (red continuous line). The 1$\sigma$ uncertainty of the fit is given by the line width. Gray dashed lines enclose the fitting range. Best-fit S\'ersic parameters are given in the text labels. The red dashed line is the MaNGA single S\'ersic fit \citep{Dominguez2021}: $n=6.49\pm0.05$, $r_{\rm e}=23.15\pm0.44\arcsec$, ${\rm SB_e}=24.34\pm0.04~g$ mag arcsec$^{-2}$. \label{fig:isopy_example}}
\end{figure*}

First, all sources apart from the galaxy of interest are masked. The masking parameters for the algorithm presented in \cite{Kluge2020} are here optimized for the shallower SDSS-DR17 images. Four masks are combined, which are created using different sets of parameters, optimized for various object sizes and brightnesses. The SB thresholds range from 24--26 $g~{\rm mag~arcsec^{-2}}$ and all masks are convolved to a diameter of up to 12\arcsec. Figure \ref{fig:isopy_example}, bottom row, second panel, shows that the automatically created mask covers well all small sources. On the other hand, the masks are less complete for large and diffuse sources. One example is the extended point-spread function (PSF) wing of the bright foreground star in the top-right corner. It is more carefully masked in the WWFI data (top row, second panel). A comparison of the measured SB profiles (right panels) demonstrates that the impact is negligible on the relevant part of the galaxy SB profile, which is brighter than the faint cut (lower gray dashed line). More details about this comparison are given in Section \ref{sec:comparison_imaging}.

Second, we fit ellipses to the galaxy isophotes and average the surface flux along these ellipses. Differently than for the BCG sample in \cite{Kluge2020}, we perform the isophote fitting now with the easier automatable {\tt python} tool {\tt photutils} \citep{Bradley2021} instead of {\tt ellfitn} \citep{Bender1987}. Apart from the different definitions of higher-order moments for deviations from perfect ellipses, the algorithms produce consistent results according to our tests. Beyond $r>3\arcsec$, we increase the S/N by averaging the surface flux in annuli around the fitted ellipses. The residual background constant is inferred as the value to which the surface flux profile converges. We determine it by the median of the outermost 10 data points after kappa--sigma clipping values that deviate by more than two times the median absolute deviation from the median. The outermost data point is defined to be 13 radial steps beyond the radius, where the initial SB profile falls below ${\rm SB}=28~g~{\rm mag~arcsec^{-2}}$. The step size is 10\%, that is, every radius is 10\% larger than the previous one. This procedure is motivated by the depth of the SDSS images. The data points must lie sufficiently far away from the detectable faint galaxy outskirts, close enough to the galaxy to avoid large-scale background inhomogeneities, and the annulus must be large enough to determine a statistically robust value.

Third, due to the limited depth of the SDSS images, we extrapolate the SB profile beyond ${\rm SB}\ge26~g~{\rm mag~arcsec^{-2}}$ to infinity using a S\'ersic function that is fitted to the intermediate SB profile between $22<{\rm SB}<26~g~{\rm mag~arcsec^{-2}}$. The central $r<2\arcsec$ are excluded because of seeing contamination (see, e.g., Figure 9 in \citealt{Kluge2020}) and possible intrinsic deviations from a S\'ersic profile \citep{Kormendy2009}. To improve the robustness of the S\'ersic-profile fitting, we iterate twice after kappa--sigma clipping outliers, which deviate by more than 2.7 standard deviations from the initial best-fit profile. We demonstrate the merging of the measured and extrapolated profiles for one exemplary galaxy in Figure \ref{fig:isopy_example}, right panels. The measured SB profile is shown by the black dots, and the red line is the best-fit S\'ersic function. It is fitted to the region enclosed by the two gray dashed lines. The final profile is created by merging the measured profile above the lower gray dashed line and the best-fit S\'ersic function below it.

Finally, we integrate the merged SB profile to calculate the structural parameters $r_{\rm e}$, SB$_{\rm e}$, and $M_{\rm tot}$. As opposed to the parametric 2D image fitting, this approach takes into account the radially varying ellipticity as well as deviations of the inner SB profile from a perfect S\'ersic function. Similar approaches were adopted in previous studies (e.g., \citealt{Cappellari2006,Nigoche-Netro2007,vonderLinden2007,Donofrio2008,Kormendy2009}). The S\'ersic index $n$ is taken from the best-fit S\'ersic function.

One further modification to \cite{Kluge2020} is that we do all analyses along the effective axis $r=\sqrt{ab}$, where $a$ is the semimajor axis radius and $b$ is the semiminor axis radius. This also affects the WWFI BCG sample, for which we refit S\'ersic functions along the effective axis. Further details are given in Appendix \ref{sec:measuncert}.

Moreover, PSF broadening by the extended PSF wings is not corrected for the normal E (SDSS) sample. Our de-broadening algorithm is not suited for relatively compact galaxies because the inner galaxy region, where the approximation of negligible scattered light is not fulfilled (see Figure 12 in \citealt{Kluge2020}), makes up a significant fraction of the galaxies. We show in Section \ref{sec:kappa} that this effect is fortunately negligible.

In total, 1849 out of 1923 measurements and fits are apparently successful. Yet, the sample needs to be cleaned of bad fits. In agreement with \cite{Meert2015}, we find that high S\'ersic indices ($n>8$) are often associated with automatic background-subtraction errors, e.g., due to neighboring bright objects or the residual stripe-pattern along the SDSS drift-scanning direction. Since the sample is large enough, we discard all 308 galaxies with such problems. Furthermore, we discard 89 galaxies with uncertainty $\delta\log(r_{\rm e})>0.12$\,dex, 23 galaxies with velocity dispersion $\sigma_0<70$\,km\,s$^{-1}$, two galaxies with contaminated spectra, the three faintest outliers ($M>-18\,g\,{\rm mag}$), and four BCGs, which overlap with our BCG sample. Finally, we end up with a clean sample of 1420 galaxies.

\subsection{Comparison to Deeper Images and Literature} \label{sec:comparison_imaging}

A possible source of systematic error is the limited SDSS survey depth of about 26--27 $g$ mag arcsec$^{-2}$ for extended sources. For instance, if the SB profile slope, that is, the S\'ersic index depends on the imaging depth, then the effective radius and average effective SB also depend on it via parameter correlations. We examine this dependency for the 54 normal Es, which are coincidentally in the field of view of the WWFI BCG imaging survey. The latter has a deeper SB limit of 30 $g'$ mag arcsec$^{-2}$ for low-redshift BCGs.

Figure \ref{fig:isopy_example} shows a comparison for the galaxy MCG+07-34-043 between the WWFI data (top row) and SDSS data (bottom row). All WWFI masks are taken from \cite{Kluge2020}, visually examined in an area $\pm400\arcsec$ around the target galaxies, and manually improved if necessary. The measured SB profiles agree well inside the SDSS fitting region $22<{\rm SB}<26~g~{\rm mag~arcsec^{-2}}$. Moreover, the WWFI data points follow the best-fit S\'ersic profile (continuous red line) smoothly down to the limiting WWFI SB of ${\rm SB}<29~g'~{\rm mag~arcsec^{-2}}$ (lower gray dashed line). Again, we fit the SB profiles only inside of the SB region constrained by the gray dashed lines and extrapolate it beyond the faint cut. We have chosen a brighter limiting SB for the smaller-sized ellipticals compared to the more extended BCGs because the flux S/N is lower in the outermost isophotes.

The best-fit S\'ersic index $n$ and the structural parameters $r_{\rm e}$ and ${\rm SB_e}$, which are calculated by integrating the (extrapolated) SB profiles, are annotated in the figure labels. Not all of them agree within the quoted $1\sigma$ uncertainties due to intrinsic variations in the SB profile and strong covariances between the parameters. However, the comparison between 54 galaxies in Figure \ref{fig:devsdss_mysersic_wwfi} shows that the shallower SDSS SB limit introduces no systematic bias on the structural parameters.

The profile of the best-fit {\tt GALFIT} S\'ersic model of MCG+07-34-043 \citep{Fischer2017,Dominguez2021} is overplotted as the red dashed line in Figure \ref{fig:isopy_example}. It agrees well in the inner region but slightly overestimates the SB at larger radii. The assumption of a pure S\'ersic profile is not fully satisfied when the nuclear region is included in the fit. Moreover, the contaminating bright, neighboring star can lead to an overestimation of the outer galaxy halo.

In Figure \ref{fig:devsdss_mysersic_manga}, we compare our structural parameters to those obtained by \cite{Fischer2017} and \cite{Dominguez2021}. Both samples overlap for 245 galaxies. The authors fit both a S\'ersic model and a combination of one S\'ersic plus one exponential model directly to the SDSS-DR14 images using the software {\tt GALFIT} \citep{Peng2002,Peng2010}. All models are seeing-convolved. As recommended, we use the provided catalog parameter {\tt FLAG\_FIT} to choose the preferred of the two models. Uncertainties of the combined parameters are only provided for the single S\'ersic fits.

There is a good agreement for the total brightnesses. Effective radii and effective SBs scatter with a very slight tendency for larger values in the MaNGA catalog. As expected, the correlation between offset direction and S\'ersic index almost vanishes.

\begin{figure*}
	\includegraphics[width=\linewidth]{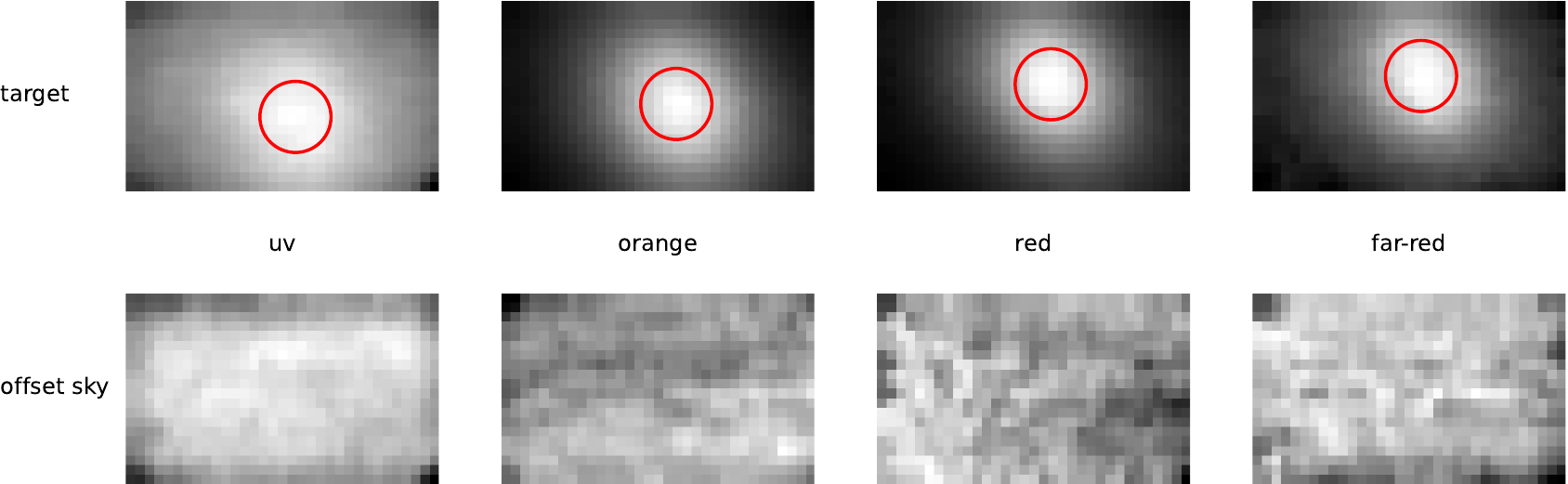}
	\caption{Collapsed LRS2 data cubes of the BCG in A1982. The top row shows the target pointings, and the bottom row shows the parallel sky pointings. The channel names are given by the text labels. All images have different flux scaling to visualize structures. The manually chosen galaxy center is marked by the red circle in each panel. It also corresponds to the aperture with 1.5\arcsec~radius, in which the spectra are averaged. Each image has a size of $6\arcsec\times12\arcsec$. \label{fig:obs}}
\end{figure*}

\section{Spectroscopic Observations} \label{sec:obsstrat}

The observations were obtained with the low-resolution spectrograph LRS2 \citep{Chonis2014,Chonis2016} on the 11\,m Hobby Eberly Telescope at the McDonald Observatory, Texas between 2017 April and 2020 April. The instrument contains four IFUs, which cover different wavelength ranges (``uv", ``orange", ``red", ``far-red"; see also Table \ref{tab:intrumentalresolution}). We refer to them as ``channels". All IFUs observe simultaneously, but the uv and orange channels point 100\arcsec~offset in the negative azimuthal direction from the red and far-red channels. Hence, offset-sky exposures (Figure \ref{fig:obs}, top panels) are obtained in the red and far-red channels, while the uv and orange channels point on the target (Figure \ref{fig:obs}, bottom panels) and vice versa. The IFU field of view is 12\arcsec~by 6\arcsec~on the sky. Such a small field rarely overlaps with bright sources.

On-target and offset-sky integration times are $\sim$10+10\,minutes for each channel. No small-scale dithering is required because 98\% of the field of view is covered by the 280 fibers per channel. Each fiber covers a circle with 0.6\arcsec~diameter on the sky. The resulting data cubes have a spaxel size of 0.4\arcsec.

All IFUs are oriented approximately parallel to the horizon. Hence, the position angle with respect to the equatorial coordinate system varies for each target. 

The mean PSF FWHM is determined using the standard stars, which were observed at different times during the nights. Gaussian fits and object-by-object scatter give $2.04\pm0.56\arcsec$ for the uv, $1.95\pm0.57\arcsec$ for the orange, $1.91\pm0.50\arcsec$ for the red, and $2.00\pm0.56\arcsec$ for the far-red IFUs. Seeing variation due to different airmasses is negligible because the telescope observes at an almost fixed elevation of $55\pm4\degr$ \citep{Hill2004}.

\begin{deluxetable}{llcccc}[!t]
	\tabletypesize{\scriptsize}
	\tablecaption{Instrumental resolutions of the observed and template spectra. \label{tab:intrumentalresolution}}
	\tablehead{
		\colhead{Instrument} & \colhead{Channel} & \colhead{FWHM} & \colhead{$\sigma$} & \colhead{$\lambda_{\rm min}$} & \colhead{$\lambda_{\rm max}$} \\
		\colhead{} & \colhead{} & \colhead{(\AA)} & \colhead{(km\,s$^{-1}$)} & \colhead{(\AA)} & \colhead{(\AA)}
	}
	\startdata
	LRS2       & uv     & 1.70 & 53  & 3690$^{*}$ &  4550$^{*}$ \\
	LRS2       & orange & $4.26-4.66$ & 104 & 4760$^{*}$ &  6185$^{*}$ \\
	LRS2       & red    & 3.21 & 56  & 6450  &  8240$^{*}$ \\
	LRS2       & far-red & 3.84 & 53  & 8275  & 10325$^{*}$ \\
	SDSS       & \ldots & $2.52\pm0.16$ & $70\pm4$ & 3780  &  5400  \\
	\hline
	X-shooter  & uvb    & 0.45 & 13  & 2990  &  5560  \\
	X-shooter  & vis    & 0.50 &  8  & 5338  & 10200  \\
	MILES      & \ldots & 2.54 & $78-63$  & 3500  &  7429  \\
	\enddata
	\tablecomments{Values marked with an * refer to the observed frame instead of rest frame. Observed spectra are redshifted by $z$, so the resolution improves with $1/(1+z)$. The instrumental resolution $\sigma$ [km\,s$^{-1}$] is calculated in the wavelength interval where the template is fitted. Two values are given for LRS2 orange, because the spectrum is split for the X-shooter uvb and vis stellar libraries. SDSS spectra are cropped to $\lambda<5400$\,\AA~for velocity dispersion measurements. For line strength measurements, we use the full spectra with $\lambda<9200^*$\AA.}
\end{deluxetable} \vspace{-4.375mm}

The instrumental wavelength resolution (Table \ref{tab:intrumentalresolution}) is determined by broadening high-resolution X-shooter template spectra \citep{Arentsen2019,Gonneau2020} to fit the Lick standard stars spectra, which we observed with the LRS2 (Section \ref{sec:licks}). Thereby, we assume that the intrinsic line broadening is negligible compared to the instrumental broadening. The instrumental broadening of the template stars is quadratically added to the measured broadening. For SDSS spectra, we use 678 random stars with ${\rm S/N}>50$. These spectra are critically sampled, which means logarithmic wavelength pixels are often larger than the standard deviation of the Gaussian instrumental broadening, given in the FITS file extension {\tt wdisp}. This pixelation additionally broadens the spectra \citep{Law2021}. We simulate realistic scenarios by convolving the stellar spectra to $\sigma_0=70{\rm~km s^{-1}}$ in order to mimic the lowest velocity dispersions in our sample. A correction ${\tt FWHM}_{\rm new} = 0.78{\tt~FWHM_{old}}+0.62$ provides consistent results.

\subsection{Data Reduction} \label{sec:datareduction}

The raw data are reduced with the Panacea pipeline\footnote{\url{https://github.com/grzeimann/Panacea}} developed by Greg Zeimann. It includes master-bias subtraction, cosmic-ray masking, fiber extraction, air-wavelength calibration, flux calibration, and corrections for atmospheric differential refraction. For our analysis, we use the calibrated data cubes for the target and sky pointings. The remaining steps are manual galaxy center selection, aperture fiber averaging, and sky subtraction using the software pPXF \citep{Cappellari2004,Cappellari2017}. 

Galaxy centers are selected manually. They vary from object to object by $\sim$1\arcsec~because of pointing inaccuracy. Moreover, there is also a channel-by-channel variation of the galaxy centers on the same order, partly because atmospheric differential refraction is corrected within each data cube but not between the four independent channels. In Figure \ref{fig:obs}, we indicate the selected galaxy centers for the BCG in A1982 by the red circle. This circle is also the aperture, in which the science spectra are averaged. Its 1.5\arcsec~radius is chosen to be consistent with the SDSS comparison sample.

As mentioned in Section \ref{sec:obsstrat}, the target and offset-sky pointings were not observed simultaneously. Therefore, the sky spectrum needs to be rescaled. The dominating reason is the varying effective mirror size of the telescope along an observation track. The mirror is fixed in elevation but the fiber pick off moves to keep track of the target motion on the sky and, thereby, observes a varying effective mirror size \citep{Hill2004}. Minor effects are also spatial and time variations of the sky brightness. These different scalings propagate into sky-subtraction errors. A convenient solution is to leave the sky scaling factor as a free parameter during template fitting (see Section \ref{sec:templatefitting}). Strongly varying night-sky emission lines are automatically clipped by the $\kappa\sigma$ feature in {\tt pPXF}. All sky spectra inside one data cube are averaged and then subtracted from the aperture-averaged galaxy spectrum. Our tests showed that averaging all sky spectra, not only those that fall inside the galaxy aperture, improves the S/N so much such that it outweighs the introduction of systematics from spatially varying line profiles. Occasionally appearing objects in the sky pointings are manually masked before averaging the sky spectra.

\subsection{Template Fitting}\label{sec:templatefitting}

Our goal is to measure robust stellar velocity dispersions of the galaxies. This is done by finding a linear combination of template star spectra, which resembles the galaxy spectrum well. This template is then convolved with the a priori unknown LOSVD. It is approximated by a Gaussian with mean $v$, standard deviation $\sigma_0$, and two higher-order Gauss--Hermite terms $h_3$ and $h_4$ to include the skewness and kurtosis, respectively (e.g., \citealt{Bender1994}). The fit is performed by iterating the weighting coefficients of the template spectra, LOSVD parameters, sky spectrum scaling, and multiplicative polynomials to match the continua of templates and science spectra.

The calibrated, aperture-averaged science spectra are fitted using pPXF. As templates, we choose the X-shooter stellar library DR2 \citep{Arentsen2019,Gonneau2020}. We only fit a linear combination of individual stars instead of stellar population models, because template mismatch is much smaller in the first case.

All X-shooter template star spectra are resampled to the same rest-frame wavelength grid using cubic spline interpolation. This is necessary because their radial velocities are known but they have not been red- or blueshifted to zero velocity on the pixel grid. Next, all template spectra are convolved by Gaussian kernels (and oversampled by a factor of 2) to match the fixed LRS2 or variable SDSS instrumental resolution (see Section \ref{sec:obsstrat} and Table \ref{tab:intrumentalresolution}). The instrumental resolution improves with (1 + $z$), because the spectra get stretched with increasing redshift $z$.

In the optical, the noise of the science and sky spectra increases blueward. Weighting these regions less improves the robustness of the fit and allows for cleaner outlier clipping. This is done by providing {\tt pPXF} with a noise spectrum. It is determined for each science spectrum by fitting it with template spectra and measuring the median absolute deviation of the residuals in 10 wavelength bins. Linear interpolation results in a smooth noise spectrum. With this noise estimate, the galaxy spectrum is fitted again using the template and sky spectra. Outliers, especially night-sky emission line residuals, are clipped iteratively if they deviate by more than 4.5 times the local noise from the best-fit template.

We mask the region between $4123{\rm \AA} < \lambda < 4173{\rm \AA}$ (observed frame) because of a time-independent, systematic flux-calibration error. Moreover, the NaD lines are excluded from the fits, because they can be contaminated by absorption in the interstellar medium and therefore may not be reliable tracers of the stellar kinematics (e.g., \citealt{Bender2015}).

Mismatch between the continua of the science and template spectra is corrected by multiplying the template spectra with fourth-order multiplicative polynomials. The coefficients are fitted simultaneously with the template weights. For a discussion of our choice of the polynomial order and rejection of additive polynomials, see Appendix \ref{sec:robustness}.

As mentioned in Section \ref{sec:obsstrat}, the effective mirror size of the Hobby Eberly Telescope varies along an observation track. The sky spectrum is taken at a different time than the science spectrum. Therefore, the sky spectrum is different between both. Conveniently, {\tt pPXF} allows fitting the scaling factor for the sky spectrum simultaneously with the template weights and polynomials.

All UV and OR spectra are shown in Appendix \ref{sec:spectra} as black lines. The red lines are the best-fit templates. Many BCGs show strong gas-emission lines at their nucleus (green peaks). Prominent examples are A262, A1795, A2052, and A2199. Those lines are modeled assuming independent gas kinematics. Both narrow and broad AGN components are fitted simultaneously with the other parameters.

For line strength measurements (see Section \ref{sec:licks}), we also make use of the red and infrared channels. Telluric absorption severely impacts the spectra at these wavelengths. We correct for it using white dwarf spectra, which were observed ideally during the same night. They are well approximated by a blackbody, depending only on surface temperature and normalization. We fit a black body function to the spectral regions devoid of white dwarf or telluric absorption lines and interpolate the affected regions using the fit. The obtained function needs to be normalized such that it can be used as a telluric absorption correction. We remove the continuum by dividing it by the best-fit blackbody function. Finally, all science and sky spectra are divided by the correction function calculated from the nearest white dwarf observation.

\subsection{Lick Indices}\label{sec:licks}

Lick line indices are absorption line strengths measured typically in units of angstrom equivalent width (e.g., \citealt{Worthey1994}). The lines have exactly defined pseudo-continua blueward and redward of them, to which a linear function is fitted. The spectral region in consideration is then divided by this function, and the flux inside a defined region bracketing the line is integrated. We utilize the python package {\tt pyphot} \citep{Fouesneau2022} for this procedure.

Lick indices must be measured at a fixed, but wavelength-dependent resolution of $\sigma_{\rm Lick}\approx210{\rm \,km\,s^{-1}}$. This avoids biases when broadened line wings leak differently into the regions where the pseudo-continua are normalized.

The common approach for galaxies with lower velocity dispersions than the Lick resolution is to convolve the galaxy spectrum to the Lick resolution, while considering the galaxy velocity dispersion and instrumental resolution. For galaxies with higher dispersion, an empirical correction must be applied. 

We follow a different approach by handling the galaxies with higher $\sigma_0$ exactly the same way as those with lower $\sigma_0$ than the Lick resolution. The reason is that a small systematic effect comes from the quantized resolution degradation by logarithmically rebinning of the spectra to 30\,km\,s$^{-1}$ wide velocity bins. Our tests showed that the effect is on the order of $\lesssim1\%$. Although it is small, it is nevertheless preferable to measure lick indices directly on the linearly binned spectra.

First, we take the linearly binned, best-fit template $T$. After it is convolved to the LRS2 instrumental resolution, we refer to it as $T*R$. It is then convolved again to (\textit{not by!}) the Lick resolution. We refer to it now as $T*R*L$. The reference line index $I_{\rm lin}(T*R*L)$ is measured on it. This template was not broadened by the stellar velocity dispersion and therefore provides a reference value.

Next, $T*R$ is logarithmically rebinned to a velocity grid of 30\,km\,s$^{-1}$ and convolved with the measured velocity dispersion of the galaxy. We refer to it as $T*I*\sigma_0$. The line index $I_{\rm log}(T*R*\sigma_0)$ is now also measured for the broadened template spectrum. The correction factor $C$ is calculated as

\begin{equation}
	C(T,\sigma_0) = \frac{I_{\rm lin}(T*R*L)}{I_{\rm log}(T*R*\sigma_0)}.
\end{equation}

All line indices, which are measured on the galaxy spectra, are multiplied by a factor defined in this way. It is determined for each line individually.

Fitted gas-emission lines are subtracted beforehand. Especially important in this context is the [N I] $\lambda\lambda5197.9,5200.4$ line doublet, which can severely contaminate the Mg$_{\rm b}$ absorption line by 0.4--2\,\AA~\citep{Goudfrooij1996}. Furthermore, gas-emission line subtraction is especially delicate for hydrogen lines, because there can be overlapping gas emission and stellar absorption. Wrongly fitted gas emission can compensate for template mismatch in the age-sensitive hydrogen lines. We examine this effect by comparing all H$\beta$ emission and absorption line strengths of the SDSS galaxy sample. If gas emission was introduced to compensate for template mismatch, we would expect an anticorrelation between gas emission and absorption line strengths. In reality, we measure a very low Pearson correlation coefficient of $R=0.08$ and conclude that template mismatch compensation by adjusting emission lines is negligible.

Line index uncertainties are calculated using 100 Monte Carlo simulations. Random evaluations of the measured noise (see Section \ref{sec:templatefitting}) are added to the best-fit template spectrum $T*R*\sigma_0$. In a few unfortunate cases, residuals of strong sky emission lines contaminate the line index measurements. These residuals are added to the noise estimate beforehand if they are at least three times larger than the local noise. Finally, the indices are measured on these 100 spectra, and their standard deviation is used as the line index uncertainty.

All measured line strengths are listed in Appendix \ref{sec:apptables}.

\section{Fundamental Plane} \label{sec:fp}

\subsection{Parameters}

The FP relates the effective radius $r_{\rm e}$, the average effective surface brightness $\langle{\rm SB_e}\rangle$ inside $r_{\rm e}$, and the aperture-corrected (see Sec. \ref{sec:sigmae8}) central stellar velocity dispersion $\sigma_{\rm e8}$:

\begin{equation}
	\log(r_{\rm e}) = \alpha \log(\sigma_{\rm e8}) + \beta \log\langle I_{\rm e}\rangle + \gamma, \label{eq:fp}
\end{equation}

where $\langle I_{\rm e}\rangle = -0.4 \langle SB_{\rm e}\rangle $ and $\langle SB_{\rm e}\rangle$ has units of $g'$ mag arcsec$^{-2}$, $r_{\rm e}$ has units of kiloparsecs, and $\sigma_{\rm e8}$ has units of kilometers per second. The effective radius $r_{\rm e}$ is measured along the effective axis $\sqrt{ab}$ with $a$ being the semimajor axis radius and $b$ being the semiminor axis radius. The average effective surface flux inside $r_{\rm e}$, $\langle I_{\rm e}\rangle$, is also measured along the effective axis.

WWFI $g'$-band magnitudes are equivalent to SDSS $g$-band magnitudes \citep{Kluge2020}. We have verified this again using the SB profiles of the 54 galaxies, which overlap in our WWFI and SDSS samples (Section \ref{sec:lowermass} and Figure \ref{fig:isopy_example}). The zero-point differences scatter by 0.04\,mag.

The FP fitting is performed by minimizing the weighted squared orthogonal residuals around the plane. The weights include the measurement uncertainties and covariances between them, the intrinsic scatter of the FP, and a luminosity dependence. Both the luminosity-dependent weighting and orthogonal sample cuts in velocity dispersion and luminosity are applied to mitigate sample selection bias. The full procedure is detailed in Appendix \ref{sec:fitting3d} and all galaxy parameters are listed in Appendix \ref{sec:apptables}.

\subsection{Correcting the Velocity Dispersion to $r_{\rm e}/8$} \label{sec:sigmae8}

The light-weighted central velocity dispersion $\sigma_0$ is measured inside a constant aperture on the sky with radius $r_{\rm ap}=1.5\arcsec$. Hence, the corresponding physical aperture increases with distance. Because of intrinsic gradients in $\sigma(r)$, this introduces an unintended distance dependency to $\sigma_0$. To eliminate it, a small, empirical correction by \cite{Jorgensen1995} is commonly applied (e.g., \citealt{Desroches2007,vonderLinden2007,Liu2008,LaBarbera2010b,Saglia2010,Samir2020}). It converts the aperture velocity dispersion $\sigma_0$ consistently into a physical value $\sigma_{\rm e8}$, the velocity dispersion at one-eighth of the effective radius $r_{\rm e}$:

\begin{equation}
	\sigma_{\rm e8} = \sigma_0 \left( \frac{r_{\rm ap}}{r_{\rm e}/8}\right)^{0.04}.
\end{equation}

We perform this correction for all galaxies, but we emphasize that it does not suffice for very large extrapolations if $r_{\rm e}$ is very large. For local BCGs, $r_{\rm e}/8>r_{\rm ap}$ is almost always the case \citep{Kluge2020}. Hence, the correction predicts negative velocity dispersion gradients of the order of $\sim$6\%. BCGs do have negative gradients in the inner $\sim$10\,kpc \citep{Bender2015,Mehrgan2019}, like normal Es. This is the scale where we perform the $\sigma_0$--$\sigma_{\rm e8}$ correction. We conclude that the correction is also adequate for the BCGs. Again, we emphasize that the correction is small. Doubling $r_{\rm e}$ decreases the correction factor for $\log(\sigma_0)$ by 0.012\,dex. From here on, we refer to $\sigma_{\rm e8}$ when we mention the central velocity dispersion.

\subsection{Faber--Jackson Relation} \label{sec:faberjackson}

The FJ relation \citep{Faber1976} is a projection of the FP, similar to the projection shown in Appendix \ref{sec:luminositydependence}, but it is not quite edge-on. It relates the total stellar luminosity $L$ with the central stellar velocity dispersion $\sigma_{\rm e8}$

\begin{equation}
	L \propto \sigma_{\rm e8}^{\zeta}. \label{eq:fjoriginal}
\end{equation}

The exponent $\zeta$ is large for bright ellipticals. In this case, the confidence intervals become highly asymmetric. That behavior is better captured by fitting the inverse of Equation (\ref{eq:fjoriginal}) and then inverting the result. The inverted FJ relation is

\begin{equation}
	\sigma_{\rm e8} \propto L^{1/\zeta}
\end{equation}

or expressed differently,

\begin{equation}
	\log(\sigma_{\rm e8}) = 1/\zeta \cdot ( \log(L) - \log(L_\times) ) + \log(\sigma_{\rm e8,\times}), \label{eq:fj}
\end{equation}

where $1/\zeta$, $L_\times$, and $\sigma_{\rm e8,\times}$ are fitted parameters. In our case, $L$ is measured in solar luminosities L$_{\odot,g'}$ in the $g'$ band, and $\sigma_{\rm e8}$ is measured in kilometers per second.

Bright ellipticals follow a steeper FJ relation than fainter ellipticals (see Section \ref{sec:results_faberjackson}). This is driven by the change from wet to dry mergers as the main formation channel for bright ellipticals. We therefore model the distribution of normal Es with a broken slope

\begin{equation}
	\log(\sigma_{\rm e8})_{\rm bs} = \begin{cases}
		\log(\sigma_{\rm e8},\zeta=\zeta_1), & \text{if $L<L_\times$},\\
		\log(\sigma_{\rm e8},\zeta=\zeta_2), & \text{if $L>L_\times$}. \label{eq:fj_brokenslope}
	\end{cases}
\end{equation}

The 2D fitting procedure is equivalent to the 3D procedure described in Appendix \ref{sec:fitting3d}. The only change is that we explicitly include galaxies with high velocity dispersions (see Appendix \ref{sec:samplecuts}).

\subsection{Mg$_b$--Luminosity Relation}

Analogous to the FJ relation, we define the fitting formula for the relation between the Mg$_{\rm b}$ absorption line strength and the total luminosity $L_{\rm tot}$ using a single slope:

\begin{equation}
	\log({\rm Mg_b}) = 1/\eta \cdot ( \log(L) - \log(L'_\times) ) + \log({\rm Mg_{b,\times}}), \label{eq:lm}
\end{equation}

and using a broken slope:

\begin{equation}
	\log({\rm Mg_b})_{\rm bs} = \begin{cases}
		\log({\rm Mg_b},\eta=\eta_1), & \text{if $L<L'_\times$},\\
		\log({\rm Mg_b},\eta=\eta_2), & \text{if $L>L'_\times$}. \label{eq:lm_brokenslope}
	\end{cases}
\end{equation}

\vspace{2mm}
\section{Results} \label{sec:results}

\subsection{Fundamental Planes of BCGs and Normal Ellipticals}

\begin{figure*}
	\centering
	\includegraphics[width=0.49\linewidth]{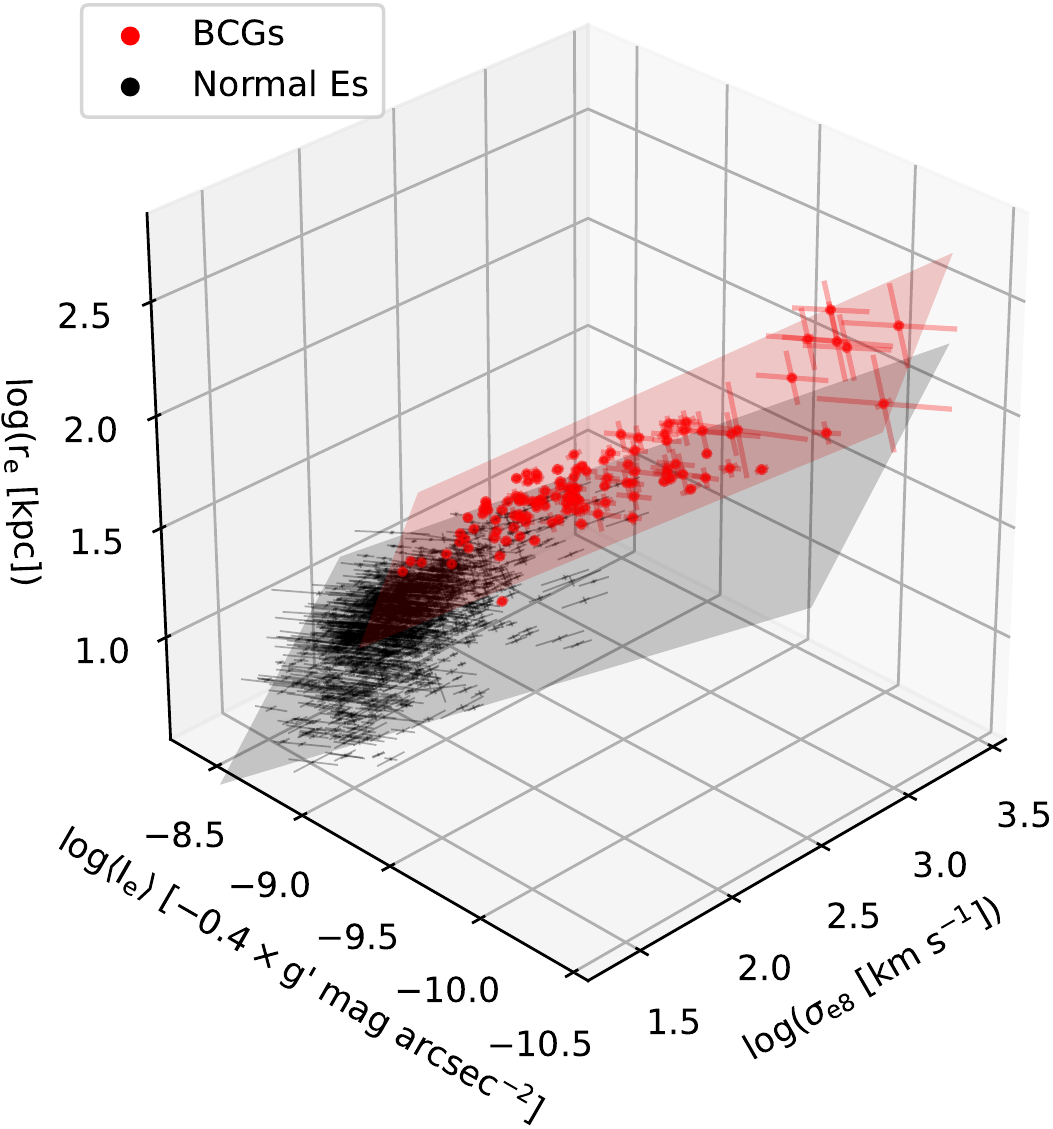}~
	\includegraphics[width=0.49\linewidth]{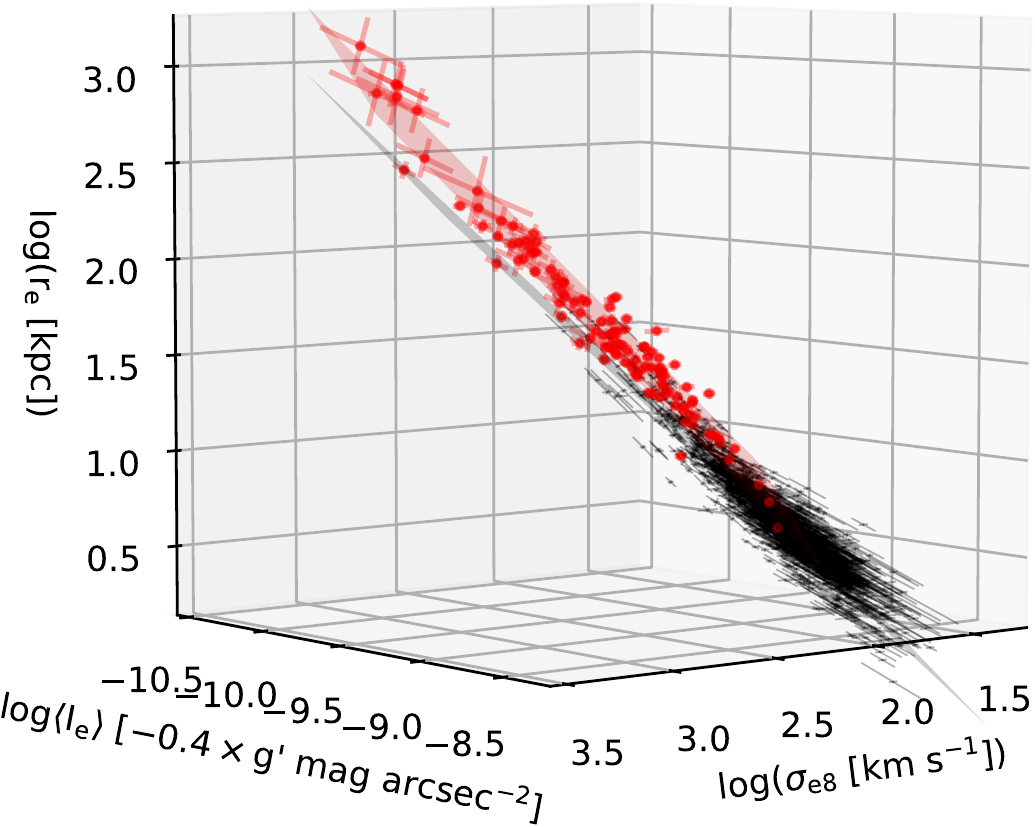}
	\caption{3D views of the FPs for BCGs (red) and normal Es (black). \label{fig:fundamental_plane_3d}}
\end{figure*}

In Figure \ref{fig:fundamental_plane_3d}, we present two 3D views of the FPs. Red and black surfaces are the FPs for BCGs and normal Es, respectively. The data points correspond to the individual galaxies. It is apparent that BCGs populate a very different region in parameter space. They are located at larger effective radii and fainter average effective surface intensities compared to the normal Es. 

\begin{figure*}[t]
	\includegraphics[width=\linewidth]{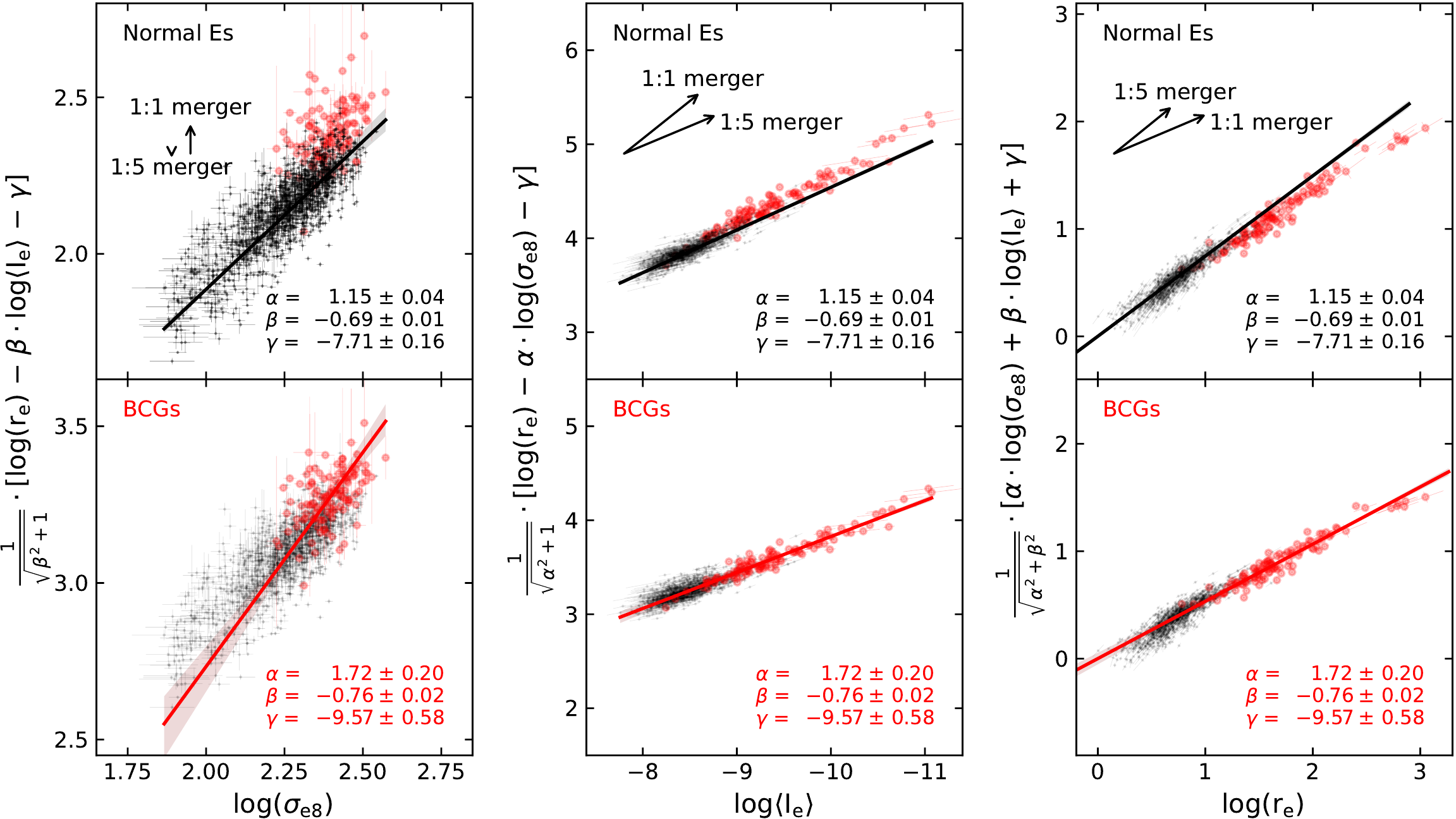}
	\caption{Top row: edge-on projections of the FP for normal Es (black). Bottom row: edge-on projections of the FP for BCGs (red). The $y$-axes are scaled such that orthogonal distances remain orthogonal. Shaded regions give the $\pm1\sigma$ confidence intervals around the projected planes (black and red lines). Best-fit slopes are given as text labels for both planes, which are defined by $\log(r_{\rm e})=\alpha\log(\sigma_{\rm e8})+\beta\log\langle I_{\rm e}\rangle+\gamma$. Arrows point in the directions where 1:1 and 1:5 mass ratio mergers would shift the data points \citep{Hilz2013}. The 1:5 merger arrow is almost parallel to the plane. \label{fig:fundamental_plane_projected}}
\end{figure*}

\begin{figure}
	\includegraphics[width=\linewidth]{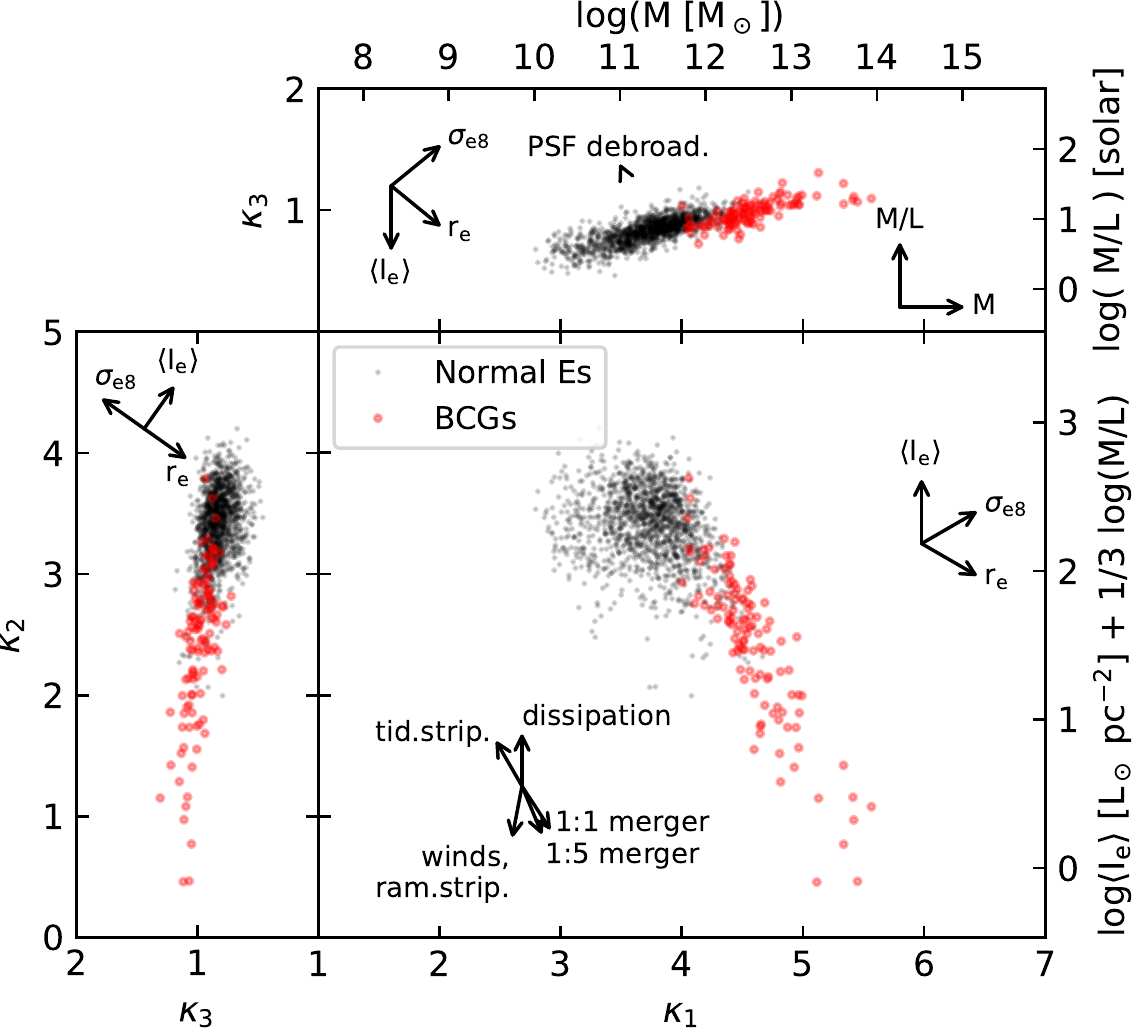}
	\caption{FP projections in $\kappa$ space. Arrows point in the direction where the data points would move as the corresponding parameter increases. Different formation processes shift the data points in different directions in the face-on view (bottom-right panel), as indicated by the arrows \citep{Bender1992}. We have updated the merger prediction using simulations from \cite{Hilz2013}. The arrow marked as ``PSF debroad." has almost no length and shows the correction for PSF de-broadening according to \cite{Kluge2020}. \label{fig:fundamental_plane_kappa}}
\end{figure}

In Figure \ref{fig:fundamental_plane_projected}, we show edge-on projections of the FPs (thick lines). All axes are scaled such that orthogonal distances remain orthogonal in the projections. The top panels show projections along the normal E FP (black line), whereas the bottom panels show projections along the BCG FP (red line). In particular the top-middle and top-right panels visualize that BCGs are offset from the normal E FP. At the median location of the BCGs, the two planes are $0.144\pm0.020$ dex apart in the $\log(r_{\rm e})$ direction. In other words, the effective radii of BCGs are $39\pm5\%$ larger than expected from an extrapolation of the FP parameters of normal Es. This offset is also apparent in the 3D view (Figure \ref{fig:fundamental_plane_3d}) and $\kappa$ projection (Figure \ref{fig:fundamental_plane_kappa}). We introduce the $\kappa$ space in Section \ref{sec:kappa}.

\begin{figure*}
	\includegraphics[width=\linewidth]{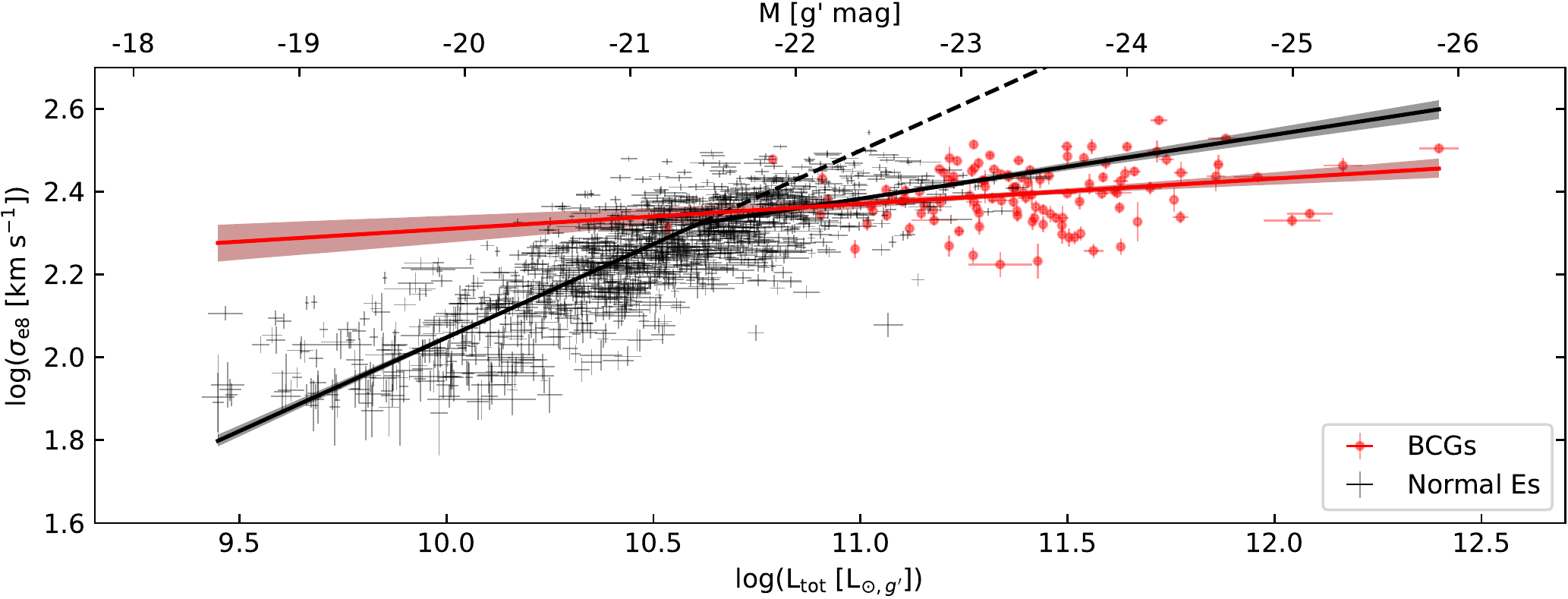}
	\caption{FJ relation for BCGs (red) and normal Es (black). The black continuous line is the linear relation with a broken slope. The black dashed line is the extrapolation of the fit to lower-luminosity ellipticals. Shades indicate the $\pm1\sigma$ confidence intervals. The fits are performed by minimizing the weighted, squared orthogonal residuals to the line. For the normal E sample, low-velocity-dispersion galaxies are discarded using an approximately orthogonal cut, and the remaining data points are weighted by the inverse of the luminosity function. \label{fig:faberjackson}}
\end{figure*}

Moreover, the left panels in Figure \ref{fig:fundamental_plane_projected} confirm the saturation in $\log(\sigma_{\rm e8})$. BCGs have only mildly higher central velocity dispersions than the luminous normal Es. That effect is even more visible in the FJ relation (Figure \ref{fig:faberjackson}), which we discuss in Section \ref{sec:results_faberjackson}. As a consequence, the $\sigma_{\rm e8}$-dependent FP slope $\alpha$ increases for BCGs by $50\pm18\%$. The uncertainty is large, because the dynamic range of the BCG sample (std$(\log(\sigma_{\rm e8}))=0.07$\,dex) is comparable to the orthogonal scatter from the FP $\delta_{\rm obs,orth}=0.05$\,dex (see Table \ref{tab:fpparams}).

The $\log\langle I_{\rm e}\rangle$-dependent slope $\beta$ is significantly more negative for BCGs by $10\pm4\%$. This change in slope resembles the broken slope (e.g., \citealt{Kluge2020}) in the Kormendy relation \citep{Kormendy1977} between BCGs and normal Es. Note that $\log(r_{\rm e})\propto\beta\log\langle I_{\rm e}\rangle$.

\begin{deluxetable*}{lccccccccc}
	\tabletypesize{\footnotesize}
	\tablecaption{Best-fitting Fundamental Plane Parameters. \label{tab:fpparams}}
	\tablehead{
		\colhead{Sample} & \colhead{$\alpha$} & \colhead{$\beta$} & \colhead{$\gamma$} & \colhead{$\delta_{\rm m,\sigma}$} & \colhead{$\delta_{\rm m,I}$} & \colhead{$\delta_{\rm m,r}$} & \colhead{$\delta_{\rm m,orth}$} & \colhead{$\delta_{\rm obs,orth}$} & \colhead{$\delta_{\rm in}$}
	}
	\colnumbers
	\startdata
	~~BCGs                & 1.717 $\pm$ 0.203 & -0.760 $\pm$ 0.023 & -9.57 $\pm$ 0.58 & 0.013 & 0.013 & 0.009 & 0.014 & 0.050 & 0.049 \\
	~~Normal Es (SDSS refit)   & 1.147 $\pm$ 0.037 & -0.691 $\pm$ 0.015 & -7.71 $\pm$ 0.16 & 0.014 & 0.042 & 0.017 & 0.015 & 0.057 & 0.055 \\
	~~Normal Es (SDSS original)  & 1.103 $\pm$ 0.040 & -0.688 $\pm$ 0.015 & -7.59 $\pm$ 0.18 & 0.004 & 0.008 & 0.010 & 0.008 & 0.056 & 0.055 \\
	\enddata
	\tablecomments{Best-fit FP parameters for the two galaxy samples. We also list the results for the normal Es when using the original published SDSS galaxy parameters (see Section \ref{sec:lowermass} and Appendix \ref{sec:sigmacomparison}). The fits are performed by minimizing the weighted, squared orthogonal residuals to the plane. The slopes in columns (2), (3), and (4) refer to the FP equation $\log(r_{\rm e})=\alpha\log(\sigma_{\rm e8})+\beta\log\langle I_{\rm e}\rangle+\gamma$. Typical measurement uncertainties $\delta_{\rm m}$ are given for each parameter in columns (5), (6), and (7) and for the orthogonal component of the typical measurement uncertainty ellipsoid in column (8). Column (9) gives the typical observed orthogonal scatter around the plane, and column (10) gives the intrinsic scatter $\delta_{\rm in} = \sqrt{\delta_{\rm obs,orth}^2 - \delta_{\rm m,orth}^2}$.}
\end{deluxetable*}
\vspace{-8.75mm}

\subsection{Comparison to Literature}

Our best-fit FP slopes are $\alpha=1.72\pm0.20$ and $\beta=-0.76\pm0.02$ for the BCGs and $\alpha=1.15\pm0.04$ and $\beta=-0.69\pm0.02$ for the normal Es. All parameters are summarized in Table \ref{tab:fpparams}. The BCG FP slopes are steeper with a $\sim2.5\sigma$ significance. We compare our results to the literature in Table \ref{tab:fpcomp}.

For our normal E sample, we notice that $\alpha$ is lower than most published values, even when we use the original SDSS photometry and dispersions. One reason is our sample selection. Faint ellipticals are weighted higher in our analysis (see Appendix \ref{sec:luminositydependence}), high-$\sigma_{\rm e8}$ galaxies are discarded (see Appendix \ref{sec:samplecuts}), and our sample extends to a lower redshift $z<0.05$ than most studies (that is, it includes a greater number of fainter ellipticals in relative numbers). By deactivating the former two options, we get a more consistent $\alpha=1.527\pm0.045$ when using SDSS photometry and dispersions instead of our own measurements. Confirming our findings, \cite{Gargiulo2009} showed that the slope $\alpha$ can decrease from $\alpha=1.35\pm0.11$ to $\alpha=1.06\pm0.06$ when the galaxy sample is extended to low-dispersion galaxies ($\sigma_{\rm e8}<100{\rm\,km\,s^{-1}}$). Hence, the typical variation $1.2 \lesssim \alpha \lesssim 1.4$ of published slopes can be explained by sample selection effects. Another effect, which increases $\alpha$ and $\beta$ is the choice of wave band. When using redder wave bands, the slopes become closer to the virial expectation of $\alpha=2$ and $\beta=-1$, because red light is less influenced by metal absorption lines \citep{Jorgensen1996,Pahre1998a,Bernardi2003,Samir2020}. \cite{Samir2020} found that $\beta$ increases in absolute value from $-0.700$ to $-0.775$ when photometric parameters are measured in the $z$ band instead of the $g$ band.

\begin{deluxetable*}{lcccc}[t]
	\tabletypesize{\small}
	\tablecaption{Comparison of the Fundamental Plane Slopes with Selected Literature. \label{tab:fpcomp}}
	\tablehead{
		\colhead{Author} & \colhead{$\alpha$(BCGs)} & \colhead{$\beta$(BCGs)} & \colhead{$\alpha$(Normal Es)} & \colhead{$\beta$(Normal Es)}
	}
	\startdata	
	This work               & $1.717\pm0.203$ & $-0.760\pm0.023$ & $1.147\pm0.037$ & $-0.691\pm0.015$ \\
	\cite{vonderLinden2007} & $1.96\pm0.10$ & -0.8! & $1.61\pm0.07$ & -0.8! \\
	\cite{Bernardi2007a}    & \ldots & \ldots & $1.307$ & $-0.763$ \\
	\cite{Bernardi2020}     & \ldots & \ldots & $1.349\pm0.023$ & $-0.737\pm0.012$ \\
	\cite{LaBarbera2010b}   & \ldots & \ldots & $1.384\pm0.024$ & $-0.7875\pm0.0025$ \\
	\cite{Hou2015}          & \ldots & \ldots & $1.337\pm0.004$ & $-0.751\pm001$ \\
	\cite{Samir2020}        & $1.61\pm0.04$ & $-0.80\pm0.0075$ & $1.40\pm0.04$ & $-0.700\pm0.0175$
	\enddata
	\tablecomments{The slopes $\alpha$ and $\beta$ refer to the FP equation $\log(r_{\rm e})=\alpha\log(\sigma_{\rm e8})+\beta\log\langle I_{\rm e}\rangle+\gamma$.}
\end{deluxetable*}
\vspace{-8.75mm}

In general, there is agreement that BCGs have steeper FP slopes. \cite{vonderLinden2007} found $\alpha=1.96\pm0.10$ for the BCGs and $\alpha=1.61\pm0.07$ for normal Es. Contrary to our results, they note that predominantly the small, low-velocity-dispersion BCGs deviate. However, these results must be interpreted with caution, because \cite{vonderLinden2007} restricted their analysis to the bright inner galaxy regions with ${\rm SB}<23\,r\,{\rm mag~arcsec^{-2}}$.

For their BCG sample, \cite{Samir2020} obtained $\alpha=1.61\pm0.04$ and $\beta=-0.800\pm0.001$. These values are formally consistent with our results within 1.7$\sigma$. However, our fitted plane is offset by 0.18\,dex toward larger $r_{\rm e}$. Most likely, this discrepancy arises from the de Vaucouleurs model fitting, which is used by the SDSS photometric pipeline and underestimates the contribution from the ICL (see Sections \ref{sec:lowermass} and \ref{sec:discussion_offset}).
	
The offset between BCGs and normal Es is $0.144\pm0.020$\,dex in our $g$-band data at the median location of the BCGs. BCGs are located at larger $r_{\rm e}$ (above) the normal Es. There is no consensus about this in the literature. \cite{Samir2020} confirmed our result only for the $i$ and $z$ bands. For bluer wave bands, the offset is in the opposite direction. Also, \cite{vonderLinden2007} found a negative offset. \cite{Liu2008} found negative offsets only for bright isophotal limits between 22 and $23~r~{\rm mag~arcsec^{-2}}$. This vanishes for fainter (but still relatively bright) isophotal limits between 24 and $25~r~{\rm mag~arcsec^{-2}}$. \cite{Bernardi2007a} claimed that BCGs, which are well fitted by a single de Vaucouleurs profile, lie on the same FP as the bulk of the early-type galaxy population. The others are slightly offset by $-0.025$\,dex. \cite{Hou2015} found that BCGs and satellites lie on the same FP only in high-mass clusters; $\alpha$ decreases to zero for BCGs in low-mass clusters. We explore the origin of the offset further in Section \ref{sec:discussion_offset_2}.

The measurement errors are usually smaller than the intrinsic scatter of the FP. We have measured the intrinsic scatter $\delta_{\rm in}$ after subtracting the typical orthogonal measurement uncertainties $\delta_{\rm obs,orth}$ in quadrature from the typical observed orthogonal scatter $\delta_{\rm obs,orth}$ (see Appendix \ref{sec:intrinsicscatter} and Table \ref{tab:fpparams}). We find $\delta_{\rm in}=0.055$ ($\delta_{\rm in}=0.049$) and for the normal E (BCG) sample. These values compare well to $\delta_{\rm in}=0.06$ ($\delta_{\rm in}=0.03$) \citep{Samir2020}, $\delta_{\rm in}=0.070$ \citep{Gargiulo2009}, $\delta_{\rm in}=0.053$ \citep{Hou2015}, $\delta_{\rm in}=0.056$ \citep{Bernardi2003}, and $\delta_{\rm in}=0.031$ \citep{Bernardi2020}. \cite{Jorgensen2006} pointed out that longer star-formation time scales for lower-mass ellipticals \citep{Thomas2005} predict a larger scatter in $M/L$ for those galaxies. We confirm a slight decrease of the intrinsic scatter $\delta_{\rm in}=0.051$ when including bright galaxies $M<-21.4\,g$\,mag (see also \citealt{Hyde2009b,Nigoche-Netro2011,Bernardi2020}).

\begin{deluxetable*}{lccccccccc}
	\tabletypesize{\small}
	\tablecaption{Best-fitting Faber--Jackson Relation Parameters. \label{tab:fjparams}}
	\tablehead{
		\colhead{Sample} & \colhead{$\zeta_1$} & \colhead{$\zeta_2$} & \colhead{$\log (L_\times)$} & \colhead{$\log(\sigma_{\rm e8,\times})$} & \colhead{$\delta_{\rm m,L}$} & \colhead{$\delta_{\rm m,\sigma}$} & \colhead{$\delta_{\rm m,orth}$} & \colhead{$\delta_{\rm obs,orth}$} & \colhead{$\delta_{\rm in}$}
	}
	\colnumbers
	\startdata	
	BCGs   & $16.4^{+9.5}_{-4.4}$ & \ldots & 0 & $1.70 \pm 0.25$ & 0.0069 & 0.0130 & 0.0127 & 0.067 & 0.066 \\
	Normal Es & $2.215^{+0.080}_{-0.074}$ & $6.48^{+0.72}_{-0.59}$ & $10.610 \pm 0.026$ & $2.323 \pm 0.009$ & 0.0186 & 0.0111 & 0.0124 & 0.072 & 0.071
	\enddata
	\tablecomments{Best-fit FJ relation parameters for the two galaxy samples. The slopes in columns (2), (3), (4), and (5) refer to Equation \ref{eq:fj_brokenslope}. Typical measurement uncertainties $\delta_{\rm m}$ are given for each parameter in columns (6) and (7), and for the orthogonal component of the typical measurement uncertainty ellipse in column (8). Column (9) gives the typical observed orthogonal scatter around the plane, and column (10) gives the intrinsic scatter $\delta_{\rm in} = \sqrt{\delta_{\rm obs,orth}^2 - \delta_{\rm m,orth}^2}$.}
	\vspace{-2.5mm}
	
\end{deluxetable*}
\vspace{-8.75mm}

It has been suggested that the intrinsic scatter is induced by (more or less) random walks of individual galaxies in parameter space over the course of their evolution \citep{Donofrio2017,Donofrio2021}. Past merger mass ratios and impact parameters vary strongly for individual galaxies. Different configurations have different effects on luminosities and velocity dispersions \citep{Naab2009,Hilz2012}, which can shift galaxy data points above or below the FJ relation.

\subsection{Projections in $\kappa$ Space} \label{sec:kappa}

The FP can be projected almost edge-on such that one axis $\kappa_1$ is proportional to the mass $M$ and another axis $\kappa_3$ is proportional to the mass-to-light ratio $M/L$ \citep{Bender1992,Bender1993,Burstein1997}:

\begin{align}
	\kappa_1 &\defeq \frac{2\log(\sigma_{\rm e8}) + \log(r_{\rm e})} {\sqrt{2}} &&\propto \mathcal{M}, \nonumber \\
	\kappa_2 &\defeq \frac{2\log(\sigma_{\rm e8}) + 2\log\langle I_{\rm e}\rangle - \log(r_{\rm e})} {\sqrt{6}} &&\propto (\mathcal{M}/L) \langle I_{\rm e}\rangle^3, \nonumber \\
	\kappa_3 &\defeq \frac{2\log(\sigma_{\rm e8}) - \log\langle I_{\rm e}\rangle - \log(r_{\rm e})} {\sqrt{3}} &&\propto \mathcal{M}/L. \label{eq:kappa}
\end{align}

This view reveals the tilt of the FP relative to the virial expectation $r_{\rm e}\propto \sigma_{\rm e8}^{2} \langle I\rangle ^{-1}$. No tilt would imply constant $\kappa_3$. In reality, $\kappa_3$ increases with $\kappa_1$, that is, the $M/L$ increases gradually for higher $M$ (see Figure \ref{fig:fundamental_plane_kappa}, top panel). For the normal E sample, we find $\kappa_3\propto (0.268\pm0.006) \kappa_1$ and consequently $M/L\propto M^{0.328\pm0.007}$. For the BCG sample, the slope is consistent with $\kappa_3\propto (0.245\pm0.022) \kappa_1$ and $M/L\propto M^{0.301\pm0.027}$. However, the BCGs are offset toward lower $M/L$ ratios than expected from an extrapolation of the relation for normal Es. This does not necessarily mean that BCGs really do have lower $M/L$ ratios. It is more likely that the central velocity dispersion underestimates the circular velocity systematically stronger for BCGs.

\begin{deluxetable}{lcc}
	\tabletypesize{\small}
	\tablecaption{Comparison of the Faber--Jackson Relation Slopes with Selected Literature. \label{tab:fjcomp}}
	\tablehead{
		\colhead{Author} & \colhead{$\zeta$ (Faint)} & \colhead{$\zeta$ (Bright)}
	}
	\startdata
	This work              & $2.215^{+0.080}_{-0.074}$ & $16.4^{+9.5}_{-4.4}$ \\
	\cite{Davies1983}      & $2.4$                     & $4.2$ \\
	\cite{Gerhard2001}     & $1.96\pm0.45$             & \ldots \\
	\cite{Matkovic2005}    & $2.01\pm0.36$             & \ldots \\
	\cite{Choi2007}        & $2.7\pm0.2$               & $4.6\pm0.4$ \\
	\cite{Desroches2007}   & $3$                       & $4.5$ \\
	\cite{Lauer2007a}      & $2.6\pm0.3$               & $6.5\pm1.3$ \\
	\cite{vonderLinden2007}& $3.93\pm0.21$             & $5.32\pm0.37$ \\
	\cite{LaBarbera2010b}  & $5.81\pm0.62$             & \ldots \\
	\cite{Kormendy2013}    & $3.74\pm0.21$             & $8.33\pm1.24$ \\
	\cite{Samir2020}       & $2.63\pm0.14$             & $6.25\pm0.39$ \\
	\cite{Donofrio2021}    & $2.70\pm0.11$             & $5.11\pm0.95$ \\
	\enddata
	\tablecomments{Different authors use different definitions for the faint and bright galaxy samples, e.g., core Es vs. coreless Es \citep{Lauer2007a,Kormendy2013}, normal Es vs. BCGs (this work; \citealt{vonderLinden2007,Samir2020}), or splitting the samples at $M=-21.5\,V$\,mag \citep{Donofrio2021}. For this work, we list the faint-end slope $\zeta=\zeta_1$ of the normal Es for the faint subsample and the BCG FJ slope for the bright subsample. There is consensus that the FJ slope becomes steeper for higher galaxy luminosities.}
\end{deluxetable}
\vspace{-4.375mm}

The face-on projection $\kappa_1$--$\kappa_2$ stores information about the involved formation processes \citep{Bender1992}. Different processes move galaxies in different directions along the FP, as illustrated by the arrows. Compared to \cite{Bender1992}, we have updated the merger directions using the dissipationless simulations by \cite{Hilz2013} for 1:5 and 1:1 mass ratios. The central velocity dispersion is assumed to stay constant (see Figure \ref{fig:faberjackson}) when the merger orbit has sufficient angular momentum. Given this assumption, the merger arrows agree with the direction of the BCGs. This implies that mergers are the main driver of BCG buildup.

Moreover, BCGs reside near cluster centers. Their light profiles cannot get truncated by tidal forces from the cluster potential. Instead, at large radii, BCGs accumulate tidally stripped material from satellite galaxies. This form of accretion is the opposite phenomenon compared to what occurs to satellite galaxies by tidal stripping. Hence, BCGs are shifted in the opposite direction from the tidal stripping arrow.

\subsection{Faber--Jackson Relation} \label{sec:results_faberjackson}

Our results are shown in Figure \ref{fig:faberjackson} and Table \ref{tab:fjparams}. We confirm the consensus of the literature that the FJ relation becomes flatter for brighter ellipticals (see Table \ref{tab:fjcomp}). The central velocity dispersion increases more slowly with luminosity. For our data, the relation changes from $L\propto\sigma_{\rm e8}^{2.2}$ to $L\propto\sigma_{\rm e8}^{6.5}$ with increasing luminosity of the normal Es and reaches $L\propto\sigma_{\rm e8}^{16}$ for the BCGs. The point where the slope breaks is at $\sigma_{\rm e8}=210.4\pm4.2$\,km\,s$^{-1}$ and $M=-21.41\pm0.07~g'$ mag or $L=10^{10.610\pm0.026} {\rm L}_{\odot,g'}$.

A comparison with literature values (Table \ref{tab:fjcomp}) shows consistency with our results, but the scatter is large. The best-fit slope $\zeta$ depends strongly on the luminosity range of the sample, which makes a stringent comparison difficult. We name the samples ``faint" and ``bright" because different authors use different brightnesses or even different parameters to split their samples. No study had specifically targeted the brightest BCGs yet. Therefore, our slope for the BCGs is higher than the bright-end slopes of previous studies.

By comparing to simulated galaxies in the Illustris simulation, \cite{Donofrio2020} found $\zeta=2.71$, formally consistent with our result. However, the slope breaks at lower dispersion $\sigma\approx100$\,km\,s$^{-1}$ and bends toward the opposite direction. In general, a comparison of the FJ relation to simulations is complicated by the different radial apertures, in which the stellar velocity dispersion is averaged. In simulations, it is usually calculated from all stellar particles inside $0.5r_{\rm e}$ (\citealt{Remus2017}; Magneticum simulation) or all luminosity weighted particles inside $r_{\rm e}$ (\citealt{Zahid2018}; Illustris simulation). In contrast, observations are usually restricted to the galaxy centers, especially for BCGs. Positive radial velocity dispersion gradients in BCGs are found both in simulations (\citealt{Marini2021}; DIANOGA simulation) and observations \citep{Dressler1979,Carter1981,Ventimiglia2010,Newman2011,Toledo2011,Arnaboldi2012,Melnick2012,Murphy2014,Bender2015,Barbosa2018,Loubser2018,Spiniello2018,Veale2018,Gu2020}. This can possibly explain why numerous BCGs in simulations have velocity dispersions higher than $\sigma\gtrsim300$\,km\,s$^{-1}$ \citep{Donofrio2020,Marini2021}.

\begin{deluxetable*}{lccccccccc}
	\tabletypesize{\small}
	\tablecaption{Best-fitting Mg$_{\rm b}$--$L_{\rm tot}$ Relation Parameters. \label{tab:lmparams}}
	\tablehead{
		\colhead{Sample} & \colhead{$\eta_1$} & \colhead{$\eta_2$} & \colhead{$\log (L'_\times)$} & \colhead{$\log({\rm Mg_{b,\times}})$} & \colhead{$\delta_{\rm m,L}$} & \colhead{$\delta_{\rm m,Mg_b}$} & \colhead{$\delta_{\rm m,orth}$} & \colhead{$\delta_{\rm obs,orth}$} & \colhead{$\delta_{\rm in}$}
	}
	\colnumbers
	\startdata
	BCGs   & $49^{+93}_{-19}$ & \ldots & 0 & $0.42 \pm 0.15$ & 0.0065 & 0.0088 & 0.0088 & 0.067 & 0.067 \\
	Normal Es & $9.68^{+0.62}_{-0.55}$ & $13.3^{+3.1}_{-2.1}$ & $10.69 \pm 0.19$ & $0.622 \pm 0.018$ & 0.0200 & 0.0155 & 0.0159 & 0.059 & 0.057 \\
	\enddata
	\tablecomments{Best-fit Mg$_{\rm b}$--$L_{\rm tot}$ relation parameters for the two galaxy samples. The slopes in columns (2), (3), (4), and (5) refer to Equation \ref{eq:fj_brokenslope}. Typical measurement uncertainties $\delta_{\rm m}$ are given for each parameter in columns (6) and (7), and for the orthogonal component of the typical measurement uncertainty ellipse in column (8). Column (9) gives the typical observed orthogonal scatter around the plane, and column (10) gives the intrinsic scatter $\delta_{\rm in} = \sqrt{\delta_{\rm obs,orth}^2 - \delta_{\rm m,orth}^2}$.}
	\vspace{-6mm}
\end{deluxetable*}
\vspace{-8.75mm}

The intrinsic scatter of the FJ relation $\delta_{\rm in}=0.071~(0.066)$ for normal Es (BCGs) is $\sim$32\% larger than the intrinsic scatter of the FP (see Table \ref{tab:fpparams}) because the FJ relation is not an edge-on projection (see Appendix \ref{sec:luminositydependence}).

\subsection{Line Strengths} \label{sec:linestrenghts}

\begin{figure}
	\includegraphics[width=\linewidth]{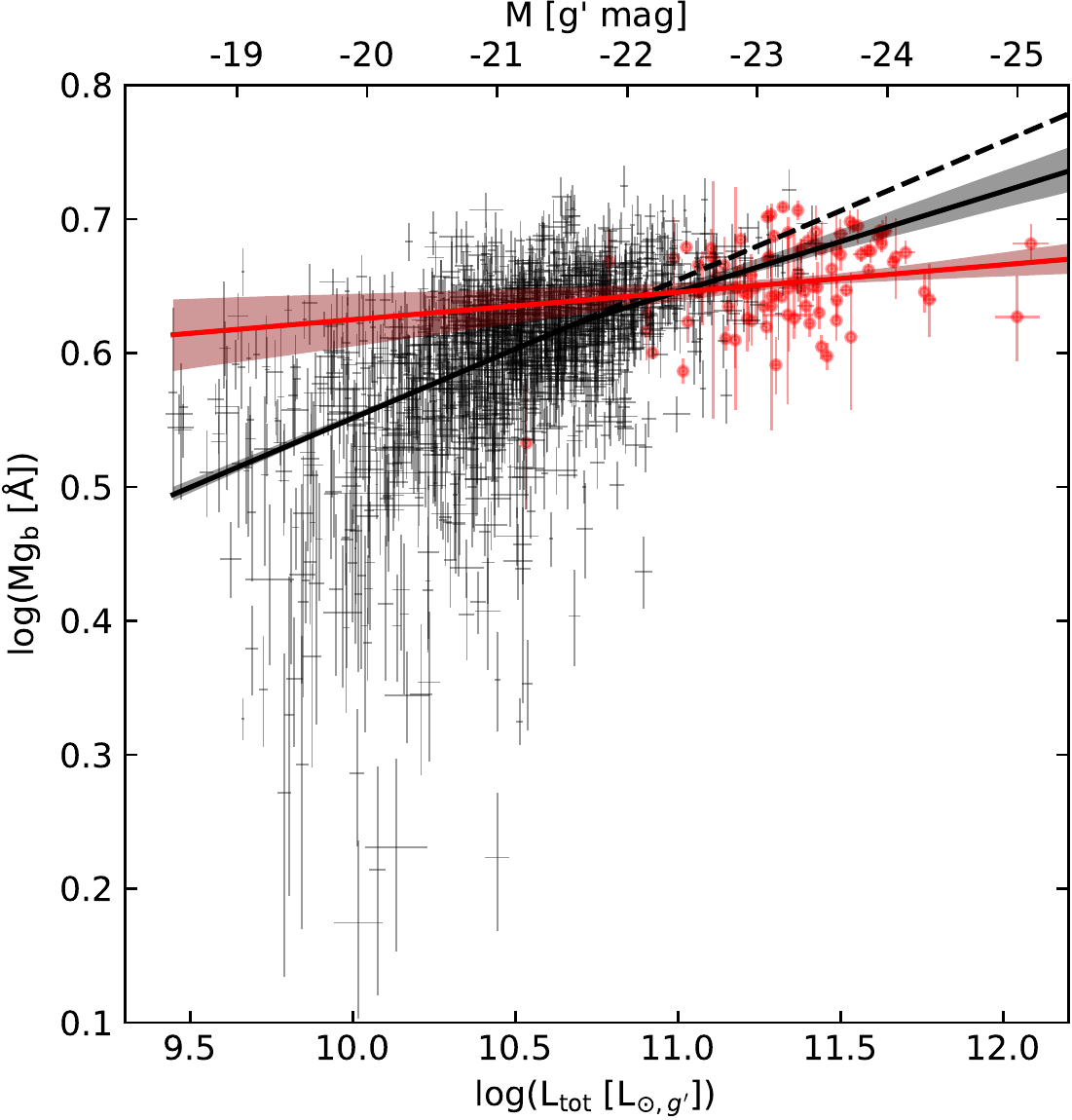}
	\caption{Mg$_{\rm b}$ absorption line strength versus total luminosity relation for BCGs (red) and normal Es (black). The best-fit lines are analogous to those in Figure \ref{fig:faberjackson}. The axes are not spaced equally, because the dynamic range is much smaller in Mg$_{\rm b}$ than in $L_{\rm tot}$. \label{fig:mgb_ltot}}
\end{figure}

In analogy to the FJ relation (Section \ref{sec:faberjackson}), we plot central Mg$_{\rm b}$ line strengths against galaxy luminosities (Figure \ref{fig:mgb_ltot} and Table \ref{tab:lmparams}). The correlation is weaker than the FJ relation. In fact, the scatter in Mg$_{\rm b}$ is comparable to the dynamic range. Moreover, the dynamic range is much larger for $\log(L_{\rm tot})$ than for $\log({\rm Mg_b})$. Therefore, a fit to the data depends strongly on how the axes are scaled. We consider this uncertainty for the calculation of the EL fractions (Section \ref{sec:discussion_totalicl}) but continue here consistently with the FP and FJ relations by performing an orthogonal fit to logarithmic parameters.

The behavior of the Mg$_{\rm b}$--$L_{\rm tot}$ relation is very similar to the FJ relation. The slope $\eta$ becomes steeper with increasing luminosity: from $L\propto{\rm Mg_b}^{9.7}$ to $L\propto{\rm Mg_b}^{13}$ for the normal Es and it reaches $L\propto{\rm Mg_b}^{49}$ for the BCGs. The point where the slope breaks is at ${\rm Mg_b}=4.17\pm0.17$\,\AA~and $M=-21.61\pm0.48~g'$ mag or $L=10^{10.69\pm0.19} {\rm L}_{\odot,g'}$, consistent with the FJ relation.

\begin{figure*}
	\includegraphics[width=\linewidth]{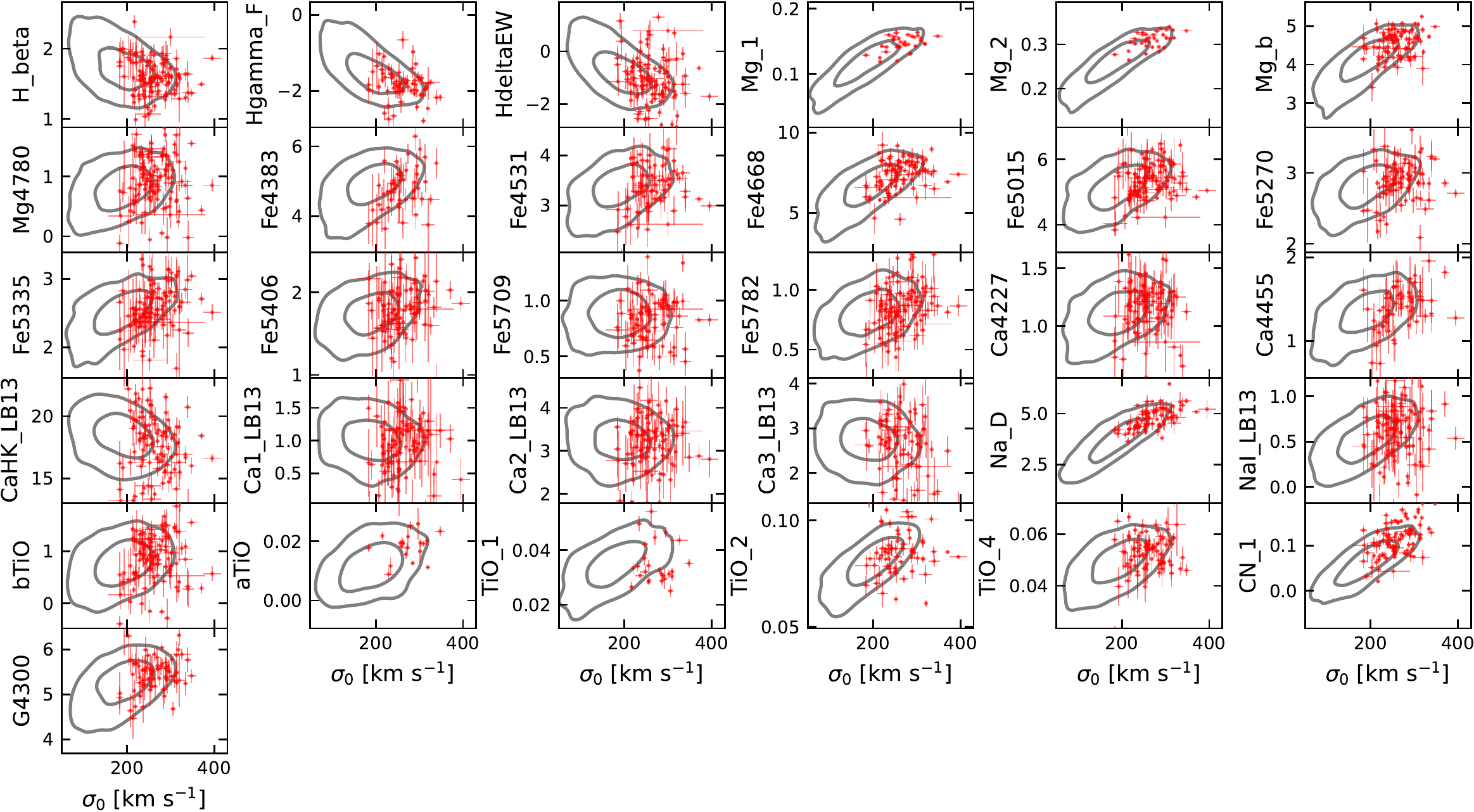}
	\caption{Lick indices against central velocity dispersion for BCGs (red) and normal Es (gray). The gray contours mark the locations where the kernel density estimate drops to 50\% and 10\% relative to the peak. The uncorrected dispersion $\sigma_0$ instead of $\sigma_{\rm e8}$ is used to enlarge the sample size to 1908 galaxies. The wavelength intervals for the indices can be found at \url{https://mfouesneau.github.io/pyphot/licks.html}. \label{fig:linestrengths}}
\end{figure*}

In Figure \ref{fig:linestrengths}, we compare all measured absorption line strengths between BCGs (red points) and normal Es (gray contours), apart from those that overlap with bright night-sky emission lines $\lambda\lambda$5577,5889,6300 and telluric absorption around $\lambda\lambda$7185,7275,7620. The negative correlations for the hydrogen lines (first three panels) imply that the stellar populations become older and/or more metal rich with increasing velocity dispersion. For all metal lines besides Ca, the correlations are positive. This implies that the stellar populations are more metal-rich for higher-velocity dispersions. In contrast, we find that CaHK and the Ca II triplet lines Ca1\_LB13, Ca2\_LB13, and Ca3\_LB13, decrease with increasing velocity dispersion \citep{Saglia2002,Cenarro2003,Cenarro2004,Smith2009,Worthey2011}, whereas Ca4227 and Ca4455 increase \citep{Cenarro2004,Smith2009,Worthey2011}.

Most line strength--velocity dispersion relations are consistent for BCGs and luminous normal Es. In particular, the Mg$_{\rm b}$--$\sigma_0$ relation agrees well. An equally good agreement is seen for the NaD line. However the NaD line strength must be interpreted cautiously because it is sensitive to interstellar absorption.

On the other hand, some Fe line strengths are systematically lower for BCGs than for normal Es at fixed velocity dispersion. The inconsistency for the same element can be explained by the fact that some Fe lines are more contaminated by other elements, others less. From Table 1 in \cite{Worthey1994}, we identify Fe5335 as the cleanest, Fe4383, Fe5270, and Fe5406 as relatively clean, and Fe4531, Fe4668, Fe5015, Fe5709, and Fe5782 as contaminated. We find the cleanest Fe lines to show the strongest deviations.

The standard interpretation of a lower Fe abundance is early star formation quenching before SN Ia could strongly enrich the interstellar medium with Fe (e.g., \citealt{Faber1992,Thomas1999}). As BCGs reside near the cluster center, they are prone to gas heating by AGN feedback and, therefore, efficient quenching of star formation (e.g., \citealt{DeLucia2007}).

\section{Discussion} \label{sec:discussion}

\subsection{BCGs Contain a Normal E in Their Center}

We compare the inner regions of BCGs and normal Es via their velocity dispersions, line strengths, and SBs. For this, we select for each BCG a set of normal Es with similar central velocity dispersions ($\delta\sigma_{\rm e8}=\pm10$\,km s$^{-1}$) or similar Mg$_{\rm b}$ line strengths ($\pm0.05$\,\AA). We refer to those normal Es as ``luminous normal Es".

Beginning with the central stellar velocity dispersion, we find very consistent values for BCGs and luminous normal Es in the $g'$-band luminosity range $10.5\lesssim \log(L_{\rm tot} [{\rm L}_{\odot,g'}])\lesssim 11$. The agreement is apparent in the $\sigma_{\rm e8}$-projection of the FP (Figure \ref{fig:fundamental_plane_projected}, left panels) and in the FJ relation (Figure \ref{fig:faberjackson}).

The central absorption line strengths are also consistent as discussed in Section \ref{sec:linestrenghts}. When plotted against velocity dispersion, the line strengths of the BCGs overlap with the contours for the normal Es (Figure \ref{fig:linestrengths}). Furthermore, in the Mg$_{\rm b}$--$L_{\rm tot}$ relation, BCGs have equivalent line strengths as luminous normal Es (Figure \ref{fig:mgb_ltot}).

\begin{figure}
	\includegraphics[width=\linewidth]{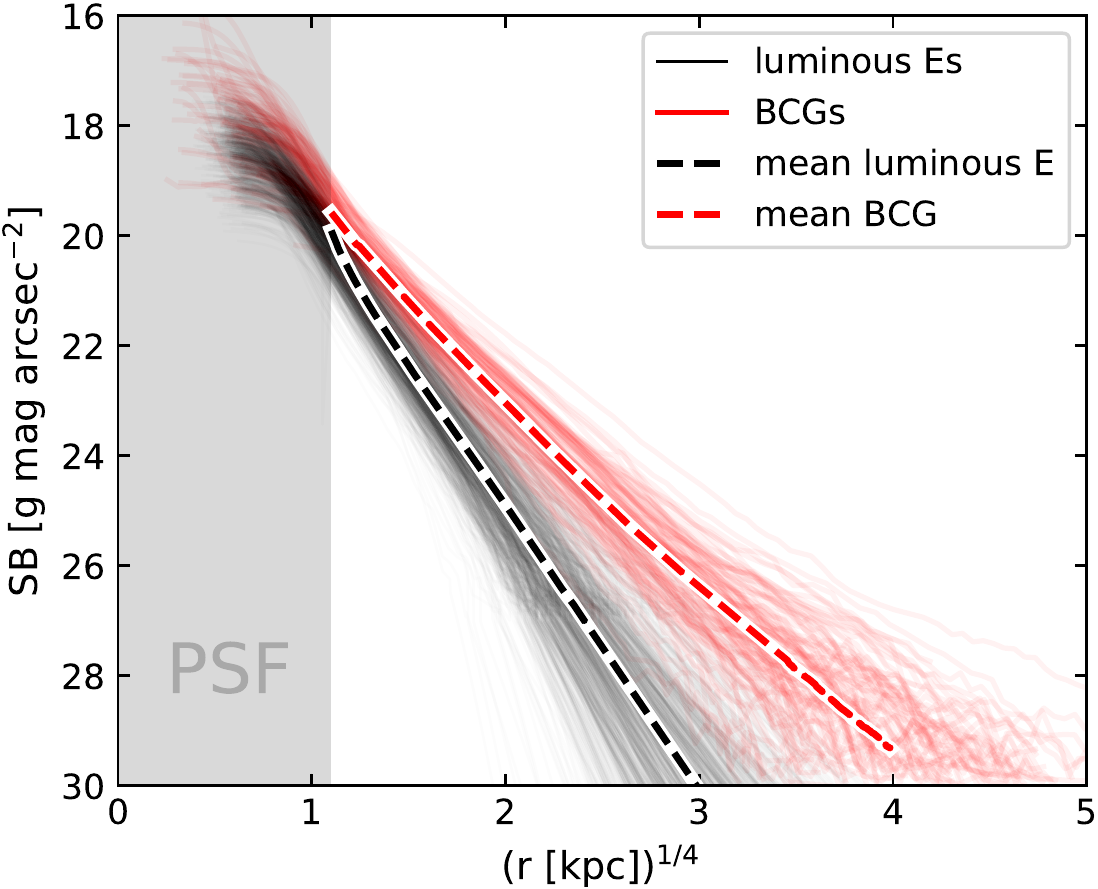}
	\caption{All BCG (red) and luminous normal E (black) SB profiles along the effective axis $r=\sqrt{ab}$. For each luminous normal E exists at least one BCG with similar central stellar velocity dispersion. The SB profiles of the luminous normal Es are extrapolated beyond ${\rm SB}>27~g~{\rm mag~arcsec^{-2}}$ using the best-fit S\'ersic function. The inner region is marked with gray shading, where the profiles are not comparable. Contrary to the BCGs, the luminous normal E profiles are measured on seeing-limited data. The dashed lines are the mean SB profiles. \label{fig:allprofiles}}
\end{figure}

\begin{figure}
	\includegraphics[width=\linewidth]{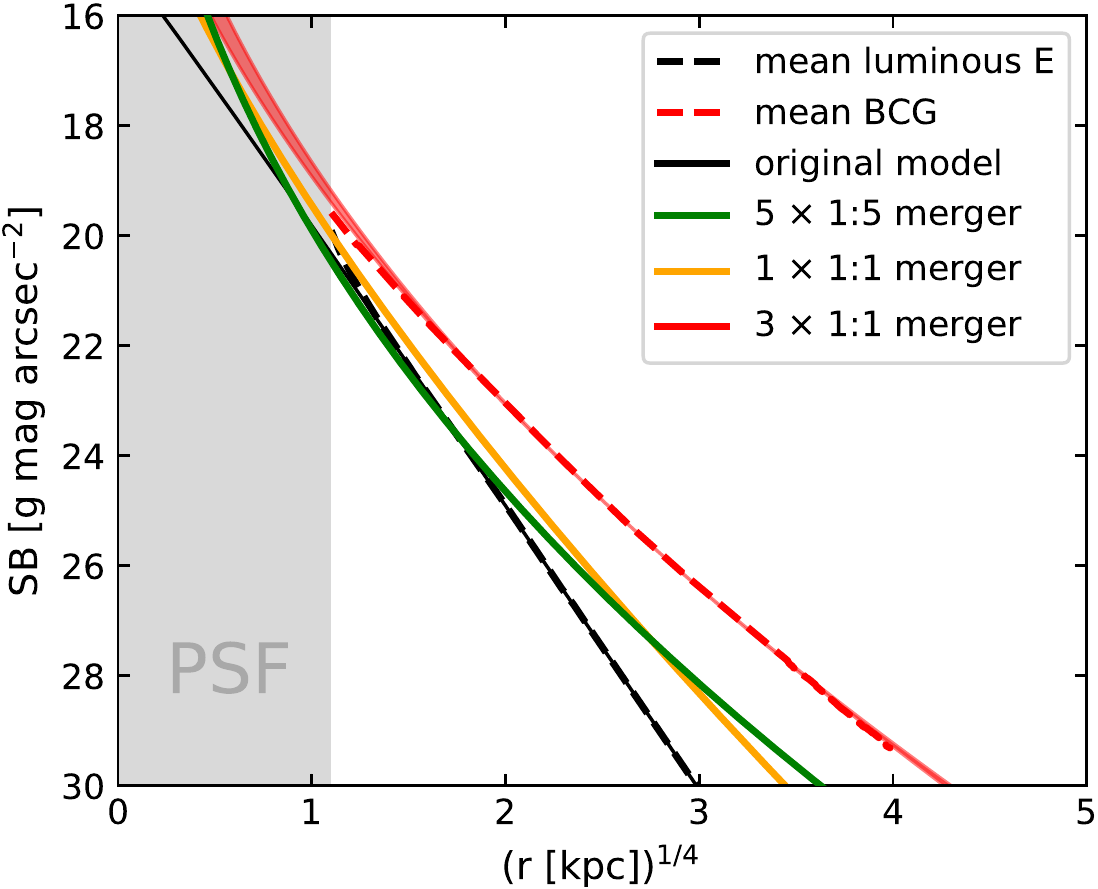}
	\caption{Evolution of simulated SB profiles (\citealt{Hilz2013}; continuous lines) for the minor and major-merger scenarios, compared to the observed profiles (dashed lines). The SB profile of the simulated pre-merger galaxy is shown by continuous black line. It is scaled to match the average profile of the luminous normal Es (black dashed line): $r_{\rm e}=7$\,kpc; ${\rm SB_e}=23~g'$ mag arcsec$^{-2}$; $n=3.9$. The green line is the SB profile after five generations of minor mergers with mass ratio 1:5 ($r_{\rm e}=33$\,kpc; ${\rm SB_e}=26.1~g'$ mag arcsec$^{-2}$; $n=9.5$). The orange line shows the resulting SB profile after one major merger with equal mass ($r_{\rm e}=13.2$\,kpc; ${\rm SB_e}=23.8~g'$ mag arcsec$^{-2}$; $n=5.7$). Here, the accreted mass is deposited also at smaller radii compared to the minor-merger scenario. The continuous red line is the extrapolated result after three generations of equal-mass mergers ($r_{\rm e}=45.6$\,kpc; ${\rm SB_e}=25.1~g'$ mag arcsec$^{-2}$; $6.9<n<8.1$; $L_{\rm tot}=7.8 L_{\rm seed}$). This agrees well with the mean SB profile of the BCGs (dashed red line). \label{fig:accretion}}
\end{figure}

\begin{figure*}
	\includegraphics[width=0.49\linewidth]{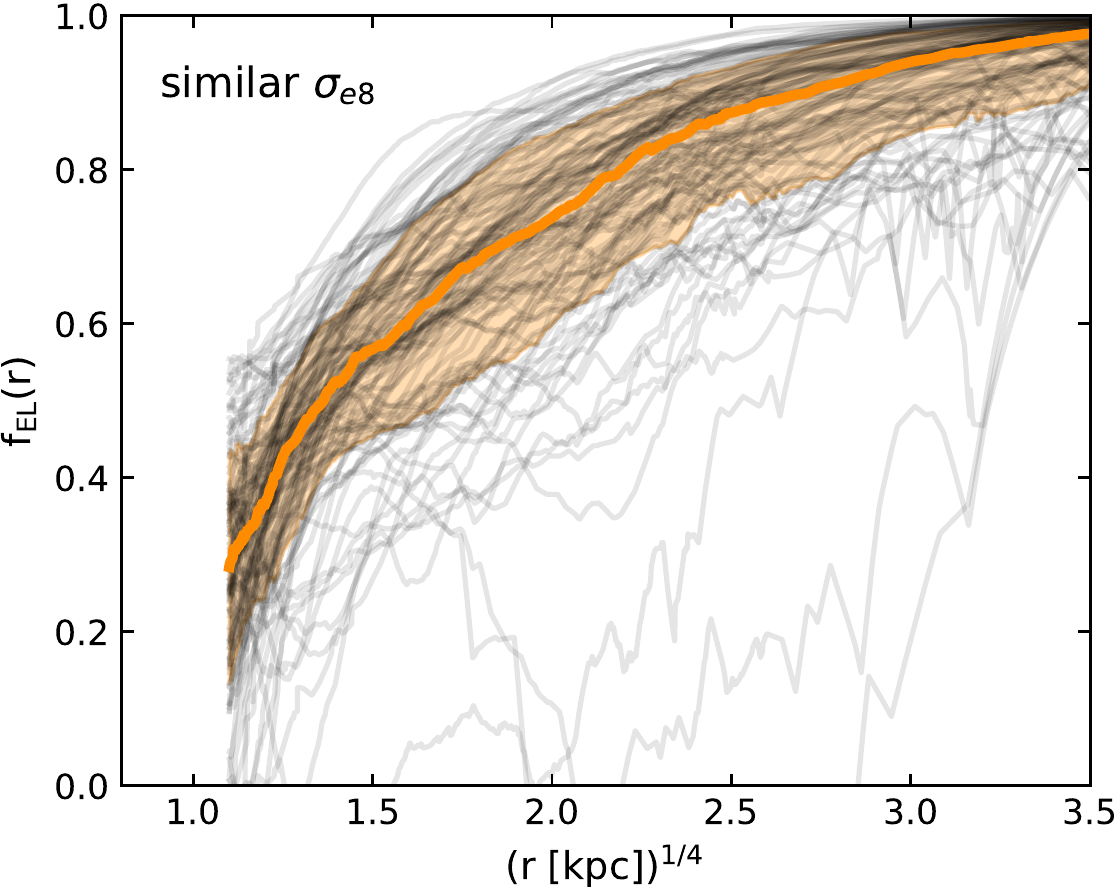}
	\includegraphics[width=0.49\linewidth]{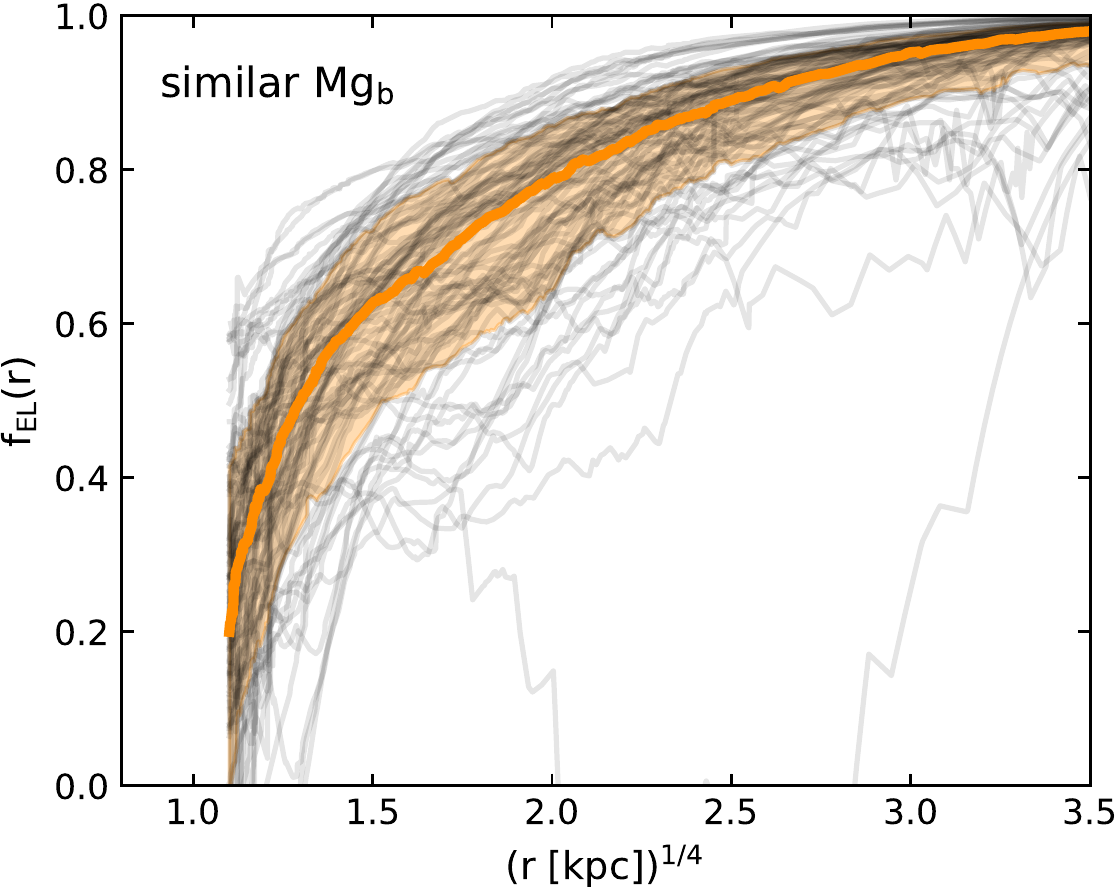}
	\caption{Radial EL fraction of all BCGs (black lines) according to Equation (\ref{eq:iclfrac}). The reference luminous normal Es are selected based on their central velocity dispersion (left panel) or central Mg$_{\rm b}$ absorption line strength (right panel). The orange line is the median profile and the orange shades encompass the 16\% and 84\% percentiles. \label{fig:iclfrac}}
\end{figure*}

Finally, we compare the SB profiles in Figures \ref{fig:allprofiles} and \ref{fig:accretion}. BCGs are everywhere brighter than luminous normal Es, but they become comparable at small radii. This similarity can also be seen by the radial EL in BCGs relative to luminous normal Es (orange line in Figure \ref{fig:iclfrac}; see Section \ref{sec:radialEl}). It approaches zero at small radii, indicating that both galaxy types have similar central SBs. Please note that the central regions inside $r<1$\,kpc cannot be compared directly because the normal E SBs are severely affected by seeing in the SDSS images. Our finding of the similar central SBs is in agreement with the flattening of the total stellar mass to central stellar mass ($r<1$\,kpc) relation for low-redshift Es with total stellar mass $M_*\gtrsim10^{10.7}$\,M$_\odot$ \citep{Chen2020,Arora2021}. The stellar mass range of our galaxy sample is consistent with the flat part of this relation.

The fact that the centers of BCGs have essentially similar properties in velocity dispersion, stellar population, and SB suggests that the BCG formation started with a luminous normal E at its center. We refer to it as the ``seed E". Subsequent mergers, major or minor, do not change these parameters much, except that the core radius will increase with every major merger due to black hole scouring (e.g., \citealt{Kormendy2009b,Thomas2014,Mehrgan2019}). EL accretion only increases the SB at larger radii and has little effect on the central region.

\subsection{Total EL Fractions} \label{sec:discussion_totalicl}

Motivated by the preceding discussion, we assume that $\sigma_{\rm e8}$ and the central line strengths remain unaffected by EL accretion. Based on that, we propose a new method to estimate EL luminosities. The seed E, which is embedded in the EL, is traced by the kinematics ($\sigma_{\rm e8}$) and stellar populations (Mg$_{\rm b}$) in the central region. We infer the luminosity of the seed E via the FJ and Mg$_{\rm b}$--$L_{\rm tot}$ relations in Equations (\ref{eq:fj}) and (\ref{eq:lm}), respectively, for low-luminosity ellipticals ($\zeta=\zeta_1$; $\eta=\eta_1$):

\begin{align}
	\log(L_{\rm E} [{\rm L}_{\odot,g'}]) &= 2.215 \cdot \log(\sigma_{\rm e8}) + 5.46,\\
	\log(L_{\rm E} [{\rm L}_{\odot,g'}]) &= 9.68 \cdot \log({\rm Mg_b}) + 4.66. \label{eq:lmlowmass}
\end{align}

The EL luminosity is ${\rm EL} = L_{\rm BCG} - L_{\rm E}$. Consequently, the total EL fraction is

\begin{align}
	f_{\rm EL} &= 1 - \frac{L_{\rm E}}{L_{\rm BCG}}.
\end{align}

In other words, the horizontal distance between the red BCG data points and the black dashed lines in Figures \ref{fig:faberjackson} and \ref{fig:mgb_ltot} measures the EL luminosity under the given assumptions.

The resulting total EL fraction is $f_{\rm EL}=73\pm15\%$, when it is inferred using the FJ relation. Alternatively, when using the Mg$_{\rm b}$--$L_{\rm tot}$ relation, we get $f_{\rm EL}=61\pm24\%$. As noted in Section \ref{sec:linestrenghts}, the fit in Equation (\ref{eq:lmlowmass}) is highly sensitive to the scaling of the Mg$_{\rm b}$ axis. When we perform the fit in linear instead of logarithmic units of Mg$_{\rm b}$, the EL fraction increases to $f_{\rm EL}=80\pm13\%$.

Both approaches are similar to the BCG/EL\footnote{We refer to the ICL fractions in \cite{Kluge2021} as EL fractions to be consistent with our current terminology (see Section \ref{sec:introduction}).} dissection method (a) in \cite{Kluge2021}, where we assume a maximum brightness of $M = -21.85~g'$ mag for normal Es. At that point, the size--brightness relation breaks its slope \citep{Kluge2020}. This method gives $f_{\rm EL}=71\pm22$. We conclude that all three methods provide consistent results.

We notice that these values are higher than those inferred using other, more conventional methods. SB profile decompositions using two S\'ersic functions find only $f_{\rm EL}=52\pm21\%$ of the BCG light in the outer S\'ersic component. The discrepancy is not surprising because, as we have already discussed previously \citep{Bender2015,Kluge2021}, two-component SB decompositions of BCGs have no physical basis (see also \citealt{Remus2017}). Similarly, an SB threshold at ${\rm{SB}}=27~g'~{\rm mag~arcsec^{-2}}$ results in only $f_{\rm EL}=34\pm19\%$ of the BCG light at fainter SBs \citep{Kluge2021}. By directly comparing SB profiles, our new approach is better suited to detect the EL, which is not present around luminous normal Es. When this definition of EL is applied, conventional methods underestimate the EL.

If we understand the EL as a lower limit for the accreted stellar component in BCGs $f_{\rm EL}<f_{\rm ex\mbox{-}situ}$, our results are consistent with numerical simulations. In the IllustrisTNG simulation, the ex situ stellar mass fraction is around $f_{\rm ex\mbox{-}situ}\approx80\%$ \citep{Pulsoni2021,Montenegro-Taborda2023} for our considered galaxy stellar mass range. Consistently, in the Magneticum simulation, the ex situ fraction is around $f_{\rm ex\mbox{-}situ}=75\pm10\%$ \citep{Remus2022}. That agrees well with our result. Finally, for the most-massive galaxies in the E-MOSAICS simulations, the ex situ stellar component dominates the surface mass density at all radii apart from the very center \citep{Reina-Campos2022}.

\subsection{Radial EL Fractions} \label{sec:radialEl}

We go one step further and calculate for each BCG the EL fraction $f_{\rm EL}(r)$ at radius $r$. Our definition of EL is the excess surface flux above an equivalent normal E with surface flux $I_{\rm E}(r)$. We approximate $I_{\rm E}(r)$ by the median of all normal E profiles with similar central velocity dispersions ($\delta\sigma_{\rm e8}=\pm10$\,km\,s$^{-1}$) or line strengths ($\delta{\rm Mg_b}=\pm0.05$\,\AA). By dividing with the BCG surface flux profile $I_{\rm BCG}(r)$, we infer the radial EL fraction

\begin{equation}
	f_{\rm EL}(r) = 1 - \frac{I_{\rm E}(r)}{I_{\rm BCG}(r)}. \label{eq:iclfrac}
\end{equation}

For simplicity, we ignore the broken slopes in the FJ and Mg$_{\rm b}$--$L_{\rm tot}$ relations (Figures \ref{fig:faberjackson} and \ref{fig:mgb_ltot}) and select all (black) reference ellipticals within the defined interval. Some of these reference galaxies deviate from the relations to higher luminosities. They probably already have low EL contamination, which will lead to a slight underestimation of $f_{\rm EL}(r)$. On the other hand, an advantage of this approach is its independence from the best-fit slope, which can strongly depend on the scaling of the axes (see Section \ref{sec:linestrenghts}).

Figure \ref{fig:iclfrac} shows the radial EL fractions for the individual BCGs (black) and also the median radial EL fraction (orange). The values within the 16th and 84th percentiles are marked by orange shading. The contamination from the EL near the BCG center gets small: $f_{\rm EL}(r<1.5\,{\rm kpc})<30\%$. This may be an upper limit, because the decreasing trend can be expected to continue toward smaller radii. The central velocity dispersions and line strengths are measured within an aperture of $r<1.5\arcsec\approx1$\,kpc. Hence, the necessary condition for our approach is sufficiently fulfilled, that is, a low contamination by the EL near the BCG center.

Going to larger radii, the EL contribution to the BCG surface flux reaches $f_{\rm EL}(r)=50\%$ already at $r_{\times,50}=3.5^{+4.8}_{-1.2}$\,kpc ($r_{\times,50}=2.8^{+2.6}_{-0.9}$\,kpc) for the FJ (Mg$_{\rm b}$--$L_{\rm tot}$) method. A contribution of $f_{\rm EL}(r)=80\%$ is reached at $r_{\times,80}=23^{+40}_{-12}$\,kpc ($r_{\times,80}=18^{+18}_{-8}$\,kpc). The errors quantify the scatter of the distributions.

The Mg$_{\rm b}$--$L_{\rm tot}$ method infers a marginally higher radial EL fraction than the FJ method. The former relation has a higher scatter. Hence, more lower-luminosity normal Es scatter into the reference region Mg$_{\rm b}\pm0.05$\,\AA~compared to $\sigma_{\rm e8}\pm10$\,km\,s$^{-1}$. This increases the radial EL fraction. Nevertheless, we conclude that both methods give consistent results.

If we understand the EL as a lower limit for the accreted stellar component, our average transition radius is consistent with the results of the Magneticum simulations $0\leq r_{\times,50}<13$\,kpc \citep{Remus2022}.

As the ICL is only a part of the EL, the EL contribution at intermediate radii is expectedly higher than the fiducial ICL contribution constrained by typical SB threshold or double S\'ersic decompositions. Those techniques predict that the ICL begins to exceed the BCG surface flux at around $r_{\times,50}\approx100$\,kpc (see Figure 1 in \citealt{Kluge2021}). Furthermore, our EL fractions can be compared to ICL fractions from semianalytic models by \cite{Contini2022}. Depending on the choice of simulation, the authors get $f_{\rm ICL}(r)=50\%$ at $r_{\times,50}=19\pm8$\,kpc or $r_{\times,50}=26\pm16$\,kpc and $f_{\rm ICL}(r)=80\%$ at $r_{\times,80}=38\pm19$\,kpc or $r_{\times,80}=43\pm26$\,kpc. These transition radii are farther out than what we observe for the EL. However, the total ICL fraction is consistent. For the mean BCG stellar mass in our sample $M\approx10^{12}\,{\rm M_{\odot}}$, \cite{Contini2022} found $f_{\rm ICL}\approx71\pm10\%$, which agrees well with our result $61\%<f_{\rm EL}<80\%$ (see Section \ref{sec:discussion_totalicl}).

\subsection{EL Formation: The Roles of Minor and Major Mergers -- SB Profiles} \label{sec:discussion_offset}

We analyze how mergers with different mass ratios transform luminous normal Es into BCGs. Therefore, we begin by comparing the observed SB profiles to simulations from \cite{Hilz2013}. Firstly, we match the SB profile of the simulated ``seed" pre-merger galaxy (Figure \ref{fig:accretion}, continuous black line) to the mean luminous normal E profile (Figures \ref{fig:allprofiles} and \ref{fig:accretion}, dashed black line) by scaling the effective radius to $r_{\rm e}=7$\,kpc and the effective SB to ${\rm SB_e}=23~g'$ mag arcsec$^{-2}$. The S\'ersic index $n$ is not scaled because it would deform the curvature of the profile. The simulated seed galaxy has $n=3.9$, which is very close to the value of the mean luminous normal E profile, for which we measure $n=3.77$. This is a very good agreement considering the scatter $\delta n=1.5$ for the individual luminous normal Es \citep{Kluge2020}.

After five generations of minor mergers with mass ratio 1:5, the total luminosity of the simulated galaxy has increased by a factor $\Delta L_{\rm sim,minor}=1.9$,\footnote{The total luminosity does not add up to $1.2^5\approx2.5$ because some particles become unbound during the mergers \citep{Hilz2013}.} and the resulting S\'ersic index is $n=9.5$. According to the scaling relation provided by \cite{Hilz2013}, the effective radius has grown by a factor $\Delta L_{\rm sim,minor}^{2.4}\approx4.7$. The corresponding SB profile is shown in Figure \ref{fig:accretion} by the green line. It does \textit{not} reproduce the mean BCG profile well (dashed red line). The curvature is too strong considering the relatively low S\'ersic index of the mean BCG profile $n=5.93$ (intrinsic scatter $\delta n=2.6$) and the brighter SB around $r\approx10$\,kpc is not reached either.

We now contrast this to major mergers. After one merger with mass ratio 1:1, the simulated galaxy has grown in luminosity by a factor of $\Delta L_{\rm sim,major}=2$, the effective radius has increased by a factor $\Delta L_{\rm sim,major}^{0.91}\approx1.9$, and the S\'ersic index has increased from $n=3.9$ to $n=5.7$. The resulting profile is shown by the orange line in Figure \ref{fig:accretion}. It is also not a good approximation to the mean BCG profile. However, we notice that contrary to minor mergers, major mergers promisingly predict surface flux growth in the inner $r<10$\,kpc. Similar growth is also seen in the evolution from luminous normal Es to BCGs. Can multiple major mergers predict the correct mean BCG profile?

\cite{Hilz2013} only provided data for two generations of major mergers. However, the luminosity growth from the mean luminous normal E to the mean BCG is a factor $\Delta L_{\rm obs}=7$. Hence, we need to extrapolate the simulations by one more major merger, resulting in a total luminosity growth of the simulated galaxy by $\Delta L_{\rm sim,major}=7.8$. According to the scaling relation provided by \cite{Hilz2013}, the effective radius grows by a factor $\Delta L_{\rm sim,major}^{0.91}\approx6.5$. For the S\'ersic index $n$, we assume both a constant $n=6.9$ and a linear extrapolation to $n=8.1$ (see Figure 5 in \citealt{Hilz2013}). The width of the red line in Figure \ref{fig:accretion} encompasses both values of $n$. It shows that the profile is rather insensitive to the exact choice. This profile provides a very good approximation to the mean BCG profile, especially around $r\approx 10$\,kpc, where the minor-merger scenario predicts no increase in SB. Our predicted number of three major mergers is consistent with the Illustris TNG300 simulation, in which BCGs have experienced, on average, three major mergers since $z=2$ \citep{Montenegro-Taborda2023}. However, the merging history is not entirely comparable because the authors include lower mass ratios $>$1:4, while we refer to 1:1 mass ratios only.

The radial growth with increasing mass $M$ predicted by \cite{Hilz2013} is $r_{\rm e}\propto M^{0.91}$ for 1:1 mergers and $r_{\rm e}\propto M^{2.4}$ for 1:5 mergers. In our observed data, we find $r_{\rm e}\propto L^{0.69\pm0.02}$ for the full normal Es sample and $r_{\rm e}\propto L^{1.63\pm0.10}$ for the BCG sample, where $L$ is the $g'$-band luminosity. While the former exponent is more consistent with the major-merger scenario, the latter exponent is between the two predictions. That means both types of mergers are relevant for BCG growth.

\subsection{EL Formation: The Roles of Minor and Major Mergers -- FP Offset} \label{sec:discussion_offset_2}

Figures \ref{fig:fundamental_plane_3d} and \ref{fig:fundamental_plane_projected} reveal an offset between the FPs of BCGs and normal Es. BCGs have systematically larger effective radii than expected from an extrapolation of the FP of normal Es (Figure \ref{fig:fundamental_plane_projected}, top-right panel). The offset is also visible in the $\kappa$-space projection (Figure \ref{fig:fundamental_plane_kappa}, top panel). BCGs \textit{appear} to have lower mass-to-light ratios than the extrapolation of normal Es predicts (top panel). This result can be interpreted as the contribution from the EL. It is worth investigating further.

We begin by interpreting the saturation in velocity dispersion (Figure \ref{fig:faberjackson}) as the main cause for the offset. Shifting the BCG data points by $+0.11$\,dex in the velocity dispersion direction matches both planes approximately (Figure \ref{fig:fundamental_plane_3d}). Which physical process causes the central velocity dispersion to stall while growing the galaxy in size? The mass ratio of merging galaxies is important in this context. Minor mergers with mass ratio 1:10 leave the central velocity dispersion constant \citep{Hilz2012}. Contrarily, one head-on major merger with 1:1 mass ratio increases the central (3D) velocity dispersion by $\sim25\%$ (Figure 17 in \citealt{Hilz2012}). We only observe a very slight increase in $\sigma_{\rm e8}$ for the BCGs relative to the luminous normal Es (Figure \ref{fig:faberjackson}). If major mergers do occur, they must have sufficient angular momentum in order to leave the central velocity dispersion almost unaffected, which in fact is not unlikely.

Each of the major- and minor-merger scenarios shift the BCG data points in Figure \ref{fig:fundamental_plane_projected} relative to the normal E data points. The directions are indicated by the arrows. Minor mergers move them along the FP, while major mergers offset them in a direction consistent with the BCGs.

\subsection{EL Formation: The Roles of Minor and Major Mergers -- General Remarks} \label{sec:discussion_offset_3}

In principle, as normal Es grow to become BCGs, 1:1 mergers become less likely. Only 7\% of present-day clusters host two similarly bright BCGs \citep{Kluge2020}. Also, the steep bright end of the galaxy luminosity function displays the rarity of bright galaxies (e.g., \citealt{Rines2008,Agulli2016,Agulli2017}). Nevertheless, we have shown in Sections \ref{sec:discussion_offset} and \ref{sec:discussion_offset_2} that a significant contribution to BCG growth must arise from major mergers, which increase the surface flux at intermediate radii $r\lesssim20$\,kpc.

The presence of age, metallicity, and color gradients (e.g., \citealt{Montes2018a,Gu2020}) disfavors the major-merger scenario. Violent relaxation would mix up different stellar populations and thereby erase or at least mitigate any gradients \citep{Contini2021}. The low observed EL metallicities require micromergers with stellar masses around $10^8-10^9$ M$_\odot$ (\citealt{Gu2020}; see also Section \ref{sec:introduction}). However, many thousands of these galaxies need to be accreted by the BCG to account for the EL total stellar mass if these dwarfs were the only contributors. For the examples of the Virgo and A2199 clusters, there exist only a few hundred of such galaxies \citep{Rines2008}. After integrating the luminosity function of A2199 for $10^8<{\rm M_\odot}<10^9$ galaxies, we find that these dwarfs make up only 10\% of the A2199 BCG luminosity. By considering the high average EL fraction $61\%<f_{\rm EL}<80\%$, we conclude that these dwarf galaxies cannot be the only contributors to the EL.

A possible reconciliation involves mass segregation (\citealt{Kim2020} and references therein). Dwarf galaxies with low metallicity are less bound and thus easily disrupted at large clustro-centric distances by the cluster tidal field or by galaxy--galaxy interactions \citep{Moore1996,Murante2007}. Massive, high-metallicity galaxies are better protected by their deeper gravitational potential. They are efficiently slowed down on their orbits by dynamical friction \citep{Chandrasekhar1943,Boylan-Kolchin2008}. When they arrive at a few tens of kiloparsecs from the BCG center, they get efficiently stripped by tidal forces \citep{Contini2018}. The released stellar material accumulates as EL at intermediate radii. The tightly bound nuclei of the progenitors remain intact for a longer time and are commonly observed as multiple nuclei in BCGs \citep{Hoessel1980,Schneider1983,Tonry1985a,Tonry1985b,Lauer1988,Kluge2020}. The lower-metallicity halos of massive galaxies are less bound than their nuclei. They already get tidally stripped at larger clustro-centric distances. This gradual stripping keeps the color gradients intact.

The proposed scenario requires further analysis of merger simulations with 1:2 or 1:3 galaxy mass ratios, similar to \cite{Hilz2012,Hilz2013}. In the Illustris TNG300 simulation, BCGs have experienced, on average, three (two) mergers with mass ratio $>$1:4 since $z=2$ ($z=1$) \citep{Montenegro-Taborda2023}. Whether these mergers can deposit enough stellar light at sufficiently small radii remains to be investigated.

\section{Summary and Conclusions} \label{sec:conclusions}

We have measured central stellar velocity dispersions from (a) new spectroscopic LRS2 observations and (b) archival SDSS data of 115 BCGs in total. Structural parameters of the BCGs are taken from \cite{Kluge2020} with minor modifications. A comparison sample of 1420 normal ellipticals (Es) was selected from the catalog by \cite{Zhu2010}. For those galaxies, new structural parameters, central velocity dispersions, and absorption line strengths have been derived from archival SDSS data to establish consistency with the BCG data. The structural parameters are obtained by directly integrating the light profiles. Analytic S\'ersic functions are only used for SB extrapolation and the derivation of the S\'ersic index $n$. We fit the FP, FJ, and Mg$_{\rm b}$--luminosity relations to normal Es and BCGs separately. Sample selection biases are mitigated by applying orthogonal sample cuts and luminosity weighting. 
\\~\\
Our main results are as follows:

\begin{enumerate}
	\item BCGs and luminous normal (cored) ellipticals have consistent central velocity dispersions, central absorption line strengths, and central SBs. However, BCGs are more luminous. This motivates a picture in which BCGs have grown from seed objects similar to present-day luminous Es. The central regions still trace the seed elliptical. 

	\item  We estimate the excess light using the FJ and Mg$_{\rm b}$--luminosity relations. It makes up 60--80\% of the total BCG luminosity. This is higher than the fiducial ICL fractions obtained with conventional SB threshold or profile decomposition methods. When we interpret the EL as a lower limit for the accreted and merged-in stellar component, our results are consistent with the IllustrisTNG, Magneticum, and E-MOSAICS simulations.

	\item We estimate the radial contribution of the BCG excess light to the BCGs by subtracting the surface flux profiles of luminous normal Es from BCGs with the same velocity dispersions or line strengths. The EL fraction is small in the center but reaches 50\% already at $r\approx3$\,kpc radius and 80\% at $r\approx20$\,kpc. This increase is much steeper than predicted by conventional BCG/ICL decomposition methods or semianalytic models. BCGs are brighter than normal Es not only at large radii, but at {\it all} radii beyond $r\gtrsim1$\,kpc.

	\item The FPs of normal Es and BCGs are offset relative to each other by 0.14\,dex. This can be explained by the shape of the radial EL profile. BCGs have $39\pm5\%$ larger effective radii than expected from an extrapolation of the FP of normal Es.

	\item The BCG FP has a $50\pm18\%$ steeper dependence on velocity dispersion $\sigma_{\rm e8}$ because of $\sigma_{\rm e8}$ saturation at high luminosities. It also has a $10\pm4\%$ steeper dependence on $\langle{\rm SB_e}\rangle$. Generally, BCGs populate different areas in the FP parameter space. As expected, they have much larger effective radii and fainter $\langle{\rm SB_e}\rangle$s than normal Es.

	\item The evolution from luminous normal Es to BCGs, and in particular the SB profile shapes, is consistent with simulations of three major mergers.
	
	\item In contrast, minor mergers deposit too few stars at intermediate radii $r\lesssim20$\,kpc and cannot reproduce the relatively low S\'ersic index $n=5.93$ of the median BCG SB profile. Nevertheless, negative age and metallicity gradients in BCGs likely require micromergers of dwarf galaxies or a more gradual stripping first of the outskirts of accreted galaxies and later of their interiors at smaller radii. A possible solution involves mass segregation. Massive galaxies are efficiently slowed down on their orbits by dynamical friction. When they arrive at a few tens of kiloparsecs from the BCG, their outskirts are efficiently stripped by tidal forces. The released stellar material accumulates as EL at intermediate and large radii.

\end{enumerate}

\section*{Acknowledgments}

We are grateful to the HET observers Steven Janowiecki and Matthew Shetrone for their helpful correspondence during the period of observations. We also wish to thank the anonymous referee for providing comments and suggestions that allowed us to improve the paper.

The Wendelstein 2m telescope project was funded by the Bavarian government and by the German Federal government through a common funding process. Part of the 2m instrumentation including some of the upgrades for the infrastructure were funded by the Cluster of Excellence ``Origin of the Universe" of the German Science foundation DFG.

This work made use of data products based on observations made with the NASA/ESA Hubble Space Telescope, and obtained from the Hubble Legacy Archive, which is a collaboration between the Space Telescope Science Institute (STScI/NASA), the Space Telescope European Coordinating Facility (ST-ECF/ESA), and the Canadian Astronomy Data Centre (CADC/NRC/CSA).

This work would not have been practical without extensive use of NASA's Astrophysics Data System Bibliographic Services and the SIMBAD database, operated at CDS, Strasbourg, France.

We also used the image display tool SAOImage DS9 developed by Smithsonian Astrophysical Observatory and the image display tool Fitsedit, developed by Johannes Koppenhoefer.

\software{Astropy \citep{Astropy2018}, Photutils \citep{Bradley2021}, pPXF \citep{Cappellari2017}, astroquery \citep{Ginsburg2019}, numpy \citep{Numpy2011}, scipy \citep{Virtanen2021}, matplotlib \citep{Hunter2007}, pyphot \citep{Fouesneau2022}.}

\facilities{HET (LRS2), Sloan, WO:2m (Wide-field camera), HST (WFPC2, ACS)}

\bibliography{bibo3}
\bibliographystyle{aasjournal}

\appendix

\section{Spectra} \label{sec:spectra}

\begin{figure*}[b]
	\centering
	\includegraphics[width=\linewidth]{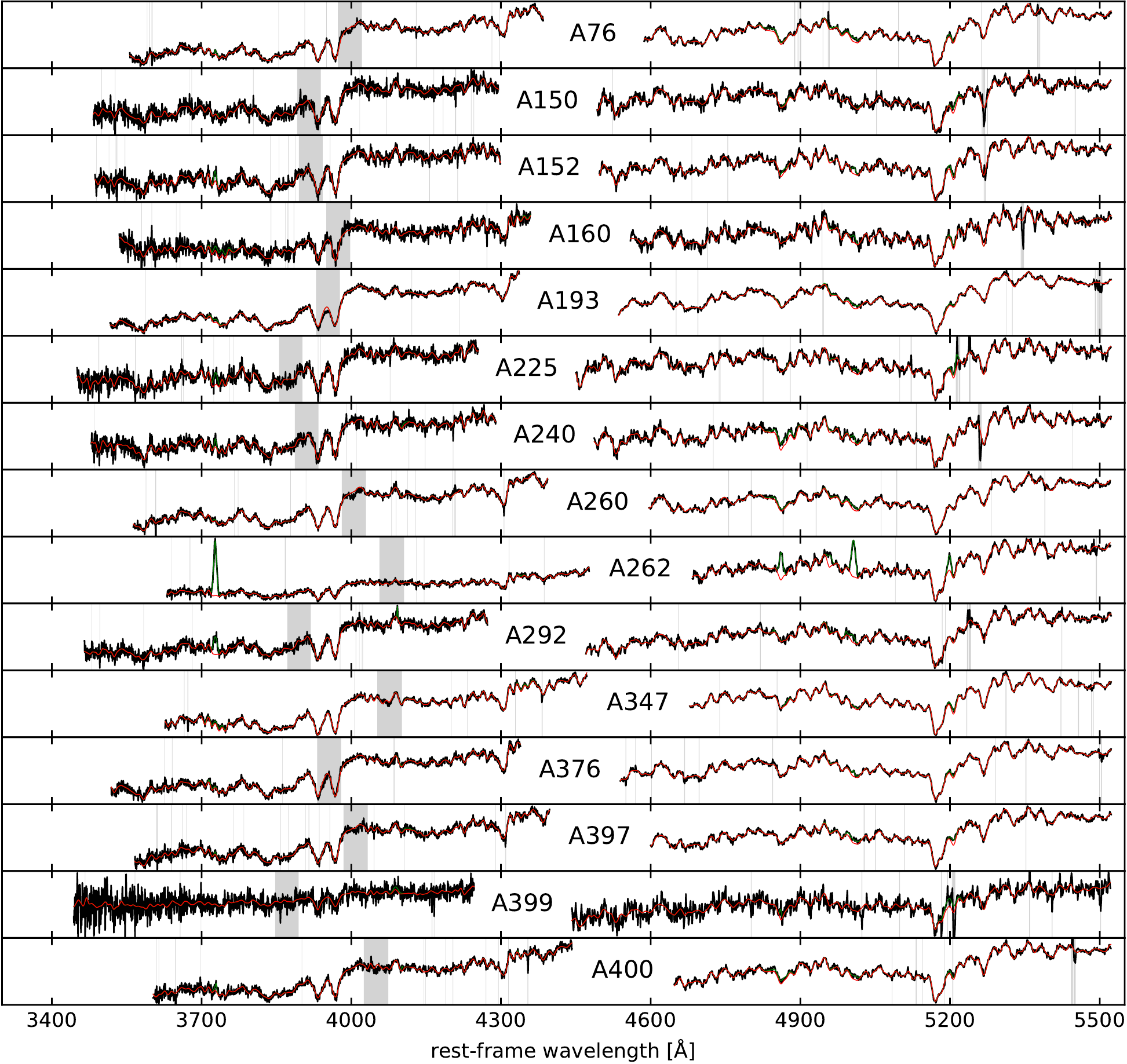}
	\caption{LRS2 uv and orange channel spectra of 75 BCGs. The LRS2 spectra are available as data behind the Figure. \label{fig:allspec}}
\end{figure*}

We present in Figure \ref{fig:allspec} the LRS2 uv and orange spectra of 75 BCGs. Black lines show the observed spectra. Red lines are the best-fit broadened X-shooter stellar templates. Green lines represent the contribution by gas emission. Gray shades are masked during the fit. For continuum matching, we have used a fourth-order multiplicative polynomial.

\begin{figure*}
	\centering
	\includegraphics[width=\linewidth]{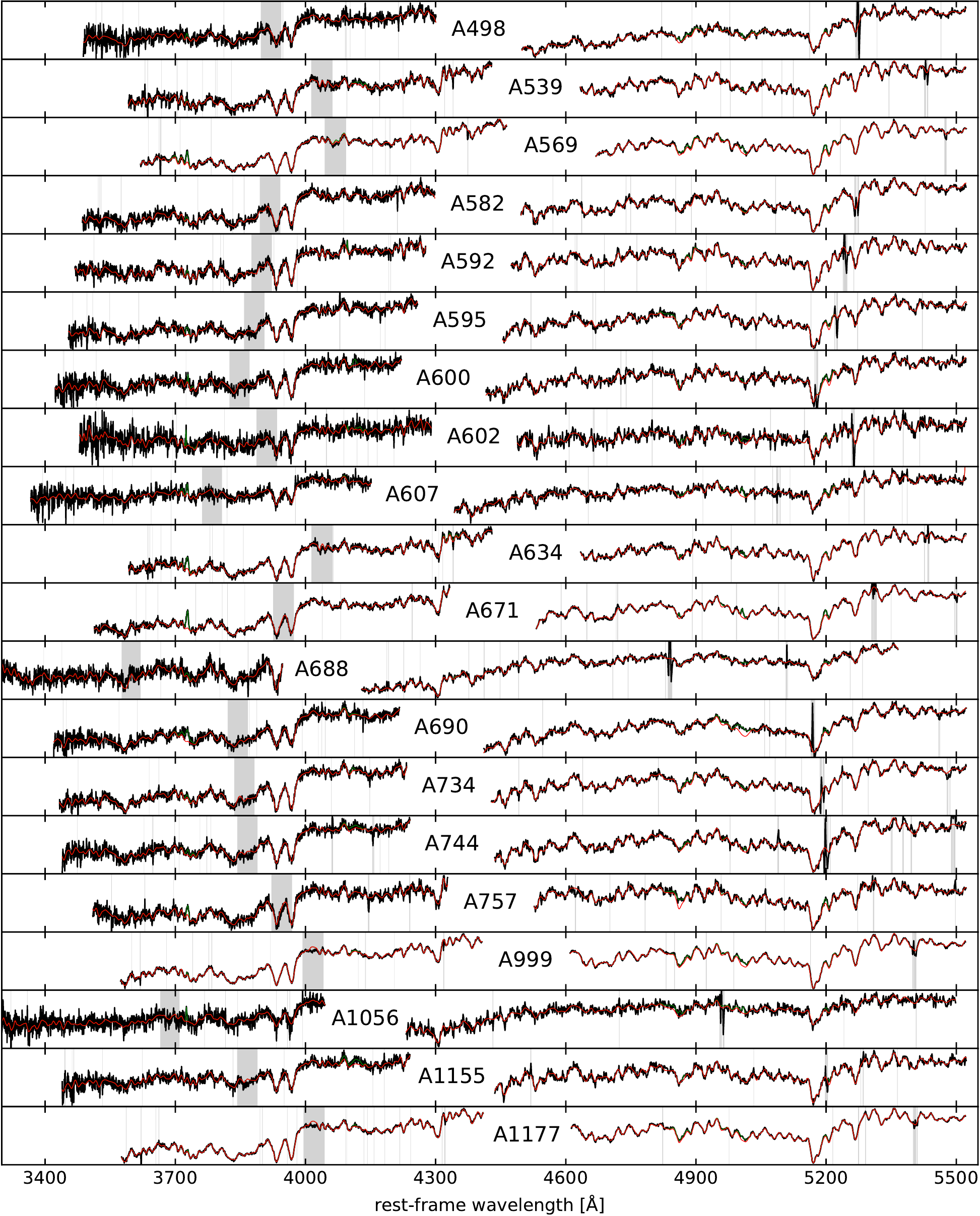}
	\textbf{Figure \ref*{fig:allspec}} \textit{(continued)}
\end{figure*}
\begin{figure*}
	\centering
	\includegraphics[width=\linewidth]{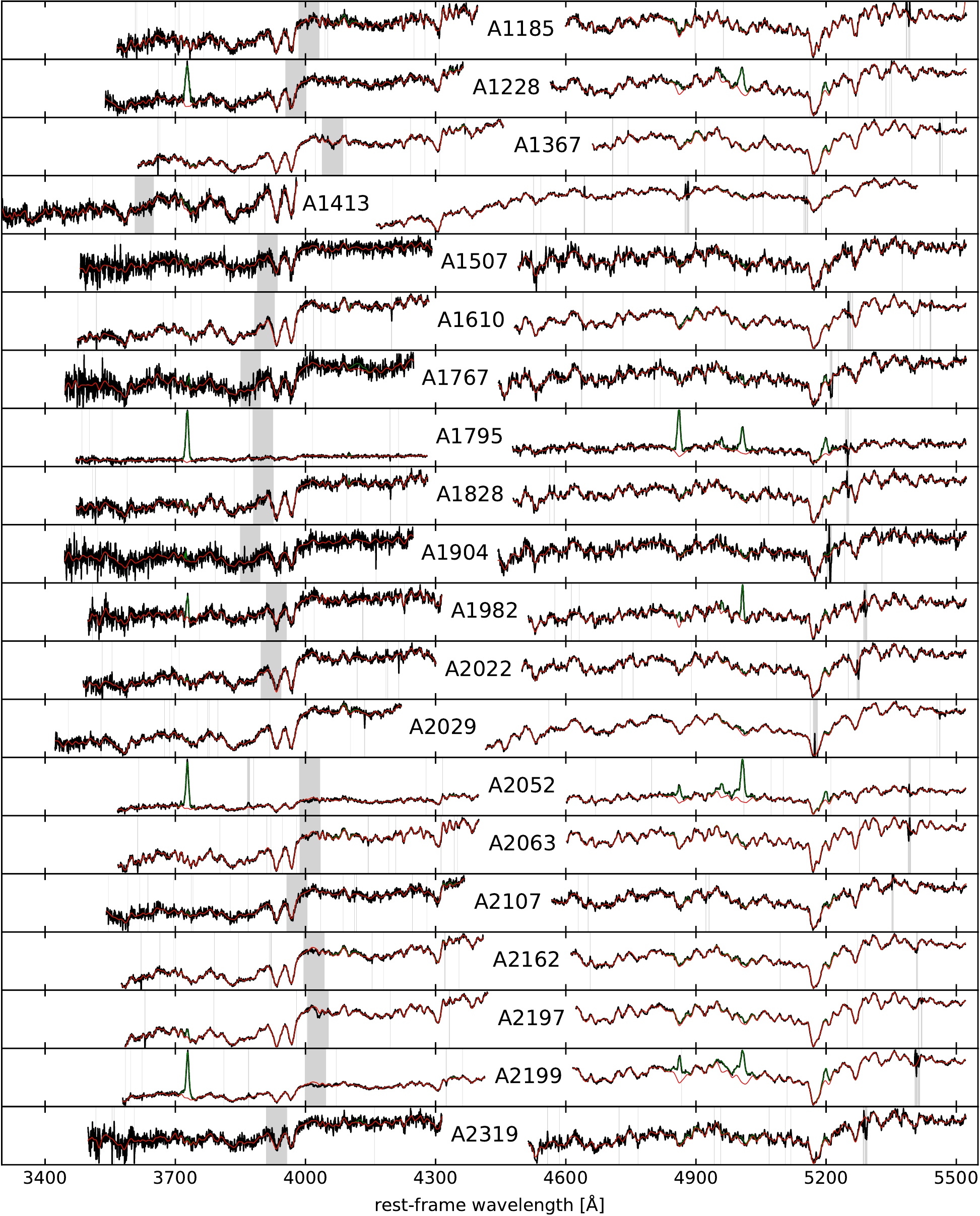}
	\textbf{Figure \ref*{fig:allspec}} \textit{(continued)}
\end{figure*}
\begin{figure*}
	\centering
	\includegraphics[width=\linewidth]{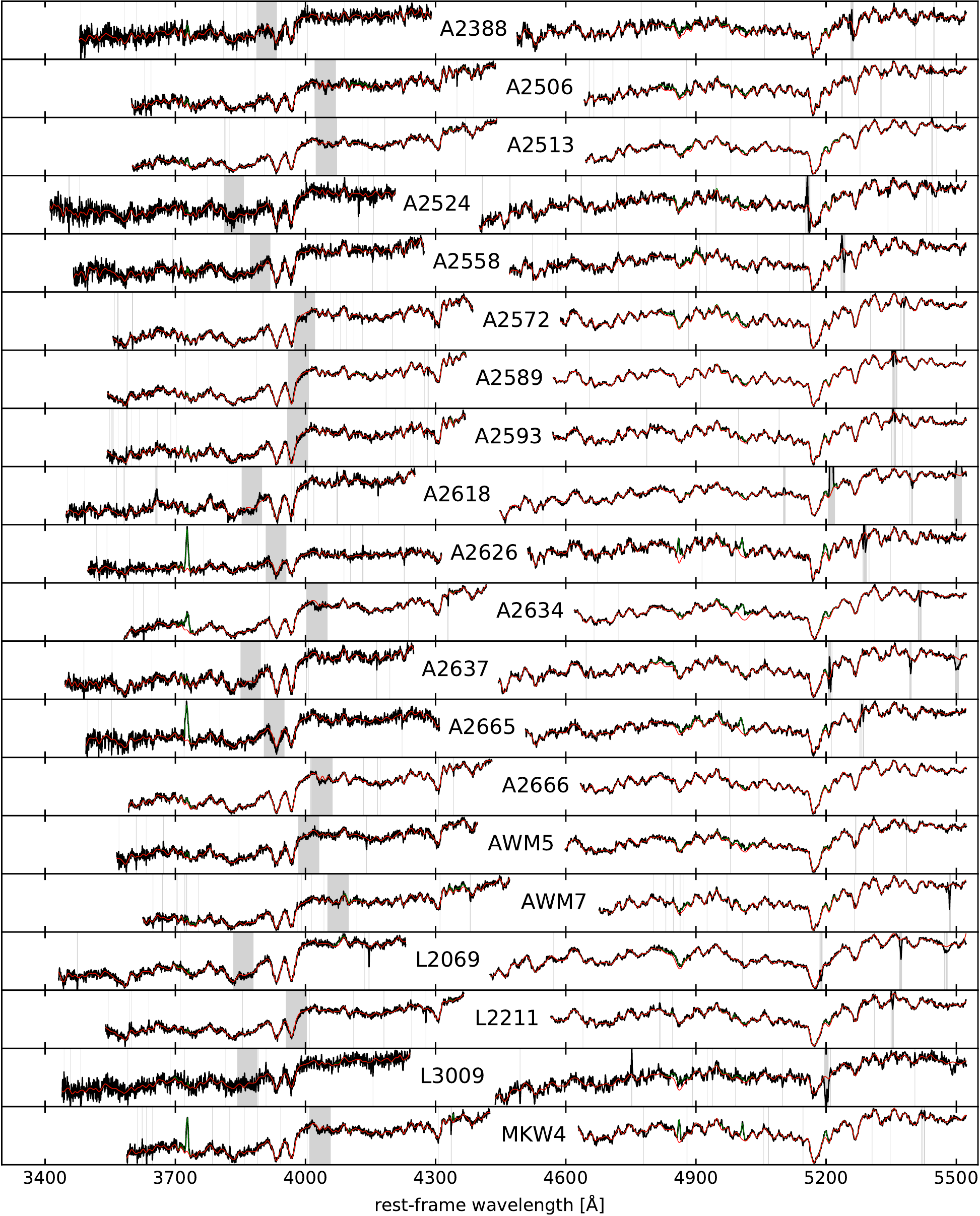}
	\textbf{Figure \ref*{fig:allspec}} \textit{(continued)}
\end{figure*}

\clearpage

\twocolumngrid

\section{Robustness of the Velocity Dispersion} \label{sec:robustness}

\begin{figure*}
	\includegraphics[width=\linewidth]{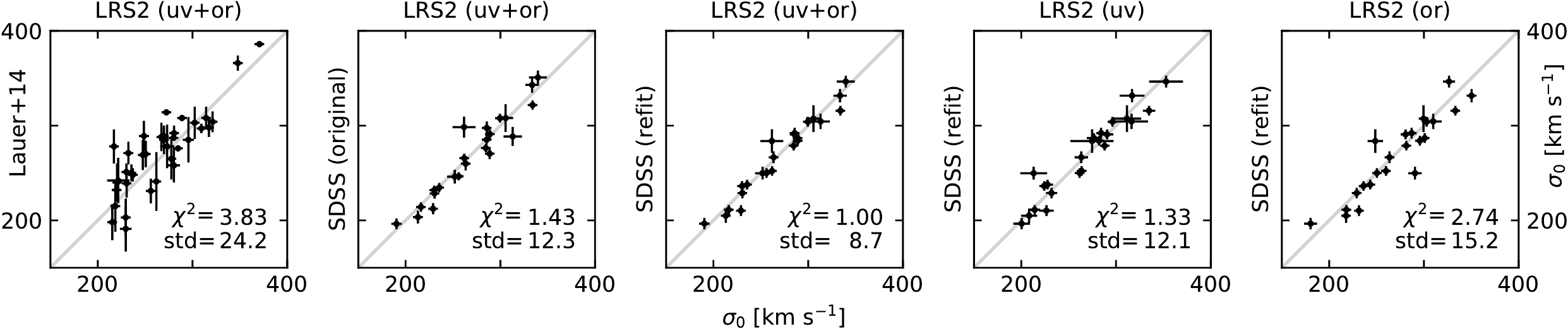}\\
	\caption{Comparisons between central velocity dispersions. Measurements on LRS2 spectra are plotted along the $x$-axis. The first three panels show mean dispersions from the uv and orange channels, whereas the fourth and fifth panels show only uv or orange channel dispersions, respectively. On the $y$-axes are plotted independent dispersions by \cite{Lauer2014} (left panel; 37 BCGs), published SDSS dispersions (second panel; 23 BCGs), and our refitted dispersions on SDSS spectra (third to fifth panels; 23 BCGs). $\chi^2$ values are given as text labels. Minimum error bars are adapted, such that $\chi^2=1$ in the middle panel. The standard deviation of the velocity dispersion differences is also given in the text labels and is in units of kilometers per second. \label{fig:sigma_lauer_kluge}}
\end{figure*}

The accuracy of the recovered LOSVD and measured line strengths depends on various factors: the fitting algorithm, template and continuum mismatch, sky subtraction, signal-to-noise ratio (S/N), and instrumental resolution. In this section, we concentrate the discussion mostly on the robustness of the velocity dispersion, because it is the most important parameter for the fundamental plane (FP) analysis. Therefore, we compare our results from the LRS2 uv and orange channels to data from \cite{Lauer2014}, SDSS \citep{Alam2015}, and our refitted SDSS spectra (Figure \ref{fig:sigma_lauer_kluge}). Effects of choosing different continuum-matching polynomials (Figure \ref{fig:dispersion_polynomials}) and a different stellar library (Figure \ref{fig:dispersion_libraries}) are also discussed.

\subsection{Comparison with Literature} \label{sec:sigmacomparison}

We begin by comparing our averaged LRS2 uv and orange channel dispersions with \cite{Lauer2014} (Figure \ref{fig:sigma_lauer_kluge}, left panel). The authors utilize the same fitting algorithm {\tt pPXF} as we do. For the overlapping 37 BCGs, the mean difference is $\Delta\sigma_0 = \sigma_{0,\rm{LRS2}} - \sigma_{0,{\rm Lauer}} =-7.8\pm4.0{\rm \,km\,s^{-1}}$ with a standard deviation ${\rm std}=24.2{\rm \,km\,s^{-1}}$. The significant offset and relatively large scatter can be explained by the different physical regions, which \cite{Lauer2014} probed in the BCG centers. Their observations were done using longslit spectrographs with 2\arcsec~ slit width and $9\pm2.5\arcsec$ extraction aperture width.

The SDSS aperture is equal to our circular aperture with 1.5\arcsec~radius. We find very good agreement between our averaged LRS2 uv and orange dispersions and the published SDSS dispersions (Figure \ref{fig:sigma_lauer_kluge}, second panel). For the overlapping 23 BCGs, the mean difference is $\Delta\sigma_0=0.9\pm2.6{\rm \,km\,s^{-1}}$ with a standard deviation ${\rm std}=12.3{\rm \,km\,s^{-1}}$. The SDSS algorithm performs a principal component analysis using 24 eigenspectra, which are trial broadened with $\Delta\sigma_{\rm trial}=25{\rm \,km\,s^{-1}}$ step size. The best-fit dispersion is identified by the minimum of the interpolated $\chi^2$ curve \citep{Aihara2011}. The agreement builds confidence that our results are robust with respect to the utilized fitting algorithm.

We also refitted the SDSS spectra using our procedure described in Section \ref{sec:templatefitting}. The comparison is shown in Figure \ref{fig:sigma_lauer_kluge}, third panel. The mean difference is $\Delta\sigma_0=1.2\pm1.8{\rm \,km\,s^{-1}}$ with ${\rm std}=8.7{\rm \,km\,s^{-1}}$. This very good agreement motivates extending our LRS2 sample of 75 BCGs with 40 more BCGs with available SDSS spectra. Moreover, Figure \ref{fig:devsdss_mysersic}, right panel, confirms that the robustness holds for the normal Es sample. Its much larger sample size reveals only a small systematic offset of $\Delta\sigma_0=2.1\pm0.2{\rm\,km\,s^{-1}}$ of the SDSS velocity dispersions (original) from our values (refit), both measured on the same spectra. This offset likely arises by our correction to the SDSS instrumental resolution (Section \ref{sec:obsstrat}).

The reduced $\chi^2$

\begin{equation}
	\chi^2 = \frac{1}{n} \sum_{i}^{n} \frac{(\sigma_{{\rm A},i} - \sigma_{{\rm B},i})^2}{\delta \sigma_{{\rm A},i}^2 + \delta \sigma_{{\rm B},i}^2} \label{eq:chi2}
\end{equation}

equals 1 when the uncertainties $\delta\sigma_{\rm A}$ and $\delta\sigma_{\rm B}$ are correctly estimated for samples A and B with velocity dispersions $\sigma_{\rm A}$ and $\sigma_{\rm B}$ of $n$ galaxies. For our comparison between the averaged LRS2 uv and orange dispersions and the refitted SDSS dispersions, we find $\chi^2=1.88$. Consequently, we conclude that the uncertainties are underestimated. By introducing a lower limit of $\delta\sigma_0\ge5.1~\rm{km~s^{-1}}$, our results become consistent with $\chi^2=1$.

Systematic errors due to template mismatch remain unnoticed when comparing spectra of the same galaxies obtained with different instruments. To quantify it, we follow three approaches: (1) varying the continuum-matching polynomials, (2) changing the stellar library, and (3) comparing dispersions measured from two different wavelength bands.

\subsection{Continuum-matching Polynomials}

The continua of the template and science spectra can vary on large scales because of dust reddening, flux-calibration uncertainties, and a so-called UV upturn (e.g., \citealt{Pipino2009,Loubser2011,Phillipps2020}). {\tt pPXF} allows compensating for this by adding or multiplying polynomials to the template spectra. The order of these polynomials is critical. If it is too high on the one hand, then real LOSVD wings can be erroneously removed. On the other hand, too low orders can leave a significant template mismatch, which the fitting algorithm tries to account for by distorting the LOSVD.

\begin{figure}
	\includegraphics[width=\linewidth]{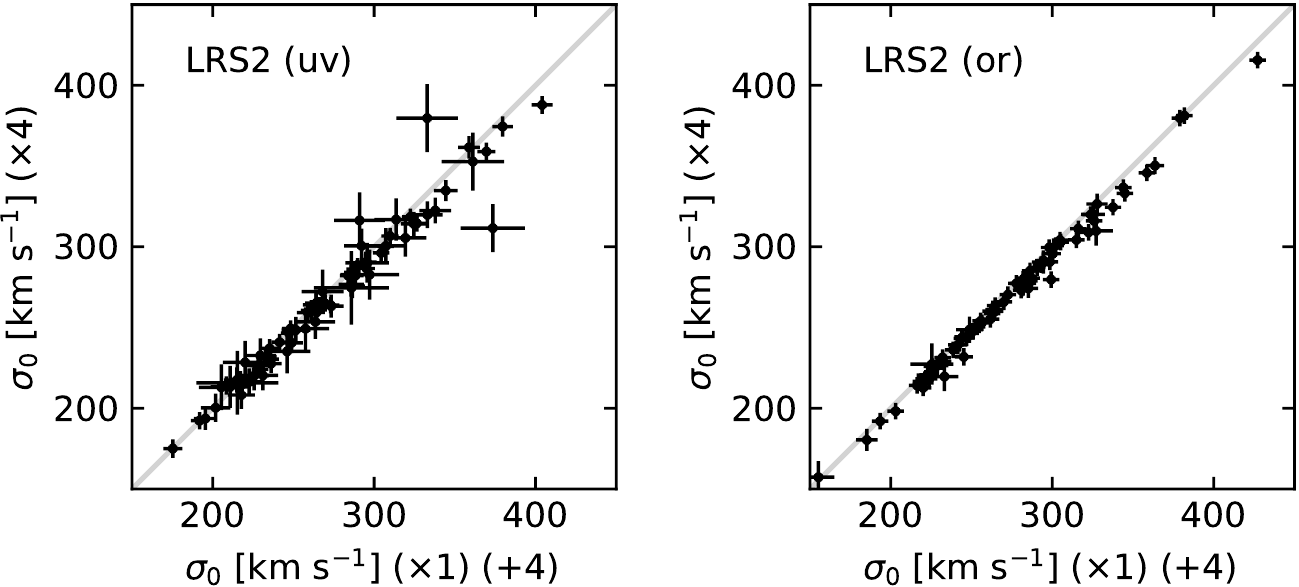}
	\caption{Comparison between the measured velocity dispersions in the uv (left) and or (right) channels, depending on the polynomial orders. The numbers after the ``$\times$" and ``$+$" symbols on the axes labels refer to the orders of the applied multiplicative and additive polynomials, respectively. That is, first-order multiplicative ($\times1$) and fourth-order additive polynomial ($+4$) on the $x$-axis and fourth-order multiplicative ($\times4$) and no additive polynomial on the $y$-axis. \label{fig:dispersion_polynomials}}
\end{figure}

In principle, either additive or multiplicative polynomials can correct continuum mismatch. Additive polynomials can partly compensate for template mismatch in absorption line strengths and sky scaling errors. Multiplicative polynomials can account for dust reddening and flux-calibration errors (e.g., \citealt{Cappellari2017}). We have tested many combinations of additive and multiplicative polynomials of different orders. A comparison in Figure \ref{fig:dispersion_polynomials} shows that the effect on the velocity dispersions is small for both LRS2 uv ($\Delta\sigma_0=4.0\pm1.3{\rm \,km\,s^{-1}}$; ${\rm std}=11.0{\rm \,km\,s^{-1}}$) and orange channels ($\Delta\sigma_0=4.8\pm0.5{\rm \,km\,s^{-1}}$; ${\rm std}=4.5{\rm \,km\,s^{-1}}$). We decide to use only fourth-order multiplicative polynomials, because they allow less freedom for the sky scaling factor. A tight constraint on that factor is important because sky-subtraction errors affect the line strength measurements.

\subsection{Changing the Stellar Library}

\begin{figure}
	\includegraphics[width=\linewidth]{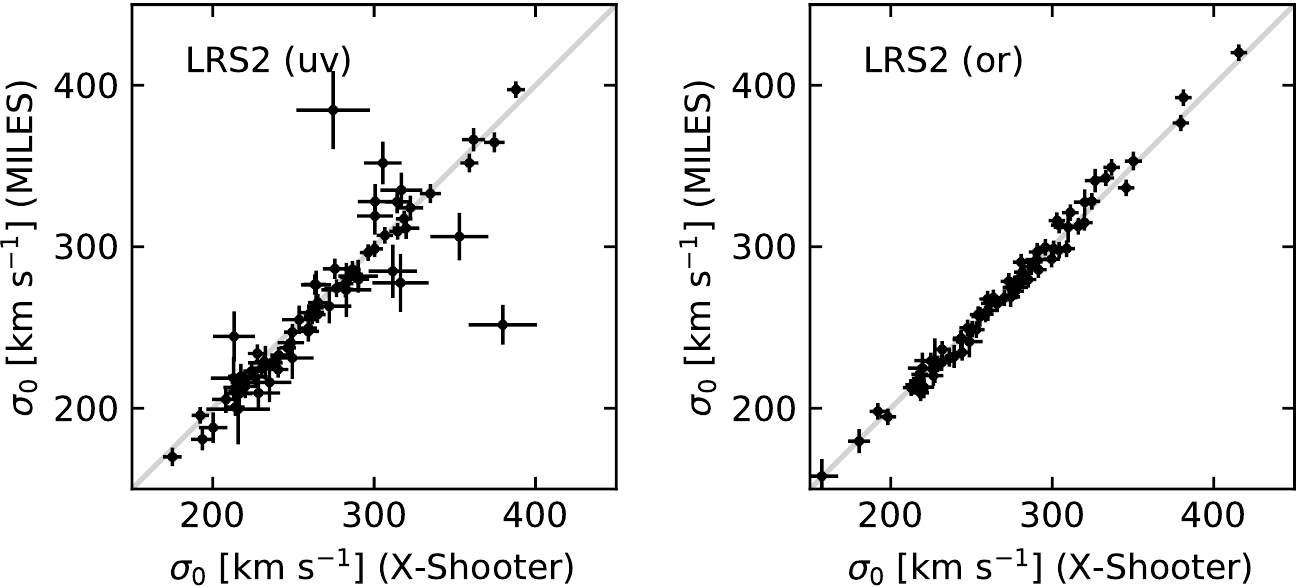}
	\caption{Comparison between the measured velocity dispersions in the uv (left) and or (right) channels, depending on the stellar library. The values on the $x$-axis correspond to the X-shooter stellar library, and the values on the $y$-axis correspond to the MILES stellar library. Fourth-order multiplicative and no additive polynomials are applied. \label{fig:dispersion_libraries}}
\end{figure}

By changing the stellar library, we estimate a lower boundary on the uncertainty due to template mismatch. A comparison between the velocity dispersions measured using stars from the X-shooter library DR2 \citep{Arentsen2019,Gonneau2020} versus stars from the MILES library v9.1 \citep{Sanchez-Blazquez2006} is shown in Figure \ref{fig:dispersion_libraries}. Again, the effect is small for the LRS2 orange channel ($\Delta\sigma_0=-0.4\pm0.7{\rm \,km\,s^{-1}}$ ${\rm std}=5.7{\rm \,km\,s^{-1}}$). It is relatively large for the LRS2 uv channel ($\Delta\sigma_0=3.5\pm2.7{\rm \,km\,s^{-1}}$; ${\rm std}=23.4{\rm \,km\,s^{-1}}$). This is due to a few outliers. The median absolute deviation is only ${\rm mad}=4.7{\rm \,km\,s^{-1}}$. These outliers have the highest redshifts in our sample and therefore low spectral coverage with the MILES library.

Moreover, the two libraries have very different instrumental resolutions (see Table \ref{tab:intrumentalresolution}). A special case occurs for the LRS2 uv spectra, which are better resolved than the MILES template spectra. Here, we degrade the LRS2 uv resolution to match the MILES instrumental resolution. The consistency of the results obtained with both libraries also confirms that instrumental broadening is correctly accounted for.

\begin{figure*}
	\includegraphics[width=\linewidth]{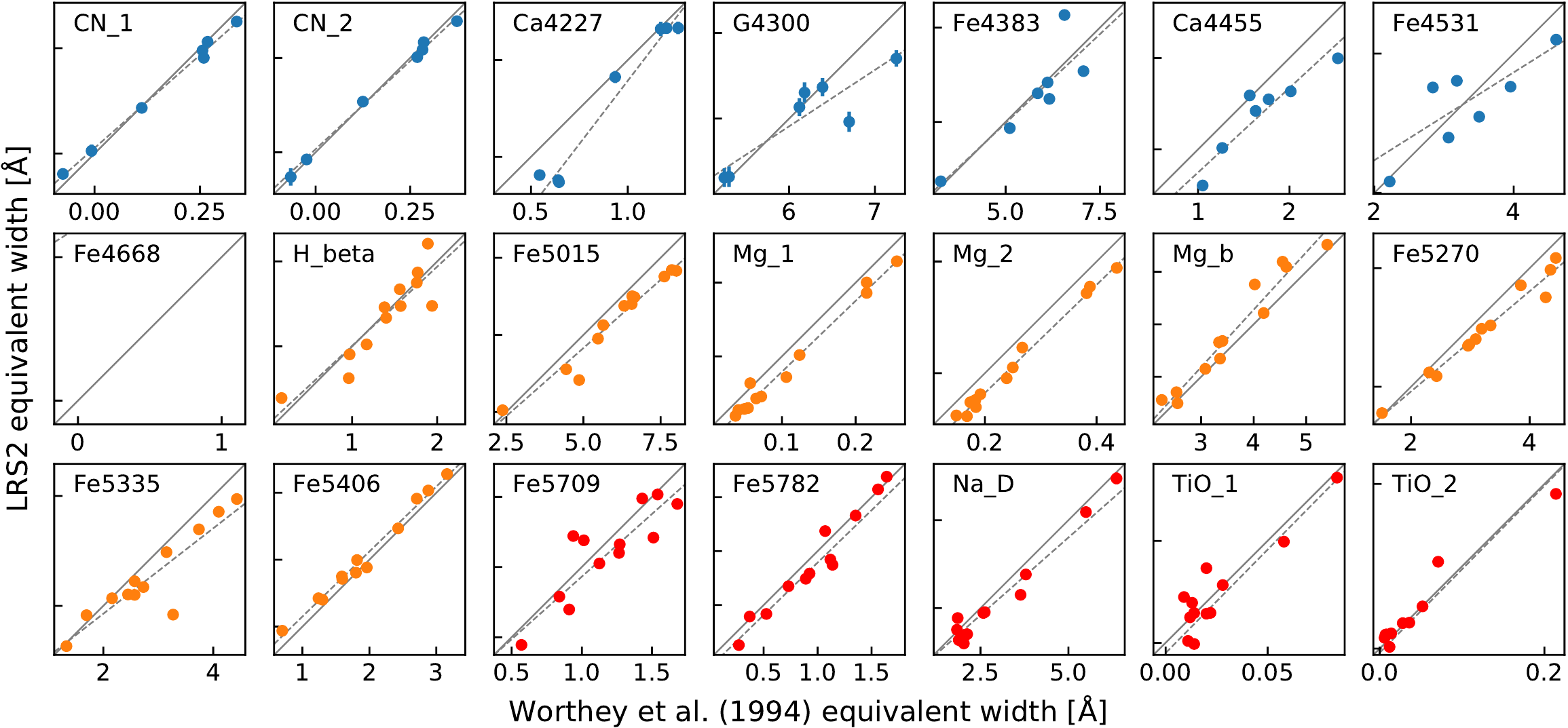}
	\caption{Comparison of Lick indices of standard stars measured by \cite{Worthey1994} ($x$-axis) and measured by us on LRS2 spectra ($y$-axis). Blue, orange, and red colors correspond to uv, orange, and red LRS2 channels, respectively. The continuous gray line shows is the 1:1 relation, and the dashed line is the fit that minimizes the squared orthogonal residuals. There is no data for the Fe4668 line, because it falls between the uv and orange channels. All units are in [\AA] apart from Mg$_1$, Mg$_2$, CN$_1$, CN$_2$, and all TiO lines, which are measured in [mag]. The compared standard stars are: HD101501, HD132142, HD64606, HR2600, HR3369, HR3418, HR3422, HR3427, HR3428, HR6136, HR6159, and HR7576. \label{fig:lickstars}}
\end{figure*}

\subsection{Comparing LRS2 uv and Orange Channels}

A comparison between the uv and orange channel velocity dispersions with the refitted SDSS velocity dispersions is shown in Figure \ref{fig:sigma_lauer_kluge}, fourth and fifth panels. The $\chi^2$ values are both larger than 1, meaning the uncertainties are underestimated. The orange channel spectra have on average 4.5 times higher signal-to-noise ratio (${\rm S/N/(30\,km\,s^{-1})}=70\pm30$) than the uv channel spectra (${\rm S/N/(30\,km\,s^{-1})}=16\pm9$) and they provide significantly smaller error bars in $\sigma_0$ (see Figures \ref{fig:dispersion_libraries} and \ref{fig:dispersion_polynomials}). Surprisingly, the standard deviations of the two datasets from the refitted SDSS dispersions are similar. That means a gain in S/N does not lower the deviations. This puts an upper limit on the effect of template mismatch of $\delta\sigma_0\le1.48;~{\rm std}\approx15{\rm \,km\,s^{-1}}$. However, we notice that unweighted averaging of the uv and or channel spectra results in a smaller scatter of the dispersions (${\rm std}=12.3{\rm \,km\,s^{-1}}$) than weighting them by their inverse squared uncertainties (${\rm std}=8.7{\rm \,km\,s^{-1}}$). This means that some of the template mismatch averages out when combining both dispersions. For completeness, we also mention that there could possibly exist distinct bluer and redder stellar populations with different intrinsic kinematics.

In summary, the impact of template mismatch is lowest when we average the LRS2 uv and orange dispersions without weighting them. Their uncertainties are also not weighted when we quadratically combine them, but we introduce a lower limit on the individual and combined uncertainties of $\delta\sigma_0\ge5{\rm \,km\,s^{-1}}$. This is motivated by comparing the measured uncertainties with deviations from refitted SDSS velocity dispersions. On the same order of magnitude are systematic effects by the choice of continuum-matching polynomials ($\Delta\sigma_0\approx5{\rm~km~s^{-1}}$) and stellar library ($\Delta\sigma_0\approx4{\rm~km~s^{-1}}$).

\section{Lick Indices} \label{sec:applick}

To verify our absorption line strength measurements, we have observed with the LRS2 12 standard stars, whose line strengths were measured by \cite{Worthey1994} and which span large dynamic ranges. We compare the results in Figure \ref{fig:lickstars}. Blue, orange, and red colors correspond to uv, orange, and red LRS2 channels, respectively. The continuous gray line shows the 1:1 relation and the dashed line is the fit that minimizes the squared orthogonal residuals. The agreement is satisfactory, given that we only aim to compare galaxies and are not interested in absolute values.

\section{Fundamental Plane Fitting Algorithm} \label{sec:fitting3d}

We use the python orthogonal distance regression algorithm {\tt scipy.odr} \citep{Boggs1989} for all fits. Our application uses the free parameters $\alpha$, $\beta$, and $\gamma$ to minimize the sum of the squared orthogonal residuals between the observed data points $X_i=(\log(\sigma_{\rm e8})_i,\log\langle I_{\rm e}\rangle_i,\log(r_{\rm e})_i)$ and the plane $P$, which is expressed by an implicit definition

\begin{equation}
	P(X,\alpha,\beta,\gamma) = \alpha \log(\sigma_{\rm e8}) + \beta \log\langle I_{\rm e}\rangle + \gamma - \log(r_{\rm e}).
\end{equation}

The shortest distance from the $i$th observed data point $X_i$ to the plane, which is simultaneously its orthogonal residual $\Delta_{{\rm obs,orth},i}$, is

\begin{equation}
	\Delta_{{\rm obs,orth},i} (X_i, \alpha,\beta,\gamma) = \frac{P(X_i,\alpha,\beta,\gamma)}{\sqrt{\alpha^2 + \beta^2 + 1^2}} \label{eq:deltaobsorth}
\end{equation}

This can be shown using a coordinate transformation and finding the closest point on the plane from the origin.

Additionally, the 3D residuals of each data point are weighted using the measurement uncertainties and the intrinsic scatter (see Appendix \ref{sec:intrinsicscatter}). The strong covariances between $\log(r_{\rm e})$ and $\log\langle I_{\rm e}\rangle$ are also considered (see Appendix \ref{sec:covariances}).

\subsection{Measurement Uncertainties $\delta_{\rm m}$} \label{sec:measuncert}

The estimation of the velocity dispersion measurement uncertainties is described in Appendix \ref{sec:robustness}. It is the direct output of {\tt pPXF} with a lower limit of $\delta_{\rm m}\sigma_0\ge5{\rm~km~s^{-1}}$. The photometric measurement uncertainties in $\delta_{\rm m}\langle I_{\rm e}\rangle$ and $\delta_{\rm m} r_{\rm e}$ are calculated from the uncertain extrapolation of the S\'ersic profiles for normal Es (Section \ref{sec:lowermass}). For the BCGs, this extrapolation is formally too certain, because the profiles are constrained to deeper SBs. Similarly to \cite{Kluge2020}, we extrapolate the SB profiles to intrinsic ${\rm SB}=30~g'~{\rm mag arcsec}^{-2}$ and to infinity, and take the average value.\footnote{It is to be expected that the cut at ${\rm SB}=30~g'~{\rm mag~arcsec^{-2}}$ gives an underestimation and the extrapolation to infinite radius gives an overestimation for the structural parameters. The truth is most likely in between, because the SB profiles are expected to drop off steeply near the splashback radius below ${\rm SB}=30~g'~{\rm mag~arcsec^{-2}}$. \citep{Diemer2014,Deason2021,Gonzalez2021}.} The standard deviation of both results is considered as the $3\sigma$ measurement uncertainty of $r_{\rm e}$ and $\langle I_{\rm e}\rangle$.

Not linear, but logarithmic parameters go into the FP. The linear error bars become asymmetric in logarithmic space. Because our FP fitting algorithm assumes Gaussian errors of the data points, we symmetrize them by

\begin{equation}
	\delta_{\rm m}\log(Y) = \frac{\log(Y+\delta_{\rm m} Y) + \log(Y-\delta_{\rm m} Y)}{2},
\end{equation}

where $Y$ stands for $\sigma_{\rm e8}$, $r_{\rm e}$, and $\langle I_{\rm e}\rangle$.

\subsection{Covariances between Photometric Parameters} \label{sec:covariances}

When fitting SB profiles, $r_{\rm e}$ and $\langle I_{\rm e}\rangle$ are strongly correlated. The bias by these covariances on the FP slopes can be corrected a posteriori using Monte Carlo realizations \citep{LaBarbera2010b}. We directly include them in the measurement covariance matrices $C_{\rm m}$ before fitting the FP:

\begin{equation}
	\boldsymbol{C_{\rm m}} = \begin{bmatrix}
		(\delta_{\rm m}\log(\sigma_{\rm e8}))^2 & 0 & 0 \\
		0 & (\delta_{\rm m}\log\langle I_{\rm e} \rangle)^2 & C_{rI}\\
		0 & C_{rI} & (\delta_{\rm m}\log(r_{\rm e})^2) \label{eq:cov_m}
	\end{bmatrix},
\end{equation}

where

\begin{equation}
	C_{rI} = \frac{\sum_k^n (\log(r_{\rm e})_k - \overline{\log(r_{\rm e})})(\log\langle I_{\rm e}\rangle_k - \overline{\log\langle I_{\rm e}\rangle})}{n-1}.
\end{equation}

The term $C_{rI}$ is calculated numerically while fitting the SB profiles from \cite{Kluge2020} along the effective axis. One-thousand random, but correlated, combinations are sampled from the (three single- or six double-) S\'ersic profile parameters. The resulting SB profiles are integrated numerically using the inner measured profile and the outer extrapolated S\'ersic profile. As mentioned in Appendix \ref{sec:measuncert}, two sets of parameters are averaged for the BCGs: the ones derived from an integration down to ${\rm SB}=30~g'$ mag arcsec$^{-2}$ and those derived from an integration out to an infinite radius. This method does not provide $C_{rI}$ covariances, so we assume a Pearson correlation coefficient of 0.9; that is, $C_{rI}=0.9\cdot\delta_{m}\log(r_{\rm e})\cdot\delta_{m}\log\langle I_{\rm e}\rangle$ for the BCGs. The exact choice has a negligible impact on the FP slopes and intrinsic scatter.

\subsection{Modeling Intrinsic Scatter $\delta_{\rm in}$} \label{sec:intrinsicscatter}

For bright, low-redshift galaxies, the measurement uncertainties are usually smaller than the intrinsic scatter of the FP. More precisely, the orthogonal component $\Delta_{\rm m,orth}$ of a typical measurement uncertainty ellipsoid is smaller than the orthogonal intrinsic scatter $\delta_{\rm in,orth}$ of the FP (e.g., \citealt{Hyde2009b,Samir2020}). It is, therefore, crucial to include the intrinsic scatter in the fit (e.g., \citealt{Gargiulo2009}). If not, the dissimilar measurement uncertainties (compare columns (6), (7), and (8) in Table \ref{tab:fpparams}) for the three observables would significantly bias the slopes in a weighted fit. The effect is most severe when the dynamic range of an observable is on the order of the intrinsic scatter. This is the case for the velocity dispersion $\sigma_{\rm e8}$ of the BCGs.

Without prior knowledge about the nature of the intrinsic scatter, we assume it to be generated by a spherically symmetric scatter, described by the diagonal covariance matrix

\begin{equation}
	\boldsymbol{C_{\rm in}} = \begin{bmatrix}
		(\delta_{\rm in})^2 & 0 & 0 \\
		0 & (\delta_{\rm in})^2 & \\
		0 & 0 & (\delta_{\rm in})^2
	\end{bmatrix}.
\end{equation}

Because this matrix describes a sphere, its projection along any unit vector, including the orthonormal vector $\boldsymbol{\hat{e}_{\rm orth}}$ to the plane, is

\begin{equation}
	\delta_{\rm in,orth} = \boldsymbol{\hat{e}_{\rm orth}}^\mathsf{T} \boldsymbol{C_{\rm in}} \boldsymbol{\hat{e}_{\rm orth}} = \delta_{\rm in}. \label{eq:delta_in_orth}
\end{equation}

The orthonormal vector is the normalized normal vector $\boldsymbol{n}$, which is the cross-product of two nonparallel vectors along the plane

\begin{equation}
	\boldsymbol{n} = \begin{bmatrix} 1\\0\\\alpha \end{bmatrix} \times \begin{bmatrix} 0\\1\\\beta \end{bmatrix} = \begin{bmatrix} -\alpha\\-\beta\\1 \end{bmatrix}.
\end{equation}

Normalization gives

\begin{equation}
	\boldsymbol{\hat{e}_{\rm orth}} = \frac{\boldsymbol{n}}{|\boldsymbol{n}|} = \begin{bmatrix} -\alpha\\-\beta\\1 \end{bmatrix} / \sqrt{\alpha^2 + \beta^2 + 1^2}.
\end{equation}

\subsection{Orthogonal Fit and Calculation of Intrinsic Scatter}

In a first run, the weights for each of the galaxy parameters are set equal. The fit is done by minimizing the sum of the squared orthogonal residuals (see Table \ref{tab:fpparams}). We use this result to estimate the orthogonal intrinsic scatter of the FP, which is afterward used for optimizing the parameter weighting in a second run.

We calculate the orthogonal intrinsic scatter using the orthogonal observed scatter $\delta_{\rm obs,orth}$ reduced by the median orthogonal measurement uncertainty ${\rm median}(\delta_{\rm m,orth})$. First of all, $\delta_{\rm obs,orth}$ is calculated by the standard deviation of the orthogonal residuals (see Equation \ref{eq:deltaobsorth}).

\begin{equation}
	\delta_{\rm obs,orth} = {\rm std}(\Delta_{\rm obs,orth}).
\end{equation}

To calculate the typical orthogonal measurement uncertainties, we project all of the $i$th measurement covariance matrices $\boldsymbol{C_{\rm m}}_{,i}$ (see Equation \ref{eq:cov_m}) onto the orthonormal vector:

\begin{equation}
	\Delta_{{\rm m,orth},i} = \boldsymbol{\hat{e}_{\rm orth}}^\mathsf{T} \boldsymbol{C_{\rm m}}_{,i} \boldsymbol{\hat{e}_{\rm orth}}
\end{equation}

This step is analogous to Equation (\ref{eq:delta_in_orth}). Then, we calculate the typical orthogonal measurement uncertainty

\begin{equation}
	\delta_{\rm m,orth} = ({\rm cmedian}({\boldsymbol{\Delta_{\rm m,orth}}})).
\end{equation}

Here, ``cmedian" refers to a clipped median. The distributions of the measurement uncertainties are non-Gaussian. They have a high tail for low logarithmic measurement values where the relative uncertainty increases (e.g., see Figure \ref{fig:faberjackson}). To get a clean estimate for the typical measurement uncertainties of the sample, we discard in the calculation high measurement uncertainties, which are larger than the median by 10 times the median absolute deviation with respect to the median. Moreover, parameter regions influenced by the sample selection borders are also discarded (see Appendix \ref{sec:samplecuts}).

Finally, the orthogonal intrinsic scatter is calculated by reducing the orthogonal observed scatter by the typical orthogonal measurement uncertainty

\begin{equation}
	\delta_{\rm in} = \sqrt{\delta_{\rm obs,orth}^2 - \delta_{\rm m,orth}^2}.
\end{equation}

\subsection{Weighted Fit}

Each FP data point $i$ is weighted\footnote{More details about the weighting procedure are described in the {\tt scipy.odr} documentation at \url{https://docs.scipy.org/doc/external/odrpack_guide.pdf}.} using the reciprocal sum of the intrinsic scatter covariance matrix $\boldsymbol{C_{\rm in}}_{,i}$ and the measurement uncertainty covariance matrix $\boldsymbol{C_{\rm m}}_{,i}$ 

\begin{equation}
	\boldsymbol{w}_i \propto (\boldsymbol{C_{\rm in}}_{,i}+\boldsymbol{C_{\rm m}}_{,i})^{-1}
\end{equation}

For the extreme case of no measurement uncertainty, the minimized distances are orthogonal to the plane. For the other extreme case of no intrinsic scatter and only one parameter dominating the uncertainty budget, then the minimized distance is along the direction of the error bar.

\subsection{Luminosity Dependence} \label{sec:luminositydependence}

It has been emphasized that choosing different sample cuts in luminosity and, e.g., velocity dispersion changes the geometric shape of the galaxy distribution \citep{Nigoche-Netro2007,Nigoche-Netro2008,Nigoche-Netro2009,Hyde2009b,Sheth2012}. Galaxies in small, fixed luminosity bins populate asymmetric regions around the FP \citep{Nigoche-Netro2009}. This biases the best-fit slope coefficients. At least some of the discrepancies between various studies \citep{Donofrio2006} can be explained by that effect.

\begin{figure}
	\includegraphics[width=\linewidth]{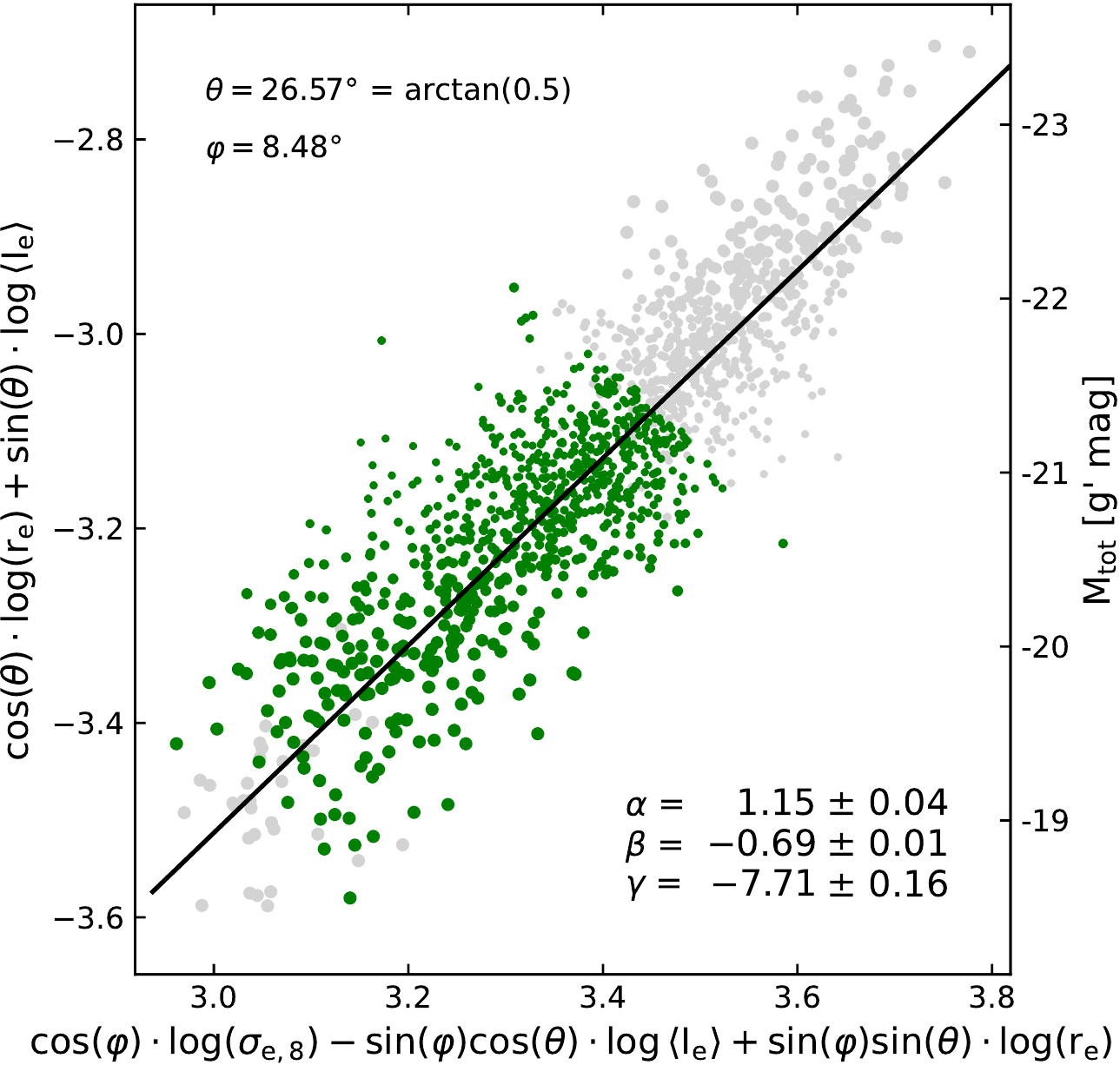}
	\caption{Edge-on projection of the FP (black line) for normal Es. Green data points are included, and gray data points are discarded before the fit. The $y$-axis is proportional to the total galaxy brightness $M_{\rm tot}$. This is achieved by rotating the 3D data point distribution by an angle $\theta=\arctan(0.5)=26.57\degr$ around the $\log(\sigma_{\rm e8})$-axis, such that the resulting vertical axis is proportional to $2\cdot\log(r_{\rm e}) + \log\langle I_{\rm e}\rangle \propto M_{\rm tot}$. Afterward, the points are rotated by an FP slope-dependent angle $\varphi$ around the new vertical axis, such that the plane is viewed edge-on. We calculate $\varphi=8.48\degr$ numerically. Point sizes are proportional to their weight, which is proportional to the inverse of the luminosity function of our sample at the given galaxy luminosity. \label{fig:fundamental_plane_Mtot}}
\end{figure}

In Figure \ref{fig:fundamental_plane_Mtot}, we show a FP projection (black line) for the normal Es, where the $y$-axis is scaled by $2\cdot\log(r_{\rm e}) + \log\langle I_{\rm e}\rangle$. We prove below that this scaling is proportional to the total brightness of the galaxies. We begin by expressing the total apparent brightness of a galaxy $m_{\rm tot}$ as

\begin{align}
	m_{\rm tot} &= -2.5\cdot\log\Big(2\cdot\langle I_{\rm e}\rangle_{\rm obs}\cdot\pi r_{\rm e} [\arcsec]^2\Big) \nonumber \\
	&= -2.5\cdot\Big(\log\langle I_{\rm e}\rangle_{\rm obs} + 2\cdot\log(r_{\rm e} [\arcsec]) + \log(2\pi)\Big). \nonumber \\
\end{align}

Using the distance modulus, we calculate the total absolute brightness $M_{\rm tot}$

\begin{align}
	M_{\rm tot} &=& -2.5\cdot\Big(\log\langle I_{\rm e}\rangle_{\rm obs} + 2\cdot\log(r_{\rm e} [\arcsec]) + \log(2\pi)\Big) \nonumber \\
	&& - (m_{\rm tot}-M_{\rm tot}) \nonumber \\
	&=&  -2.5\cdot\Big(\log\langle I_{\rm e}\rangle_{\rm obs} + 2\cdot\log(r_{\rm e} [{\rm kpc}]) + \log(2\pi)\Big) \nonumber \\
	&&   +5\cdot\log\left(\frac{\rm kpc}{\arcsec}\right) - (m_{\rm tot}-M_{\rm tot}). \nonumber \\
\end{align}

We express the effective radius in physical units using the angular scale ${\rm kpc}/\arcsec=D_{\rm A}[{\rm au}]/1000$ with the angular diameter distance $D_{\rm A}$, as well as the distance modulus $m_{\rm tot}-M_{\rm tot}=5\log(D_{\rm L}[\rm pc])-5$ using the luminosity distance $D_{\rm L}$:

\begin{align}
	M_{\rm tot} = &-2.5\cdot\Big(\log\langle I_{\rm e}\rangle_{\rm obs} + 2\cdot\log(r_{\rm e} [{\rm kpc}]) + \log(2\pi)\Big) \nonumber \\
	&+5\cdot\log(D_{\rm A}[{\rm au}]/1000) - (5\log(D_{\rm L}[{\rm pc}])-5). \nonumber \\
\end{align}

Transforming $D_{\rm L}=D_{\rm A}\cdot(1+z)^2$ and solving gives

\begin{align}
	M_{\rm tot} = &-2.5\cdot\Big(\log\langle I_{\rm e}\rangle_{\rm obs} + \log((1+z)^4) + 2\cdot\log(r_{\rm e} [{\rm kpc}])\Big) \nonumber \\
	&- 38.567 \nonumber \\
\end{align}

For $\log\langle I_{\rm e}\rangle=\log\langle I_{\rm e}\rangle_{\rm obs}+\log((1+z)^4)$, we have already applied the cosmic dimming correction. Finally, we get

\begin{equation}
	M_{\rm tot} = -2.5\cdot\Big(\log\langle I_{\rm e}\rangle + 2\cdot\log(r_{\rm e} [{\rm kpc}])\Big) - 38.567. \label{eq:fpMtot}
\end{equation}

We now project the data point distribution, such that the new $y$-axis is proportional to $M_{\rm tot}$. Therefore, the 3D distribution must be rotated by an angle $\theta$ around the $\log(\sigma_{\rm e8})$-axis, such that the $\log(r_{\rm e})$-axis is projected twice as long as the $\log\langle I_{\rm e}\rangle$-axis. That is the case for $\theta=\arctan(0.5)=26.57\degr$. The resulting distribution must then be rotated again by an angle $\varphi$ around the new axis $\cos(\theta)\cdot\log(r_{\rm e}) + \sin(\theta)\cdot\log\langle I_{\rm e}\rangle$, such that the FP is viewed edge-on. We calculate $\varphi=8.48\degr$ numerically. This depends on the FP slopes $\alpha$ and $\beta$. This procedure is quantitatively expressed by the operation

\begin{equation}
	\boldsymbol{X_{\rm new}} = \boldsymbol{R_{\varphi}} \cdot (\boldsymbol{R_{\theta}} \cdot \boldsymbol{X}),
\end{equation}

where

\begin{align}
	\boldsymbol{R_{\theta}} &=\begin{bmatrix}
		1 & 0 & 0 \\
		0 & \cos(\theta) & -\sin(\theta) \\
		0 & \sin(\theta) & \cos(\theta)
	\end{bmatrix},\\
	\boldsymbol{R_{\varphi}} &=\begin{bmatrix}
		\cos(\varphi) & -\sin(\varphi) & 0 \\
		\sin(\varphi) & \cos(\varphi) & 0 \\
		0 & 0 & 1
	\end{bmatrix},\\
	\boldsymbol{X} &=\begin{bmatrix}
		\log(\sigma_{\rm e8}) \\
		\log\langle I_{\rm e}\rangle \\
		\log(r_{\rm e})
	\end{bmatrix}.\\
\end{align}

Finally, the first and third components of the resulting vector $\boldsymbol{X_{\rm new}}$ are plotted against each other in Figure \ref{fig:fundamental_plane_Mtot}. We caution that this projection should not be done by inserting the FP Equation (\ref{eq:fp}) into the equation for the total brightness (\ref{eq:fpMtot}) (as done by \citealt{Donofrio2008,Gargiulo2009}). This solution is only strictly valid for points that fulfill the FP equation. Real data points scatter from it. Consequently, their distribution would appear horizontally compressed.

Figure \ref{fig:fundamental_plane_Mtot} shows the projection of the FP along total brightness $M_{\rm tot}$ for the normal Es. Green data points are used for fitting the FP. Gray points are discarded (see Appendix \ref{sec:samplecuts}). The luminosity function of our sample drops around $M_{\rm tot}>-20.5\,g'$\,mag. As a consequence, the orthogonal residuals are asymmetric at that region. More data points scatter orthogonally above the relation than below it. We correct for this effect by dividing the weight $w$ of each data point by the value of a kernel density estimate of the luminosity function $\phi$ of our sample \citep{Desroches2007}

\begin{align}
	w\propto \frac{1}{\phi(M_{\rm tot})}
\end{align}

The data points in Figure \ref{fig:fundamental_plane_Mtot} are sized proportionally to their weight. Capping the weight, where less than 20 galaxies are in a brightness bin of 0.1\,$g'$\,mag avoids that few galaxies dominate the fit. That condition is only fulfilled for part of the normal Es sample. The BCG sample is too sparse. Therefore, the BCG weights are only scaled by the covariance matrices, but not by the luminosity function.

Luminosity weighting the normal Es changes the slopes from $\alpha=1.211\pm0.020$ and $\beta=-0.705\pm0.009$ to $\alpha=1.287\pm0.018$ and $\beta=-0.727\pm0.009$.

\subsection{Sample Cuts and Saturation in Velocity Dispersion} \label{sec:samplecuts}

Another asymmetry of the normal Es sample arises from the cut at low velocity dispersions. Our sample is restricted to galaxies with $\sigma_0>70$\,km\,s$^{-1}$ because of the resolution limit of the SDSS spectrograph. That cut is apparent in the FP projection in Figure \ref{fig:fundamental_plane_projected}, top left panel, and in the FJ relation in Figure \ref{fig:faberjackson}. No data points exist below $\log(\sigma_{\rm e8})<1.85$. The problematic aspect is that galaxies are kept, which scatter above the relation, but other galaxies are excluded, which scatter below it. This asymmetry biases the fitted FP slopes \citep{Donofrio2008,Nigoche-Netro2008}, especially because those rare galaxies are attributed to high weight (see Appendix \ref{sec:luminositydependence}).

We correct for this effect by applying an approximately orthogonal cut to the sample. For this, we consider the 3D location on the FP $\boldsymbol{X_{l,i}}$, where the $i$th data point $\boldsymbol{X_i}$ most likely originates from.\footnote{In the absence of measurement errors, i.e., for a purely orthogonal fit, it is the location where the orthogonal vector crosses the plane. For more details, we refer to the {\tt ODRPACK} manual at \url{https://docs.scipy.org/doc/external/odrpack_guide.pdf}.} The velocity dispersion component of $\boldsymbol{X_{l,i}}$ is $\sigma_{{\rm e8},l,i}$. We introduce the condition that $\sigma_{{\rm e8},l}$ must be larger than three times the $\sigma_{\rm e8}$-scatter from the plane above the sample cut

\begin{align}
	\log(\sigma_{{\rm e8},l}) \stackrel{!}{>} & \log(70 {\rm\,km\,s^{-1}}) + \nonumber \\
	& 3\times {\rm std}_i(\log(\sigma_{{\rm e8},l,i}) - \log(\sigma_{{\rm e8},i})).
\end{align}

This cut discards only 32 galaxies. The FP slope $\alpha$ changes only slightly from $\alpha=1.287\pm0.018$ to $\alpha=1.278\pm0.019$.

A physical effect is responsible for the deviations at high velocity dispersions. The FJ relation in Figure $\ref{fig:faberjackson}$ shows that $\sigma_{\rm e8}$ saturates at high values. For those galaxies, dry mergers increase the effective radius, while not affecting the central velocity dispersion (see Section \ref{sec:discussion_offset_2}). This effect is also visible as a curvature in the edge-on projection in Figure \ref{fig:fundamental_plane_Mtot}. We discard 604 galaxies from the normal Es sample by applying, again, an approximately orthogonal cut at $\sigma_{\rm e8}<210$\,km\,s$^{-1}$ and $M_{\rm tot}>-21.41~g~{\rm mag}$, where the FJ relation breaks its slope (see Table \ref{tab:fjparams}):

\begin{align}
	\log(\sigma_{{\rm e8},l}) \stackrel{!}{<} & \log(210 {\rm\,km\,s^{-1}}), \nonumber \\
	M_{{\rm tot},l} \stackrel{!}{>} & -21.41~g~{\rm mag}. \\
\end{align}

The FP slopes change significantly from $\alpha=1.278\pm0.019$ and $\beta=-0.729\pm0.009$ to $\alpha=1.147\pm0.037$ and $\beta=-0.691\pm0.015$.

\subsection{About the Uniqueness of Orthogonality}

We make the assumption that the intrinsic scatter is orthogonal to the plane. However, the term ``orthogonal" is ambiguous when the variables have different units. In our case, only logarithms of variables are considered. Changing the units leaves the slopes unaffected and simply shifts the zero-point.

Still, there are arbitrary choices that affect the best-fit slopes by stretching the axes. For example, this is done by replacing the effective surface intensity $\langle I_{\rm e}\rangle$ by the effective surface brightness $\langle \rm{SB}_{\rm e}\rangle = -2.5 \langle I_{\rm e}\rangle$ (e.g., \citealt{LaBarbera2010b}) or by standardizing the variables.

A physical motivation for the choice of the coordinate system can be found in the virial theorem. In Equation (\ref{eq:fptheory}), we derive $\log(r_{\rm e}) = 2\log(\sigma_0) - \log\langle I_{\rm e}\rangle$. It requires stretching the $\sigma_0$ axis by a factor 2. Consequently, the slopes of the BCG FP change from $\alpha=1.717\pm0.203$ to $\alpha=0.984\pm0.132$ and $\beta=-0.760\pm0.023$ to $\beta=-0.743\pm0.019$. We see that $\alpha$ reacts sensitively. On the other hand, the slopes of the normal E FP are more robust. They change from $\alpha=1.147\pm0.037$ to $\alpha=1.090\pm0.034$ and $\beta=-0.691\pm0.015$ to $\beta=-0.709\pm0.015$. Consequently, the offset between the FPs of normal Es and BCGs (see Section \ref{sec:discussion_offset_2}) remains significant, independently of the choice of coordinate system.

\onecolumngrid

\vspace{12mm}

\section{Tables} \label{sec:apptables}

We present in Tables \ref{tab:struct_normalE} and \ref{tab:struct_BCG} the structural parameters (see Section \ref{sec:photprocedure} and \citealt{Kluge2020,Kluge2023}) and aperture-corrected velocity dispersions (see Sections \ref{sec:templatefitting} and \ref{sec:sigmae8}) of the normal Es and BCGs, respectively. Tables \ref{tab:lines_normalE} and \ref{tab:lines_BCG} list the absorption line strengths (see Section \ref{sec:licks}) and central velocity dispersions of the normal Es and BCGs, respectively.

\begin{deluxetable*}{ccccccccccc}[h]
	\tabletypesize{\footnotesize}
	\tablecaption{Structural Parameters and Velocity Dispersion of Normal Ellipticals. \label{tab:struct_normalE}}
	\tablehead{
		\colhead{ID} & \colhead{$\log(\sigma_{\rm e8})$} & \colhead{$\delta\log(\sigma_{\rm e8})$} & \colhead{$\log(r_{\rm e})$} & \colhead{$\delta\log(r_{\rm e})$} & \colhead{$\langle{\rm SB_e}\rangle$} & \colhead{$\delta\langle{\rm SB_e}\rangle$} & \colhead{$M_{\rm tot}$} & \colhead{$\delta M_{\rm tot}$} & \colhead{$n$} & \colhead{$\delta n$} \\
		\colhead{} & \colhead{(log(km\,s$^{-1}$))} & \colhead{(log(km\,s$^{-1}$))} & \colhead{(log(kpc))} & \colhead{(log(kpc))} & \colhead{($g$ mag arcsec$^{-2}$)} & \colhead{($g$ mag arcsec$^{-2}$)} & \colhead{($g$ mag)} & \colhead{($g$ mag)} & \colhead{} & \colhead{}\\
		(1) & (2) & (3) & (4) & (5) & (6) & (7) & (8) & (9) & (10) & (11)
	}
	\startdata
	$0$ & $2.1576$ & $0.0154$ & $0.922$ & $0.037$ & $21.114$ & $0.256$ & $-22.064$ & $0.079$ & $3.68$ & $0.54$\\
	\ldots & \ldots & \ldots & \ldots & \ldots & \ldots & \ldots & \ldots & \ldots & \ldots & \ldots \\
	$1922$ & $2.1271$ & $0.0165$ & $0.551$ & $0.010$ & $21.119$ & $0.052$ & $-20.205$ & $0.043$ & $2.45$ & $0.23$	
	\enddata
	\tablecomments{Galaxy parameters for the sample of 1420 normal Es. These data points are used for the FP analysis. The columns are: galaxy number in the catalog from \cite{Zhu2010} (1), logarithm of the central stellar velocity dispersion corrected to $r_{\rm e}/8$ (2) and uncertainty (3), logarithm of the effective radius $r_{\rm e}$ (4) and uncertainty (5), average effective surface brightness $\langle{\rm SB_e}\rangle=-2.5\log\langle I_{\rm e}\rangle$ (6) and uncertainty (7), total galaxy brightness (8) and uncertainty (9), and S\'ersic index $n$ (10) and uncertainty (11). The full table is available in machine-readable form.}
\end{deluxetable*}

\begin{deluxetable*}{ccccccccc}[h]
	\tabletypesize{\footnotesize}
	\tablecaption{Structural Parameters and Velocity Dispersion of BCGs. \label{tab:struct_BCG}}
	\tablehead{
		\colhead{ID} & \colhead{$\log(\sigma_{\rm e8})$} & \colhead{$\delta\log(\sigma_{\rm e8})$} & \colhead{$\log(r_{\rm e})$} & \colhead{$\delta\log(r_{\rm e})$} & \colhead{$\langle{\rm SB_e}\rangle$} & \colhead{$\delta\langle{\rm SB_e}\rangle$} & \colhead{$M_{\rm tot}$} & \colhead{$\delta M_{\rm tot}$} \\
		\colhead{} & \colhead{(log(km\,s$^{-1}$))} & \colhead{(log(km\,s$^{-1}$))} & \colhead{(log(kpc))} & \colhead{(log(kpc))} & \colhead{($g'$ mag arcsec$^{-2}$)} & \colhead{($g'$ mag arcsec$^{-2}$)} & \colhead{($g'$ mag)} & \colhead{($g'$ mag)}\\
		(1) & (2) & (3) & (4) & (5) & (6) & (7) & (8) & (9)
	}
	\startdata
	A76 & $2.4019$ & $0.0124$ & $2.088$ & $0.018$ & $25.080$ & $0.066$ & $-23.925$ & $0.008$\\
	\ldots & \ldots & \ldots & \ldots & \ldots & \ldots & \ldots & \ldots & \ldots \\
	MKW4 & $2.4358$ & $0.0131$ & $1.339$ & $0.016$ & $21.978$ & $0.056$ & $-23.284$ & $0.008$
	\enddata
	\tablecomments{Galaxy parameters for the sample of 115 BCGs. These data points are used for the FP analysis. The columns are: host cluster name as in \cite{Kluge2020} (1), logarithm of the central stellar velocity dispersion corrected to $r_{\rm e}/8$ (2) and uncertainty (3), logarithm of the effective radius $r_{\rm e}$ (4) and uncertainty (5), average effective surface brightness $\langle{\rm SB_e}\rangle=-2.5\log\langle I_{\rm e}\rangle$ (6) and uncertainty (7), and total galaxy brightness (8) and uncertainty (9). The full table is available in machine-readable form.}
\end{deluxetable*}

\begin{deluxetable*}{ccccccccccc}
	\tabletypesize{\small}
	\tablecaption{Absorption Line Strengths and Velocity Dispersions of Normal Ellipticals. \label{tab:lines_normalE}}
	\tablehead{
		\colhead{ID} & \colhead{R.A.} & \colhead{Decl.} & \colhead{$z$} & \colhead{$\sigma_0$} & \colhead{$\delta\sigma_0$} & \colhead{H\_beta} & \colhead{$\delta$H\_beta} & \colhead{...} & \colhead{G4300} & \colhead{$\delta$G4300} \\
		\colhead{} & \colhead{(J2000)} & \colhead{(J2000)} & \colhead{} & \colhead{(km\,s$^{-1}$)} & \colhead{(km\,s$^{-1}$)} & \colhead{(\AA)} & \colhead{(\AA)} & \colhead{\ldots} & \colhead{(\AA)} & \colhead{(\AA)} \\
		(1) & (2) & (3) & (4) & (5) & (6) & (7) & (8) & \ldots & (67) & (68)
	}
	\startdata
	$0$ & $0.8839$ & $-10.7447$ & $0.02995$ & $144.6$ & $5.1$ & $2.290$ & $0.117$ & \ldots & $4.936$ & $0.165$ \\
	\ldots & \ldots & \ldots & \ldots & \ldots & \ldots & \ldots & \ldots & \ldots & \ldots & \ldots \\	
	$1924$ & $170.5815$ & $32.9386$ & $0.04400$ & $128.5$ & $5.1$ & $1.713$ & $0.205$ & \ldots & $5.680$ & $0.319$
	\enddata
	\tablecomments{Spectroscopic parameters of 1906 normal Es. The columns are: galaxy number in the catalog from \cite{Zhu2010} (1), R.A. (2), decl. (3), redshift $z$ (4), central stellar velocity dispersion inside $r=1.5\arcsec$ aperture (5) and uncertainty (6), and strenghts and uncertainties of 31 absorption lines ordered as in Figure \ref{fig:linestrengths} (7--68). The full table is available in machine-readable form.}
\end{deluxetable*}

\begin{deluxetable*}{cccccccccccc}
	\tabletypesize{\small}
	\tablecaption{Absorption Line Strengths and Velocity Dispersions of BCGs. \label{tab:lines_BCG}}
	\tablehead{
		\colhead{ID} & \colhead{R.A.} & \colhead{Decl.} & \colhead{Data set} & \colhead{$z$} & \colhead{$\sigma_0$} & \colhead{$\delta\sigma_0$} & \colhead{H\_beta} & \colhead{$\delta$H\_beta} & \colhead{...} & \colhead{G4300} & \colhead{$\delta$G4300} \\
		\colhead{} & \colhead{(J2000)} & \colhead{(J2000)} & \colhead{} & \colhead{} & \colhead{(km\,s$^{-1}$)} & \colhead{(km\,s$^{-1}$)} & \colhead{(\AA)} & \colhead{(\AA)} & \colhead{\ldots} & \colhead{(\AA)} & \colhead{(\AA)} \\
		(1) & (2) & (3) & (4) & (5) & (6) & (7) & (8) & (9) & \ldots & (68) & (69)
	}
	\startdata
	A76 & $9.8596$ & $6.7344$ & W & $0.03774$ & $279.8$ & $6.7$ & $1.789$ & $0.056$ & \ldots & $6.041$ & $0.122$ \\
	\ldots & \ldots & \ldots & \ldots & \ldots & \ldots & \ldots & \ldots & \ldots & \ldots & \ldots & \ldots \\	
	MKW4 & $181.6625$ & $28.1747$ & W & $0.02840$ & $285.6$ & $5.1$ & $1.804$ & $0.082$ & \ldots & $4.960$ & $0.160$
	\enddata
	\tablecomments{Spectroscopic parameters of 115 BCGs. The columns are: host cluster name as in \cite{Kluge2020} (1), R.A. (2), decl. (3), dataset ``W" for WWFI or ``S" for SDSS (4), redshift $z$ (5), central stellar velocity dispersion inside $r=1.5\arcsec$ aperture (6) and uncertainty (7), and strenghts and uncertainties of 31 absorption lines ordered as in Figure \ref{fig:linestrengths} (8--69). The full table is available in machine-readable form.}
\end{deluxetable*}

\end{document}